\documentclass[11pt]{article}
\usepackage{amsmath,amssymb,amsthm}
\usepackage[margin=1in]{geometry}
\usepackage{graphicx} 
\usepackage{cite} 
\usepackage{hyperref} 
\usepackage{float}
\usepackage{caption}
\usepackage{pdfpages}
\usepackage{enumitem}
\usepackage{booktabs}
\usepackage{threeparttable}
\usepackage{pdflscape}
\usepackage{multirow}
\usepackage{array}
\usepackage{multicol}
\usepackage{etoolbox}
\usepackage{subcaption}
\usepackage{svg}
\usepackage[super]{nth}
\usepackage{setspace}
\hypersetup{
  colorlinks = TRUE,
  citecolor  =[HTML]{8B0000},   
  linkcolor  =[HTML]{8B0000},
  urlcolor   =[HTML]{8B0000}
}

\newtheorem{proposition}{Proposition}

\usepackage{natbib}
 \bibpunct[, ]{(}{)}{,}{a}{}{,}%
\newcommand{\initialNRep}{2,059} 
\newcommand{\finalNRep}{1,893} 
\newcommand{\finalPromptNRep}{18,560}

\newcommand{\initialNLogo}{2,067} 
\newcommand{\finalNLogo}{1,857} 
\newcommand{\finalPromptNLogo}{18,425}

\newcommand{\finalTotalUserN}{3,750} 

\title{Prompt Adaptation as a Dynamic Complement in Generative AI Systems\thanks{
  \footnotesize
  We thank Vivian Liu for early contributions to this project. The authors are also grateful to Greg Sun, Ethan Mollick, Nicholas Otis, Solene Delecourt, Rembrand Koning, Daniel Rock, Emma Wiles, Sonia Jaffe, Jake Hofman, and Benjamin Lira Luttges for their feedback. We have benefited from seminar and conference feedback at AFE 2025, the NYU-TAU Workshop on AI, MIT CODE, UC Berkeley, Microsoft, and the World Bank.
  \textbf{Author contributions:} E.J., S.S., and D.H. led, directed, and oversaw the project; J.Z. designed and built the experiment apparatus; B.S.M. led the design of the online experiment flow and Qualtrics survey; J.Z. led the prompt replay process; M.A. led the analysis of prompt text, with contributions from H.T. and E.J.; E.J. and H.T. led all other data analysis and engineering, with contributions from J.Z., D.H., and M.A.; B.S.M., S.S, and D.H. led the writing of the manuscript; and all authors contributed to designing the research and writing the manuscript and supplementary information.
  \textbf{Author declarations:} S.S. is currently an employee of Microsoft. M.A. was formerly a paid intern of and is currently an employee of Microsoft. D.H. was formerly a paid intern and visiting researcher at Microsoft, and is currently a visiting researcher at OpenAI. E.J. was supported by NSF grant \#1745640. The authors gratefully acknowledge research funding from Microsoft. All of the ``target images'' were collected from Unsplash, Reshot, Shopify, Pixabay, or Gratisography; all of these images have licenses for free use for commercial and noncommercial purposes. This study was reviewed by the UC Berkeley Committee for Protection of Human Subjects Protocol 2023-06-16480.}
}

\author{Eaman Jahani \\ University of Maryland 
\and Benjamin S. Manning \\ MIT 
\and Joe Zhang \\ Stanford University 
\and Hong-Yi TuYe \\ MIT
\and Mohammed Alsobay \\ MIT
\and Christos Nicolaides \\ University of Cyprus 
\and Siddharth Suri\thanks{To whom correspondence may be addressed. Email: \url{david.holtz@columbia.edu} or \url{suri@microsoft.com.}} \\ Microsoft Research 
\and David Holtz\footnotemark[2] \\ Columbia Business School}
\date{\today} 

\begin{document}
\vspace{-1in}
\maketitle

\begin{abstract}
\noindent 
As generative AI systems rapidly improve, a key question emerges: how do users adapt to these changes, and when does such adaptation matter for realizing performance gains? Drawing on theories of dynamic capabilities and IT complements, we study \textit{prompt adaptation}---how users adjust their inputs in response to evolving model behavior---using a common experimental design applied to two preregistered tasks with 3,750 total participants who submitted nearly 37,000 prompts. We show that the importance of prompt adaptation depends critically on task structure. In a task with fixed evaluation criteria and an unambiguous goal, user prompt adaptation accounts for roughly half of the performance gains from a model upgrade. In contrast, in an open-ended creative task where the space of acceptable outputs is effectively unbounded and quality is subjective, performance improvements are driven primarily by model capability; prompt adaptation plays a limited role. We further show that automated prompt rewriting cannot generally substitute for human adaptation: when aligned with task objectives, it can modestly improve performance, but when misaligned, it can actively undermine the gains from model improvements. Together, these findings position prompt adaptation as a dynamic complement whose importance depends on task structure and system design, and suggest that without it, a substantial share of the economic value created by advances in generative models may go unrealized.
\end{abstract}

\newpage

\onehalfspacing

\section{Introduction}

Generative AI is being integrated into work practices across the economy \citep{zhang2023generative, bright2024generative}, yielding notable productivity gains in tasks as diverse as software development, writing, and scientific research \citep{brynjolfsson2023generative, dell2023navigating, Noy2023, peng2023impact, Yu2024AIImpact}. Recent research points to even greater potential ahead, demonstrating advances in automating core scientific processes \citep{Manning2024Automated}, including tasks as complex as chemical research and proving mathematical theorems \citep{RomeraParedes2024Math, Boiko2023AutonomousResearch}. The adoption of generative AI is also occurring at an unprecedented pace, with recent research showing that approximately 28\% of U.S. workers are already using generative AI in their jobs—a rate that significantly outpaces early adoption of personal computers and internet technology at comparable points in their diffusion \citep{bright2024generative, bick2024rapid}.

As with many other general-purpose technologies, the effectiveness of generative AI depends not only on the technology itself, but on users’ ability to craft inputs that produce high-quality results. To interact with generative AI systems, users provide written instructions---or \textit{prompts}---that guide the model’s behavior. These prompts can range from simple commands (e.g., ``write a short story about a robot'') to highly detailed specifications tailored to particular outputs (e.g., a series of paragraphs instructing an AI system to implement a complete piece of software). In this way, prompting serves as a complementary skill—one that, like spreadsheet modeling in the early PC era, can determine the productivity impact of a given tool \citep{brynjolfsson2000beyond}.

Prompting has quickly become an area of active research and practice. Scholars have developed taxonomies of prompt engineering techniques \citep{oppenlaender2023}, documented recurring patterns in prompt construction \citep{schulhoff2024prompt}, and examined how developers embed prompts into software systems \citep{liang2024prompts}. Other studies have explored prompting strategies for specific applications, including image generation \citep{donyehiya2023, xie2023} and clinical documentation \citep{yao2024clinicians}. In parallel, practitioners have built prompt libraries, shared tutorials, and developed tools to support prompt design. These developments signal a growing consensus that prompting plays a meaningful role in extracting value from generative AI systems.

Yet despite this consensus, prompting remains understudied as a dynamic practice. Many prompt libraries and tutorials present effective prompts as reusable artifacts. But prompts that work well with one model version may underperform or break entirely with the next \citep{liang2024prompts, meincke2025prompting}. While recent research increasingly views prompting as an adaptive process, empirical evidence remains limited on how these strategies evolve—both as users refine prompts for a single model and as they adjust to model updates—and on how these changes ultimately affect performance. This raises a broader question for individuals and organizations investing in prompting capabilities: Are prompt strategies transferable across model versions, or must they be continually revised to match changing model behavior?

To begin exploring this question, we identify \textit{prompt adaptation}\footnote{We use the phrase ``prompt adaptation'' in reference to changes in user prompting behavior in response to evolving model capabilities. This is distinct from the broader practice of prompt engineering, which includes static best practices, libraries, and templates, among other things.} as a measurable behavioral mechanism through which user-side inputs evolve alongside technical advances. We conceptualize prompt adaptation as a \emph{dynamic complement}---that is, a user capability that adapts in response to changes in a technological system and is critical to realizing the full economic value of system improvements. In contrast to static complements (e.g., fixed training, prompt templates), dynamic complements emerge through situated use with rapid feedback, respond to model-level change, and may be enabled or suppressed by system design.

Generative AI systems support a wide variety of task structures. Some tasks are steering-oriented, where users aim to reach a clearly defined goal and success, can be measured against an objective benchmark (e.g., write code to solve a specific problem). Others are creative, where goals are open-ended, evaluation is subjective, and improvement depends more on exploration and curation than on precision (e.g., make a beautiful picture). Recent work helps to establish these regimes: \citet{vafa2025s} investigate the steerability of models, i.e., how effectively users can reach a specified target, whereas \citet{orwig2024language} and \citet{zhou2024generative} describe generative AI–assisted creation as a co-creative process of idea generation, filtering, and aesthetic evaluation. In organizational settings, steering tasks (or \emph{bounded} tasks, herein) dominate operational workflows such as classification, brand-compliant content generation, or reproducing structured materials where quality can be defined ex ante. Creative tasks (or \emph{unbounded} tasks, herein), by contrast, underpin innovation activities such as logo or campaign design, product ideation, and concept development, where success depends on novelty, appeal, and fit to evolving goals. Understanding how prompting and prompt adaptation operate across these two regimes is essential for organizations deploying generative AI at scale.

To assess the role of prompt adaptation in shaping outcomes across different types of tasks---and to separate its contribution from the direct effects of model improvement---we conducted two pre-registered online experiments with an aggregate sample of \finalTotalUserN{} participants. The first, a bounded task, asks participants to iteratively prompt one of three randomly assigned text-to-image models---DALL-E 2, DALL-E 3, or DALL-E 3 with automated large-language-model–based prompt rewriting---to reproduce a reference image as accurately as possible. It is explicitly designed to capture steering or target-matching behavior. Each participant submitted at least ten prompts and was eligible for a substantial performance-based bonus, incentivizing careful refinement. The second, an unbounded task, captures creative or open-ended behavior: participants design a logo for a hypothetical organization based on a short textual vignette without a specified target image or objective benchmark. Apart from this difference in goal structure, the two tasks, which were conducted on the same custom interface, were identical from the perspective of participants. By comparing outcomes across treatment arms, conducting post-hoc analyses that re-evaluate prompts on alternative models, and contrasting results across the two experiments, we estimate how users adapted their prompts in response to model improvements and how these adaptations contributed to overall performance across both bounded and unbounded settings.

Across both tasks, we find that participants assigned to DALL-E~3 produced significantly better outputs than those assigned to DALL-E~2, but that the sources of these gains depend on task structure. In the bounded image replication task, about half of the gains came from participants \emph{adapting} their prompts to exploit the new model’s capabilities---replaying DALL-E~2 prompts on DALL-E~3 yields only about half the total improvement. By contrast, in the open-ended logo generation task, performance gains were driven primarily by improvements in model capability, with prompt adaptation accounting for a much smaller share of the overall effect (roughly 7\%, compared to over 90\% attributable to the model). Importantly, the prompt adaptation we observe does not appear to be limited to advanced ``prompt engineers'': benefits from prompt refinement after interacting with more capable models are observable across the outcome distribution. Finally, to assess whether automation can substitute for user-side prompt adaptation, we examine automated prompt revision via GPT-4 rewriting. We find that such automation can either erode or modestly enhance performance depending on alignment with the end user’s goals: automated rewriting substantially reduced gains in the replication task, but slightly improved performance in the creative setting. Together, these findings position prompt adaptation as a dynamic complement whose importance depends on task structure and system design, shaping how advances in generative AI translate into realized value.

In terms of related literature and additional theory, our research builds on work in information systems, emphasizing the importance of dynamic, user-driven complements to digital technologies. Research on IT-enabled dynamic capabilities has shown that the value of new systems depends not only on technical infrastructure but on organizations’ ability to reconfigure routines and user behaviors in response to ongoing change \citep{bharadwaj2000resource, joshi2010changing, teece1997dynamic}. Related work on post-adoptive IT use has demonstrated that users often engage only superficially with new systems and that meaningful performance gains tend to emerge only when users experiment with and refine their interaction strategies over time \citep{jasperson2005comprehensive}. Recent research on human-AI collaboration further underscores that interface design and task structure shape the degree to which users can learn from and adapt to model behavior \citep{fugener2022cognitive}. And the concept of co-evolution has been introduced to describe how humans and generative AI systems jointly adapt over time, forming interdependent capabilities that neither could realize alone \citep{bohm2023co-evolution}. 

We also engage with work on general-purpose technologies, which has long emphasized that the productivity gains from technical advances depend on the development of new human and organizational complements \citep{brynjolfsson1993productivity, brynjolfsson2000beyond, brynjolfsson2021productivity, david1990dynamo}. We conceptualize prompt adaptation as a \emph{dynamic} complement that co-evolves with model capability, emerges through use rather than formal training, and whose importance depends on task structure and system design. In this way, prompt adaptation shapes how---and under what conditions---technical improvements translate into downstream economic value.

This paper makes three core contributions. First, we conceptualize prompt adaptation as a dynamic complement to improvements in generative AI models; one through which users actively shape how technical advances translate into realized performance. Leveraging a replay-based analysis enabled by our experimental design, we provide direct empirical evidence for this distinction by separating performance gains attributable to model capability from those arising through user-side adaptation, and show that prompt adaptation can account for a substantial share of realized performance improvements. Second, we identify task structure as a key boundary condition for the importance of prompt adaptation: it is a first-order driver of gains in bounded, steerable tasks with fixed objectives, but plays a much more limited role in open-ended creative tasks where quality is unbounded in nature. Finally, we show that automation intended to simplify prompting is not a neutral substitute for user adaptation. We find that automated prompt rewriting can either undermine or complement performance, depending on its alignment with user goals. Together, these contributions clarify when and how user adaptation amplifies model improvements, positioning prompt adaptation as a dynamic complement whose economic importance depends on task structure and system design.

The remainder of the paper is organized as follows. We begin by presenting a simple conceptual framework that characterizes how output quality evolves with improvements in model capacity and with users’ corresponding adjustments in prompting effort. We then describe our experimental design and the two task settings we study---a bounded, steerable image replication task and an open-ended logo generation task---along with the data and techniques used in our analyses. Next, we present our empirical findings, including a decomposition of the overall effect into components attributable to model improvements versus prompt adaptation, distributional effects across outcome quantiles, and the impact of automated prompt revision. We conclude by synthesizing these results and discussing their implications for organizations adopting generative AI.


\section{Conceptual Framework}
\label{sec:framework}

We first develop a stylized analytical framework to understand how overall output quality depends jointly on model capacity and on users’ prompting behavior. Our goal is not to propose a fully normative model of user–AI interaction, but rather to distinguish improvements directly attributable to the model itself from those arising through user-side prompt adaptation. This distinction motivates our experimental design and yields testable predictions about how performance gains and their distributional impacts change as models improve.

Although our empirical analysis focuses on two tasks---image replication and logo generation---the same conceptual structure applies broadly to settings in which users interact with generative models, including text generation, code assistance, and scientific research. In all such environments, output quality reflects a combination of model capability, user skill, and prompting effort, and total performance improvements can be decomposed into direct (model-driven) and behavioral (prompting-driven) components. The two tasks we study correspond to different regimes of this framework: image replication is a bounded task with clear ceiling (perfectly replicating the image pixel-for-pixel), whereas logo generation is a more open-ended creative task without an optimal outcome. We develop the core logic for the bounded case in the main text and provide complete derivations for both bounded and unbounded formulations in Appendix~\ref{sec:framework_appendix}. Nearly all of the main results carry over to the unbounded case, with minor differences arising from curvature differences across specifications.\footnote{We note that both the bounded exponential and the unbounded logarithmic forms are illustrative rather than general, and are not intended to be literal representations of the specific tasks we study. They adopt simple structures that allow us to decompose total performance gains into a ``model effect'' (improvements in capacity holding prompts fixed) and a ``prompting effect'' (improvements from user adaptation holding capacity fixed). This decomposition is a simplifying abstraction that isolates these mechanisms analytically, though in practice they may interact. The comparative statics of $Q(\theta,s,x)$ also depend on its curvature and on how model capacity and skill interact (e.g., multiplicatively versus additively). We use these tractable forms to convey intuition, not to claim functional-form invariance.}

\subsection{Notation and Problem Setting}

Let $\theta \in (0,1]$ denote the model’s capacity to translate prompts into high-fidelity outputs, $s \in (0,1]$ denote a user’s baseline skill in prompt engineering, and $x \ge 0$ denote the effort the user expends on writing and refining prompts. Each unit of effort incurs a linear cost $c(x)=kx$ for some $k>0$, reflecting the total cognitive and temporal costs of refining one's prompts. We assume $k$ is small ($k\!\approx\!0$), consistent with the low marginal cost of iterative prompting and using generative AI models more generally \citep{shahidi2025coasean}.

We begin with a bounded quality function,
\[
Q(\theta,s,x)=1-e^{-\theta s x},
\]
which is increasing in model capacity, user skill, and prompting effort, but exhibits diminishing marginal returns and has a natural upper bound on performance. We define the user's utility as 
\[
U(\theta,s,x) \;=\; Q(\theta,s,x) \;-\; k\,x.
\]
The user chooses their effort to maximize the utility. We assume \(\theta s > k\)---cost is not prohibitive to effort---which implies there is a unique interior optimum \(x^*(\theta,s)\):
\begin{equation} \label{eq:optimal_effort}
x^*(\theta, s)  \;=\;  \frac{1}{\theta s}\,\ln\Bigl(\frac{\theta s}{k}\Bigr) > 0,
\end{equation}
with an optimal quality
\begin{equation} \label{eq:q_star}
    Q^* \;=\; Q\big(\theta, s, x^*(\theta, s)\big) = 1 - \frac{k}{\theta s}  > 0.
\end{equation}

\noindent It follows immediately from Equation~\ref{eq:q_star} that optimal quality is increasing with both model capacity 
\begin{equation} \label{eq:q_star_cap}
    \frac{\partial Q^*}{\partial \theta} = \frac{k}{\theta^2 s} > 0 
\end{equation}
and user skill
\begin{equation} \label{eq:q_star_skill}
    \frac{\partial Q^*}{\partial s} = \frac{k}{\theta s^2} > 0.
\end{equation}

\subsection{Decomposition into Model and Prompting Effects}

We now consider an upgrade from a model with capacity $\theta_1$ to one with higher capacity $\theta_2$. As model capacity $\theta$ increases, Equation~\ref{eq:optimal_effort} implies that optimal effort $x^*(\theta,s)$ also rises. Thus, an improved model affects performance through two channels: it directly increases output quality for a fixed prompt, and it indirectly encourages users to invest more effort in refining their prompts. We refer to these as the \emph{model effect} and the \emph{prompting effect}, respectively.

Formally, let $x^*(\theta_1,s)$ and $x^*(\theta_2,s)$ denote the user’s optimal prompting effort before and after the upgrade. The total improvement in output quality is
\[
\Delta Q 
= Q\bigl(\theta_2, s, x^*(\theta_2,s)\bigr)
- Q\bigl(\theta_1, s, x^*(\theta_1,s)\bigr).
\]
This total gain can be decomposed as
\begin{equation}
\label{eq:total_decomp}
\begin{aligned}
\Delta Q
&= \underbrace{\Bigl(Q(\theta_2,s,x^*(\theta_1,s))-Q(\theta_1,s,x^*(\theta_1,s))\Bigr)}_{\text{Model effect }} \\
&\quad + \underbrace{\Bigl(Q(\theta_2,s,x^*(\theta_2,s))-Q(\theta_2,s,x^*(\theta_1,s))\Bigr)}_{\text{Prompting effect }}.
\end{aligned}
\end{equation}

\noindent
The first term isolates the gain from upgrading the model while holding the user’s prompt strategy fixed at its pre-upgrade optimum. The second term captures the additional improvement arising from user-side prompt adaptation. Put differently, even if model capacity improves, one may forgo performance gains if their prompts are not updated accordingly.

\subsection{Distributional Impacts of Model Improvements}

A related natural question asks whether access to a more capable model levels the playing field, or widens performance gaps between users. In the language of our framework, how does a transition from a model with capacity $\theta_1$ to one with higher capacity $\theta_2$ affects the distribution of realized output quality. In other words, does access to a more capable model tend to level the playing field, or does it widen performance gaps between users? Under the bounded formulation of the framework analyzed here, $Q(\theta,s,x)=1-e^{-\theta s x}$, improvements in model capacity imply a specific pattern in how performance gains are distributed across users.

\begin{proposition}[Equalizing Effect of Model Improvements]
\label{prop:equalizing_effect}
Under the bounded formulation, improvements in model capacity compress the distribution of output quality across users.
\end{proposition}

\noindent
This aggregate result can be understood by examining the separate distributional implications of the model effect and the prompting effect.

\begin{proposition}[Model Effect: Equalizing Channel] 
\label{prop:model_eq}
Holding prompting effort fixed, an increase in model capacity compresses the outcome distribution: the spread of outcomes narrows as performance at the lower end of the distribution rises proportionally more. 
\end{proposition}

\begin{proposition}[Prompting Effect: Expanding Channel] 
\label{prop:prompt_exp} 
Allowing users to re-optimize their prompting effort expands the outcome distribution: performance differences across users widen as the upper portion of the distribution shifts further upward. 
\end{proposition}

\noindent
The formal proofs and derivations for all three propositions are provided in Appendix~\ref{sec:framework_appendix}.

Taken together, these results imply that model improvements generate two opposing distributional forces. The direct mechanical improvement in capacity—the model effect —is equalizing, while the behavioral response—the prompting effect—is expanding. In aggregate, the equalizing channel dominates, producing the overall compression established in Proposition~\ref{prop:equalizing_effect}.\footnote{In the unbounded but concave logarithmic specification discussed in Appendix~\ref{app:unbounded_quality}, both the model and prompting channels are equalizing rather than opposing. The reversal stems from the additive interaction of capacity and skill, which makes improvements in $\theta$ more uniform across users rather than reinforcing skill differences.} Intuitively, as model capacity increases, performance near the top of the distribution is already close to the performance ceiling and experiences smaller marginal gains, while adaptive prompting raises the performance of the upper tail more sharply. 


\section{Experiment Design and Methods}

To empirically examine whether users do, in fact, adapt their prompts in response to model improvements---and how much this adaptation contributes to overall performance---we conducted two pre-registered online experiments. The first experiment (the \emph{image replication experiment,} hereinafter) was conducted on a sample of \initialNRep{} participants on Prolific between December 12 and December 19, 2023. In this experiment, participants were asked to replicate a target image as closely as possible using a generative AI model. The second experiment (the \emph{logo generation experiment,} hereinafter) was conducted on a sample of \initialNLogo{} participants on Prolific between June 30 and July 1, 2025. In this experiment, participants were asked to create a logo for a hypothetical organization described in a short textual vignette (e.g., a sign for a business or a crest for a university). In both experiments, the goal was to assess how both model capability and prompting behavior influence final outcomes.\footnote{Full pre-registration details for both experiments, including hypotheses and planned analyses, are available in an online repository. All procedures were approved by an institutional review board
(\textbf{protocol information removed to preserve author anonymity}) and participants provided informed consent. We will release anonymized data and replication code upon publication.}

Together, these two experiments are designed to span distinct but empirically important regimes of generative AI use. The image replication experiment captures a bounded, steerable task in which success can be evaluated against a fixed target and output quality has a natural upper bound. This structure enables precise measurement of performance against an objective benchmark. By contrast, the logo generation experiment captures an open-ended, creative task in which goals are underspecified, quality is inherently comparative rather than absolute, and there is no clear performance ceiling. Despite these differences in task structure, both experiments share a common experimental architecture that allows us to decompose the performance gains into model-driven (model effect) and user-driven (prompting effect) components. Studying both settings allows us to assess whether prompt adaptation operates similarly when users are steering models toward a known objective versus when they are iterating within a creative design space, and to examine how the relative contributions of model capability and user adaptation vary across these regimes.

\subsection{Experimental Setting}

In both experiments, each participant was randomly and blindly assigned to one of three model conditions: DALL-E 2, DALL-E 3 (hereafter ``DALL-E 3 (Verbatim)'' or simply ``DALL-E 3''), or DALL-E 3 with automatic prompt revision (``DALL-E 3 (Revised)''). These models differ not only in technical capability but also in whether they apply hidden large language model (LLM)-based modifications to user prompts. In the DALL-E 2 condition, participants interacted with an earlier-generation model that interprets prompts directly without intermediate rewriting. In contrast, the DALL-E 3 API, by default, forwards user prompts to GPT-4 before image generation. This intermediate GPT-4 step rewrites the prompt (typically by adding detail or restructuring language) before passing it to the image model. This behavior is intended to improve image quality, but is not observed by or known to the prompter. In the DALL-E 3 (Revised) condition, we allowed this default behavior to proceed unaltered.This condition allows us to test whether system-side prompt rewriting serves as a substitute for user-side prompt adaptation, or instead changes users’ ability to steer the model. In the DALL-E 3 (Verbatim) condition, we attempted to suppress GPT-4-based prompt rewriting by prepending a hidden system message instructing the model to leave the user’s prompt unchanged. While some rewriting still occurred, the rate and extent of modifications were substantially reduced. In all conditions, participants were unaware that their prompt might be rewritten or modified before image generation. 

Importantly, all generative models in the experiment were memoryless: each prompt was processed independently, with no carryover from prior attempts. Participants were explicitly informed of this property before beginning the task to ensure they understood that previous prompts would not influence subsequent generations. Although this setup departs from typical real-world interfaces that maintain conversational context, it enables us to isolate prompt-level adaptation and to measure users’ iterative learning without confounding from model memory or accumulated dialogue.

We designed a custom experimental interface resembling ChatGPT, which was used for both tasks.
The participant's ``objective'' (either the target image or logo brief, depending on the task) was displayed on the right side of the screen and a scrollable history of prompts and generated images on the left. 
For both experiments, participants had up to 25 minutes to submit at least 10 prompts. They were paid \$4 for completing the task, plus a \$8 bonus (a 200\% increase) if their highest-scoring image (according to the relevant outcome metric below) was in the top 20\% of participants. The median completion time in the image replication experiment was 23.5 minutes, with an a average hourly wage (including bonuses) of about \$13. The median completion time in the logo generation experiment was 26.1 minutes, with an average hourly wage (including bonuses) of about \$11.50. After they had finished submitting prompts, participants completed a demographic survey covering age, gender, education, occupational skills, and self-assessed proficiency in creative writing, programming, and generative AI. We removed from our analyses any participants who did not submit at least 10 prompts, repeated the same prompt five or more times consecutively,\footnote{This pre-registered exclusion criterion was intended to filter out low-effort behavior in which participants repeatedly reused the same prompt in an attempt to complete the task as quickly as possible. While repeating a prompt could also reflect legitimate exploration given the stochastic nature of generative AI outputs, we note that this behavior occurred rarely in our data: the criterion affected 15 participants in the image replication experiment and 32 participants in the logo generation experiment (approximately 0.75–1.75\% of participants).} or failed to complete the post-task survey. This resulted in a final sample of \finalNRep{} participants and \finalPromptNRep{} prompts in the image replication experiment, and a final sample of \finalNLogo{} participants and \finalPromptNLogo{} prompts in the logo generation experiment.\footnote{Although participants were required to submit at least 10 prompts to be included in our analysis, the final prompt count in both experiments is slightly below 10 $\times$ $n_{participants}$ because we excluded some prompts due to technical issues (e.g., safety filter triggers, duplicate attempt numbers) and limited our main analysis to each participant’s first 10 prompts to mitigate potential selection bias. Full details are provided in Appendix~\ref{sec:sample_construction}.} 

\subsection{Image Replication Experiment}

In the image replication experiment, participants were asked to replicate a ``target image'' as closely as possible using the generative model provided to them. Each participant was randomly assigned one of 15 target images in addition to their model assignment. These images were drawn from three broad categories—business and marketing, graphic design, and architectural photography—to represent common use cases for text-to-image generation.\footnote{The selection of 15 target images was guided by the goal of capturing diversity across common text-to-image use cases (business and marketing, graphic design, and architectural photography), as well as by practical considerations identified through pilot testing. While a larger or systematically pretested image set would be desirable, there is no established standard for sampling text-to-image targets, and for a laboratory experiment of this scope some pragmatic choices were necessary. Our aim was to ensure that tasks were neither trivially easy nor prohibitively difficult, and to align broadly with categories commonly represented on repositories such as PromptBase. Target images were assigned using complete randomization across model–image cells, ensuring that image difficulty was orthogonal to model assignment. In addition, our analysis accounts for variation in difficulty by normalizing performance within each target image. Images were sourced from platforms that permit free research use for commercial and noncommercial purposes (e.g., Unsplash, Reshot, Shopify, Pixabay, Gratisography).} All images were sourced from platforms that permit free research use (e.g., Unsplash, Reshot, Shopify, Pixabay, Gratisography). 

\subsubsection{Outcome Definition: Expected CLIP Embedding Cosine Similarity}

The primary outcome in the image replication experiment is the similarity between each participant-generated image and the assigned target image, measured using the cosine similarity of CLIP embeddings \citep{radford2021learning}. CLIP (Contrastive Language–Image Pretraining) is a neural network trained to jointly embed images and text into a shared latent space, such that semantically or visually similar items lie close together. By embedding both the target image and each generated image into this space and computing the cosine similarity between them, we obtain a quantitative measure of how closely the generated image matches the target along both visual and conceptual dimensions.

Because the output of each generative model is stochastic, the same prompt can yield different images across attempts. To account for this variability, we generated 10 images for each prompt and computed their cosine similarity to the target image individually. We then averaged these 10 similarity scores to produce an expected quality score for each prompt—the primary outcome variable used in our analysis. As a robustness check, we replicated all analyses using DreamSim, a recently developed alternative to cosine similarity on CLIP embeddings that is based on perceptual similarity and aligns more closely with human judgments.\footnote{DreamSim \citep{fu2023dreamsim} is designed to better capture human perceptions of image similarity than traditional embedding-based methods. Our results are robust to using DreamSim in place of CLIP cosine similarity. Full DreamSim-based analyses can be found in Appendix~\ref{sec:robustness}.} The two measures were highly correlated, and our findings were consistent across both.

\subsection{Logo Generation Experiment}

In the second experiment, each participant was asked to generate an original logo based on a short set of instructions or ``briefs.'' Each participant was randomly assigned one of 15 briefs in addition to their model assignment. One example is:

\begin{quote}
``The owners of Green Pine Brewery are seeking a logo for their family-owned craft brewery located in Vermont.
\begin{itemize}[itemsep=0pt, topsep=0pt, parsep=0pt]
    \item \textbf{Required visual elements:} pine tree, hops, barley
    \item \textbf{Context for use:} beer labels, merchandise, digital marketing
\end{itemize}
Please create the best possible design based on the client's description and the listed requirements.''
\end{quote}
All briefs followed this exact format: the name of the organization, a single sentence of background information, 2--3 required visual elements, and the anticipated context of use.

The briefs span a range of topics, from a community park sign to a soccer team crest to a university letterhead. We provide the full list of briefs in Appendix~\ref{sec:exp_design}. They were written with three goals in mind: (i) capture a wide variety of realistic use cases, (ii) include a small number of concrete required elements that provide shared reference points for evaluators in an otherwise subjective task, and (iii) allow for open-ended creative possibilities. This third feature contrasts with the image replication experiment, where the best possible performance is well defined (perfectly replicate the target image).

\subsubsection{Outcome Definition: Bradley-Terry Strength}

Because each logo-generation task was open-ended and lacked an objective reference image, we could not assess output quality using embedding-based similarity metrics (e.g., cosine similarity of CLIP representations). Instead, we employed the Bradley–Terry (BT) model \citep{bradley1952rank}, a standard approach for inferring latent quality scores when only relative judgments are available \citep{zermelo1929berechnung, ford1957solution}. At a high level, the BT model is a probabilistic model that estimates a strength parameter (which we refer to hereafter as ``BT $\beta$'') for each item based on the outcomes of many pairwise comparisons.

To operationalize this, within each logo-generation task we enumerated all possible pairwise matchups among the participant-generated images and their replayed counterparts. We then randomly sampled 20\% of these comparisons and obtained outcomes, indicating which image better satisfied the design brief, using a large language model judge. The resulting win-loss matrix was converted into continuous BT $\beta$s by fitting a maximum-likelihood model following a standard iterative procedure \citep{ford1957solution, hunter2004mm}. Random subsampling of the data and automated LLM-based evaluation were necessary, given the scale of our dataset: even after subsampling, we conducted nearly 21 million pairwise comparisons across the 15 logo tasks. A detailed description of the procedure we used to generate BT $\beta$s is found in Appendix \ref{sec:bradley-terry}.\footnote{A natural methodological question is whether BT $beta$s derived from LLM-based pairwise comparisons should be calibrated against human judgments. We therefore attempted to benchmark LLM-based evaluations against human ratings by recruiting human evaluators to assess approximately 19,000 image pairs and estimating corresponding BT $beta$s from those comparisons. Within individual logo briefs, however, we observed substantial disagreement among human raters, with brief-level Krippendorff’s alpha values frequently below 0.6 and in several cases near zero or negative. This degree of disagreement indicates that human-derived BT estimates are themselves unstable at the task level, reflecting the intrinsic ambiguity of the evaluation problem rather than noise in any particular rating source. Given this high level of subjectivity, we rely on LLM-based pairwise comparisons to obtain a consistent and scalable measure of relative quality across designs, while recognizing that quality in this task is inherently noisy and evaluative. Our approach does not treat LLM judgments as ground truth, but rather as a practical mechanism for aggregating comparative assessments in a setting where even expert human agreement is limited.}

Because quality in the logo generation experiment was defined via pairwise comparisons rather than similarity to a fixed reference, we did not resample multiple generations per prompt prior to estimating Bradley–Terry (BT) $\beta$s. In our first experiment, generating multiple images per prompt increased image generation volume (and thus direct study costs) by an order of magnitude while delivering smaller-than-anticipated gains in statistical power. Moreover, because BT $\beta$s are estimated from pairwise comparisons, increasing the number of images by an order of magnitude would have required roughly two orders of magnitude more comparisons, further increasing evaluation costs. Given these considerations, and consistent with our preregistration, we analyzed the logo generation experiment without prompt-level resampling.

\subsection{Replay Analysis for Separating Model and Prompting Effects}

A central goal of our experiment is to distinguish how much of the performance improvement in image replication stems from using a more capable model versus how much comes from users adapting their prompts. Recall that the language of our conceptual framework shows that the total improvement in output quality from upgrading a generative AI model with capacity \(\theta_1\) to a model with higher capacity \(\theta_2\) can be decomposed into two parts: the model effect, i.e., the gain from applying the same prompts to a better model,

\[
M = Q\bigl(\theta_2, s, x^*(\theta_1, s)\bigr) - Q\bigl(\theta_1, s, x^*(\theta_1, s)\bigr),
\]
and the prompting effect, i.e., the additional improvement from adapting prompts to take advantage of the more capable model,
\[
P = Q\bigl(\theta_2, s, x^*(\theta_2, s)\bigr) - Q\bigl(\theta_2, s, x^*(\theta_1, s)\bigr).
\]

To estimate the model and prompting components empirically, we conducted a replay-based analysis using prompts from participants in the DALL-E~2 and DALL-E~3 (Verbatim) conditions. In both experiments, we re-evaluated prompts under the alternative model---replaying DALL-E~2 prompts on DALL-E~3 and DALL-E~3 prompts on DALL-E~2---thereby holding prompt content fixed while varying model capacity. These cross-model evaluations provide empirical analogues of $Q(\theta_2,s,x^*(\theta_1,s))$ and $Q(\theta_1,s,x^*(\theta_2,s))$, respectively, and together with the original prompt–model pairings, allow us to implement the decomposition into model and prompting effects shown in Equation~\ref{eq:total_decomp}.  

In the image replication experiment, replay was conducted several weeks after the initial data collection, so we additionally re-generated outputs for the original prompt--model combinations contemporaneously with replay to guard against model drift \citep{chen2023gptchanging}. In the logo generation experiment, replay occurred shortly after data collection, so this additional step was unnecessary.


\section{Results}

In this section, we present our empirical findings on how increases in model capacity affect user performance across the two tasks. For each task, we first examine whether access to a more capable model improves average output quality, how users adapt their prompts in response, and how much of the overall performance gain can be attributed to model improvements versus prompt adaptation using the replay analyses. Next, we study how these gains reshape the distribution of outcomes across users and interpret these distributional impacts through the lens of our conceptual framework. Finally, we assess the impact of automated prompt rewriting using the DALL-E~3 \emph{Revised} condition.

\subsection{Image Replication Task}

We begin with the image replication task. As discussed above, this task represents a bounded, steerable application of generative AI, in which users aim to match a known target and output quality has a natural upper bound. 

\subsubsection{Overall Impact of Model Upgrades}

\begin{figure*}[t]
    \centering
    \caption{Overall Performance and Prompting Behavior for the Image Replication Task}
    \includegraphics[width=\linewidth]{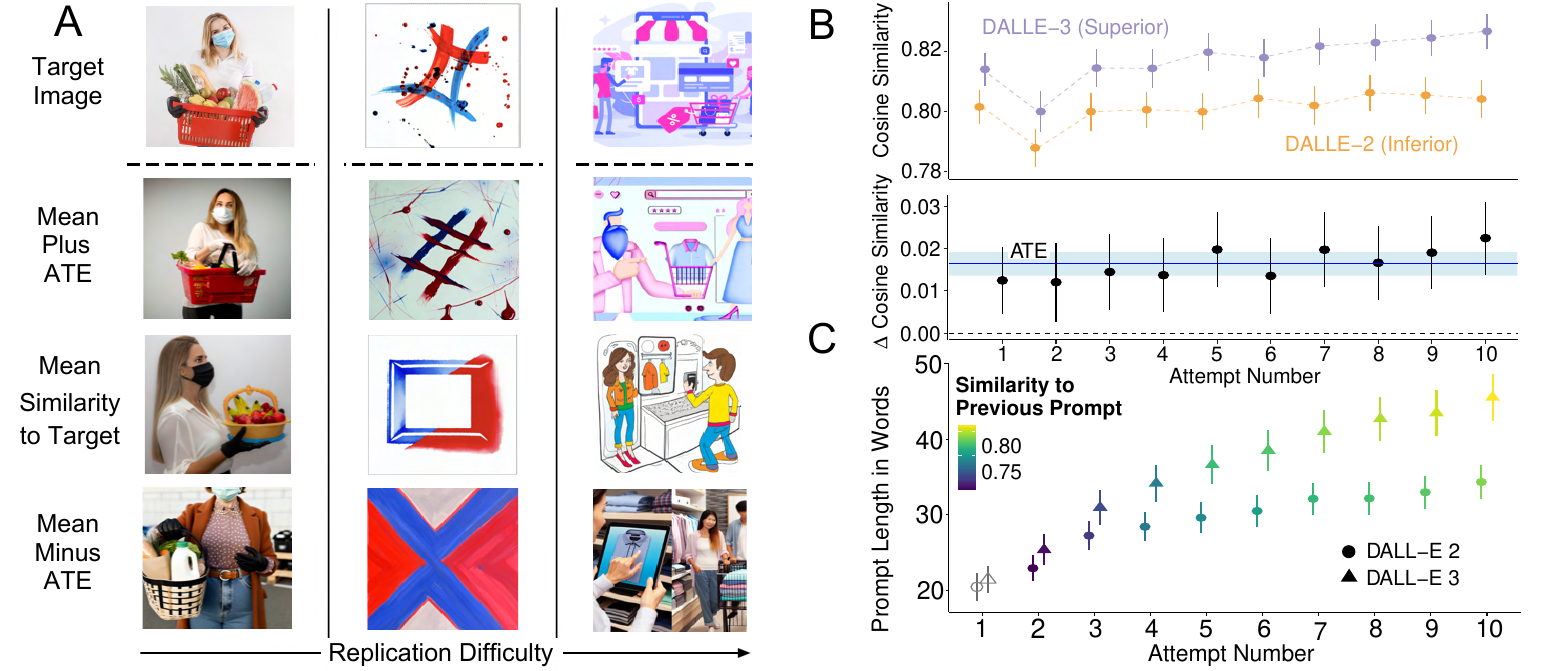}
    \caption*{\small{\textit{\emph{(A)} For three example target images, the middle row shows participant-generated images closest to the mean similarity across all prompts. The rows above and below show images of approximately one average treatment effect (ATE) more or less similar to the target, illustrating the typical performance difference between model conditions. \emph{(B)} Top: average CLIP cosine similarity to the target image by attempt, separately for DALL-E 2 and DALL-E 3 participants. Bottom: the difference between these averages, with the dark blue line indicating the overall ATE and the shaded region showing the 95\% confidence interval. \emph{(C)} Average prompt length by attempt (y-axis), with error bars representing 95\% confidence intervals. Color shading indicates the average textual embedding similarity between each prompt and the participant’s previous prompt, capturing the extent of prompt reuse and refinement over time.}}}
\label{fig:1abc}
\end{figure*}

We first examine whether participants using DALL-E 3 achieve higher performance than those using DALL-E 2, as implied by Equation~\ref{eq:q_star_cap}. Figure~\ref{fig:1abc} summarizes these findings. The top row of Panel A presents three of the target images provided to participants. Below each target, three generated images are shown, drawn from the full sample of participants across both model conditions. The middle row for each target shows the image whose cosine similarity to the target is closest to the mean similarity across all participants. The rows above (below) show images that are approximately one average treatment effect (ATE) more (less) similar to the target than the mean. This visualization provides qualitative intuition for the magnitude of the effect we estimate: the typical difference in fidelity between participants using DALL-E 2 and those using DALL-E 3 (Verbatim). Panel B shows that, across the 10 required prompt attempts, participants assigned to DALL-E 3 (Verbatim) produce images that are, on average, 0.0164 higher in cosine similarity to the target (95\% CI: [0.0104, 0.0224], \(p<10^{-5}\)). This improvement corresponds to roughly 0.19 standard deviations in performance. The gap persists across all attempts; participants using DALL-E 3 start off producing closer matches and maintain that edge through their 10th prompt.

Participants’ dynamic prompting behavior also differs substantially between the two models. As shown in panel C of Figure~\ref{fig:1abc}, those assigned to DALL-E 3 write prompts that are, on average, 24\% longer than those assigned to DALL-E 2, and this gap widens over successive attempts. Moreover, we observe that DALL-E 3 participants are more likely to reuse or refine their previous prompts (indicated by the color scale---see Appendix~\ref{sec:image-data} for details), which suggests a more exploitative approach once they discover the model’s capacity to handle detailed or complex instructions. Analyses of parts of speech confirm that these extra words likely provide additional descriptive information rather than mere filler: the proportion of nouns and adjectives—the two most descriptively informative parts of speech—is nearly identical across model conditions (48\% for DALL-E 3 vs.\ 49\% for DALL-E 2; \(p = 0.215\)), suggesting that the increase in prompt length reflects the addition of semantically rich content rather than unnecessary verbosity. Together, these patterns illustrate prompt adaptation in practice: users progressively supply more information-dense prompts as they naturally learn the model's affordances---despite no explicit instructions to do so.

\subsubsection{Replay Analysis and Decomposition of Effects}

The differences we observe in prompting behavior suggest that users are actively adapting to the capabilities of the model they are assigned. But how much of the overall performance improvement we observe for DALL-E 3 users is due to the model’s enhanced technical capacity, and how much is due to users rewriting their prompts in response to that capacity? To answer this question, we turn to the replay analysis described earlier, which allows us to isolate these two effects empirically.

\begin{figure}[t]
    \centering
    \caption{Replay Analysis and Effect Decomposition for the Image Replication Task.}
    \includegraphics[width=\linewidth]{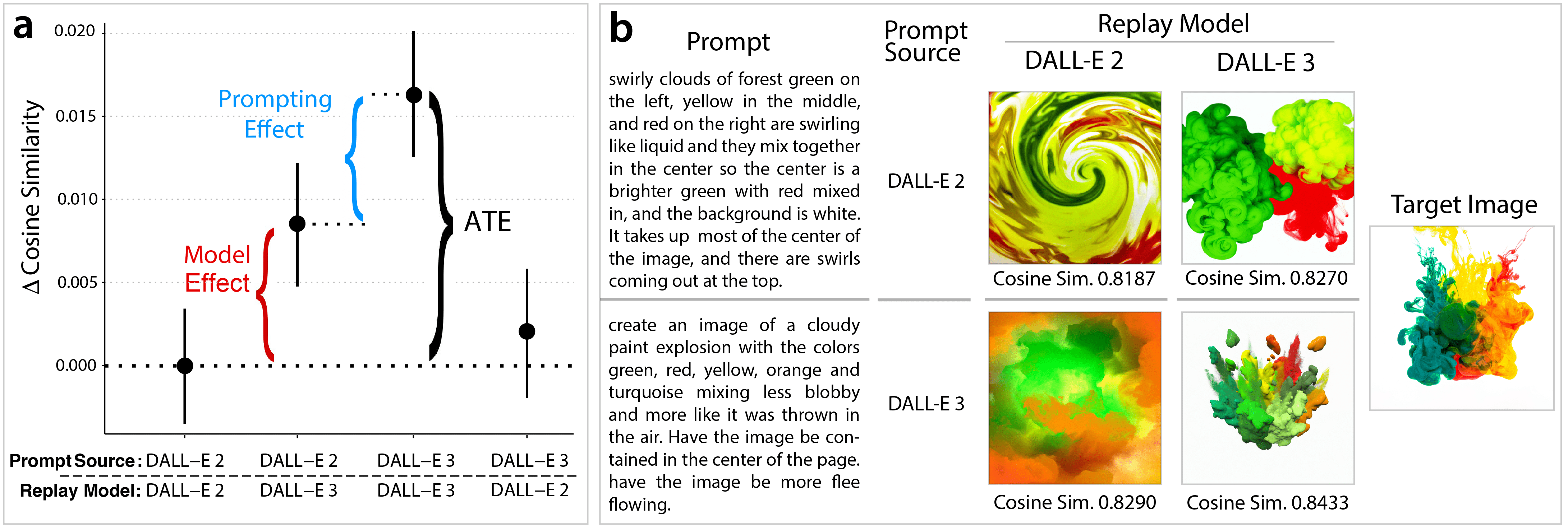}
    \caption*{\small{\textit{\emph{(a)} Average performance of prompts evaluated across four prompt-model combinations. Comparing DALL-E 2 prompts evaluated on DALL-E 2 versus DALL-E 3 isolates the model effect. Comparing reused DALL-E 2 prompts on DALL-E 3 to original DALL-E 3 prompts on DALL-E 3 isolates the prompting effect. Error bars represent 95\% confidence intervals based on bootstrapped standard errors clustered at the participant level. \emph{(b)} A single target image with two submitted prompts: one written by a DALL-E 2 participant (top row) and one by a DALL-E 3 participant (bottom row). Images show how each prompt performs on both models, illustrating the model and prompting effects qualitatively.}}}
\label{fig:fig2}
\end{figure}

Panel A of Figure~\ref{fig:fig2} presents the results. To estimate the model effect, we hold prompts fixed and compare how the same DALL-E~2 prompts perform when evaluated on DALL-E~2 versus when replayed on DALL-E~3 (Verbatim). Because these prompts were written without knowledge of DALL-E 3's capabilities, any improvement reflects the gain from using a more capable model while holding the prompt fixed. We find that performance improves by 0.0084 in cosine similarity when these prompts are evaluated on DALL-E 3 (\(p < 10^{-8}\); bootstrapped standard errors clustered at the participant level), which accounts for approximately 51\% of the total difference in performance between the DALL-E 2 and DALL-E 3 arms.

To estimate the prompting effect, we then compare the performance of these same DALL-E 2 prompts to the performance of prompts originally written by DALL-E 3 participants, both evaluated on DALL-E 3. Because both sets of prompts are evaluated on the same model, any difference reflects the effect of users adapting their prompts to the model’s capabilities. We find that this prompting effect accounts for the remaining $\approx 48$\% of the total improvement\footnote{The total, model, and prompting effects are estimated using three separate two-way fixed-effects models, as explained in Appendix \ref{sec:ate_decomposition}. Because each effect is estimated from a different subset of counterfactual prompt–model comparisons with its own fixed-effects, the resulting model and prompting effect estimates are not guaranteed to sum exactly to the total improvement.}, corresponding to an increase of 0.0079 in cosine similarity (\(p = 0.024\)). Importantly, when we apply prompts written by DALL-E 3 users to DALL-E 2, we observe no performance benefit relative to the original DALL-E 2 prompts (\(\Delta = 0.0020\); \(p = 0.56\)). This asymmetry reinforces the idea that the gains from prompt adaptation depend on the model’s capacity to act on that additional information.

Panel B of Figure~\ref{fig:fig2} illustrates these effects using a single target image. The two rows show different prompts submitted for that target, along with the images they generate when evaluated on each model. In the top row, a prompt originally written by a DALL-E 2 participant yields a higher-fidelity image when replayed on DALL-E 3, demonstrating the improvement in output quality that comes from upgrading the model while holding the prompt fixed. In the bottom row, a prompt written by a DALL-E 3 participant produces a noticeably lower-quality image when rendered by DALL-E 2, underscoring the limits of prompt adaptation when the model lacks sufficient capacity to execute the instructions effectively.

Taken together, these findings offer empirical support for our theoretical claim: prompt adaptation operates as a dynamic complement that users deploy in response to improved model capabilities—and accounts for a substantial share of realized performance gains.

\subsubsection{Distributional Effects}

\begin{table}[t]
\centering
\small
\caption{Distributional Effects of Model Upgrades in the Image Replication Task}
\begin{tabular}{@{}c@{\quad \quad \quad}c@{\quad \quad}c@{}}
\toprule
\textbf{Effect} & \textbf{Estimate (SE)} & \textbf{\emph{p}-value}\\
\midrule
\addlinespace
Total & 0.022 (0.003) & {$<$ $10^{-5}$}\\
Total $\times$ Outcome Quantile (Percentile) & $-$0.00012 (0.00005) & 0.015\\
\addlinespace
\midrule
\addlinespace
Model & 0.011 (0.002) & {$<$ $10^{-5}$}\\
Model $\times$ Outcome Quantile (Percentile) & $-$0.00006 (0.00003) & 0.021\\
\addlinespace
\midrule
\addlinespace
Prompting & 0.011 (0.003) & {$<$ $10^{-5}$}\\
Prompting $\times$ Outcome Quantile (Percentile) & $-$0.00006 (0.00005) & 0.244\\
\addlinespace
\bottomrule
\end{tabular}
\caption*{\small{\textit{Negative interactions with outcome quantile indicate that gains from model upgrades are larger at lower points in the realized outcome distribution, implying compression of performance outcomes. The total and model effects exhibit statistically significant compression, while we do not detect a statistically significant distributional pattern for the prompting effect. Note: the un-interacted total effect coefficient gives the QTE estimate at the 0th percentile.}}}
\label{table:qte}
\end{table}

Next, we examine how model improvements and prompt adaptation reshape the distribution of performance outcomes in the image replication task using a quantile treatment effects (QTE) perspective. 

Table~\ref{table:qte} reports estimates of the total, model, and prompting effects at different points of the realized outcome distribution, with outcome quantiles indexed by percentile. The total effect exhibits a negative and statistically significant interaction with outcome quantile (\(-0.000115\), \(p = 0.0152\)), indicating that performance gains from upgrading from DALL-E~2 to DALL-E~3 are larger at lower points in the outcome distribution. This pattern is consistent with the framework’s prediction that model improvements compress outcomes in bounded tasks. This compression is driven primarily by the model effect. Holding prompts fixed, the interaction between the model effect and outcome quantile is also negative and statistically significant (\(-0.000059\), \(p = 0.0210\)), consistent with diminishing returns near the performance ceiling. While the bounded formulation predicts that prompt adaptation exerts an expanding force on the outcome distribution, we do not detect a statistically significant interaction between the prompting effect and outcome quantile in this setting (\(-0.000056\), \(p = 0.2444\)). This result does not rule out the presence of an expanding prompting channel, but implies that any such effect is modest in magnitude relative to the dominant equalizing model effect. 

\subsubsection{Prompt Revision}

Finally, we examine whether automated prompt rewriting can serve as a substitute for human-driven prompt adaptation in the image replication task. In the DALL-E~3 (Revised) condition, user prompts were silently rewritten by GPT-4 before being submitted to the image model. Although participants in this condition still outperformed those using DALL-E~2 (\(\Delta = 0.0069\); \(p = 0.042\)), they achieved substantially lower performance than participants using DALL-E~3 without rewriting. On average, automated prompt revision reduced the benefit of DALL-E~3 by 58\% (95\% CI: [40\%, 76\%]). Qualitative inspection of revised prompts suggests that the rewriting process often introduced additional details or altered phrasing in ways that were misaligned with participants’ objective of replicating a specific target image as faithfully as possible. In this bounded, steerable task, such modifications reduced users’ ability to precisely control the model’s output and partially offset the performance gains achieved through direct user-side prompt adaptation.

\subsection{Logo Generation Task}

We next turn to the logo generation task. This logo generation task, however, represents an open-ended, creative application of generative AI---such as design or ideation---in which success is evaluated comparatively rather than against a known target and output quality does not have a natural upper bound. Our empirical analyses mirror that of the image replication experiment.

\begin{figure*}[t]
    \centering
    \caption{Overall Performance and Prompting Behavior for the Logo Generation Task}
    \includegraphics[width=\linewidth]{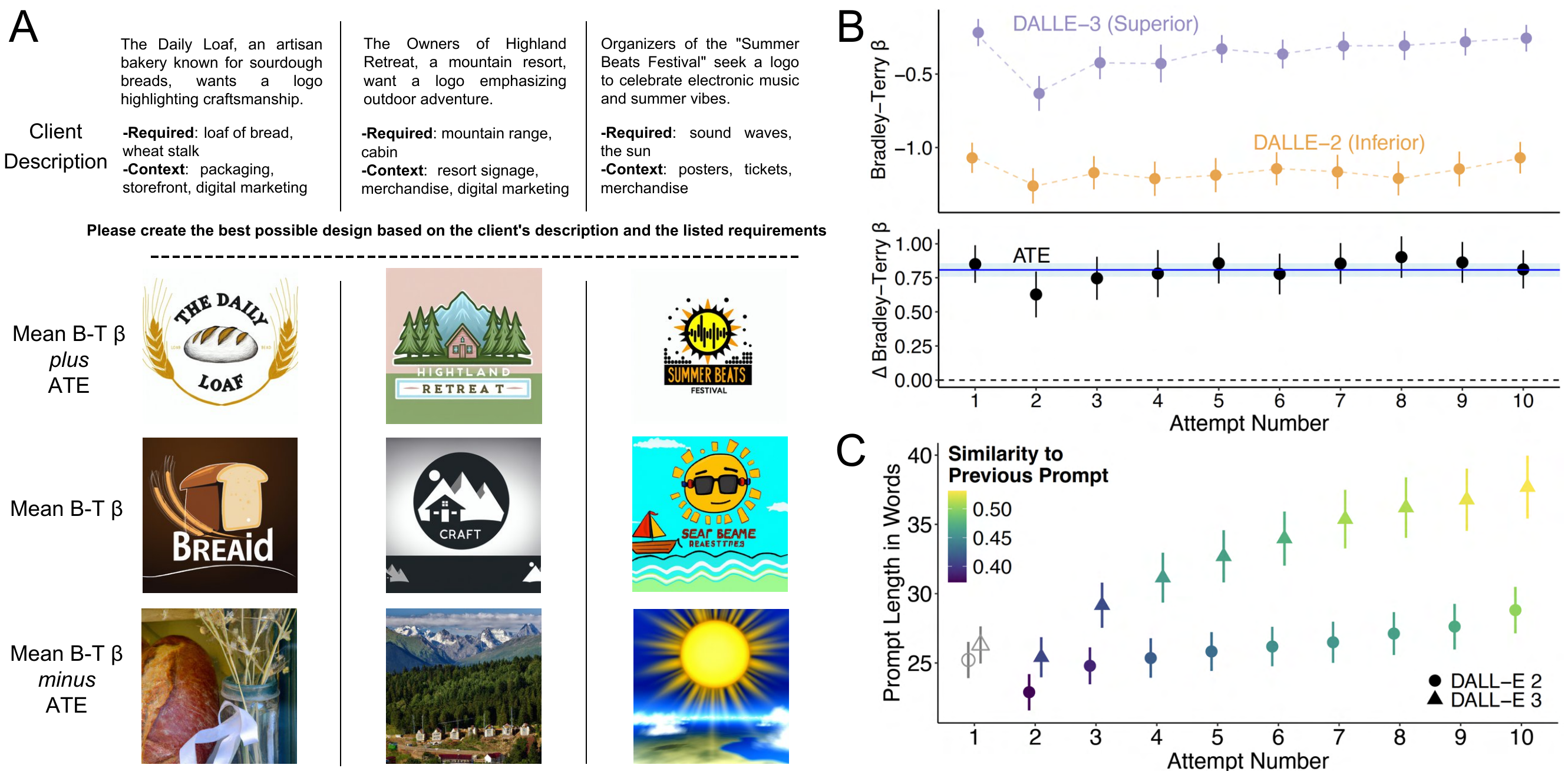}
    \vspace{-0.5em}
    \caption*{\small{\textit{\emph{(A)} For three example target images, the middle row shows participant-generated images closest to the mean Bradley-Terry $\beta$ across all prompts. The rows above and below show images of approximately one average treatment effect (ATE) higher or lower quality, illustrating the typical performance difference between model conditions. \emph{(B)} Top: average Bradley-Terry $\beta$, separately for DALL-E 2 and DALL-E 3 participants. Bottom: the difference between these averages, with the dark blue line indicating the overall ATE and the shaded region showing the 95\% confidence interval. \emph{(C)} Average prompt length by attempt (y-axis), with error bars representing 95\% confidence intervals. Color shading indicates the average textual similarity between each prompt and the participant’s previous prompt, capturing the extent of prompt reuse and refinement over time.}}}
\label{fig:1abcnew}
\end{figure*}

\subsubsection{Overall Impact of Model Upgrades}

We begin by examining whether access to a more capable model improves performance in the logo generation task. Participants assigned to DALL-E 3 produce substantially higher-quality logos than those assigned to DALL-E 2, with an estimated average treatment effect of 0.8075 in Bradley–Terry $\beta$ (95\% CI: [0.7120, 0.9030], $p < 10^{-10}$), corresponding to an improvement of approximately 0.57 standard deviations. Panel A of Figure \ref{fig:1abcnew} provides qualitative intuition for the magnitude of this effect by showing representative logos at the mean, as well as approximately one average treatment effect above and below the mean. Panel B of Figure \ref{fig:1abcnew} shows how this performance gap unfolds over successive prompting attempts. As in the image replication task, the advantage of the more capable model appears early and persists, although it does not appear to diverge across attempts. Finally, Panel C of Figure~\ref{fig:1abcnew} shows similar differences to the image replication task in prompting behavior across model conditions, with participants using DALL-E~3 prompts that are approximately 24.7\% longer on average and exhibiting greater iteration-to-iteration similarity, consistent with more exploitative refinement. The proportion of nouns and adjectives is also once again statistically indistinguishable across model conditions (43\% for DALL-E 3 vs. 42\% for DALL-E 2; $p$ = 0.12)

Together, these patterns show that in an open-ended, creative task without a fixed target or performance ceiling, increased model capability leads to large and persistent performance gains.
These gains are accompanied by systematic changes in user prompting behavior as users adapt their prompts in response to the model’s perceived capabilities. These results largely mirror those from the image replication task.

\subsubsection{Replay Analysis and Decomposition of Effects}

\begin{figure}[t]
    \centering
    \caption{Replay Analysis and Effect Decomposition for the Logo Generation Task.}
    \includegraphics[width=\linewidth]{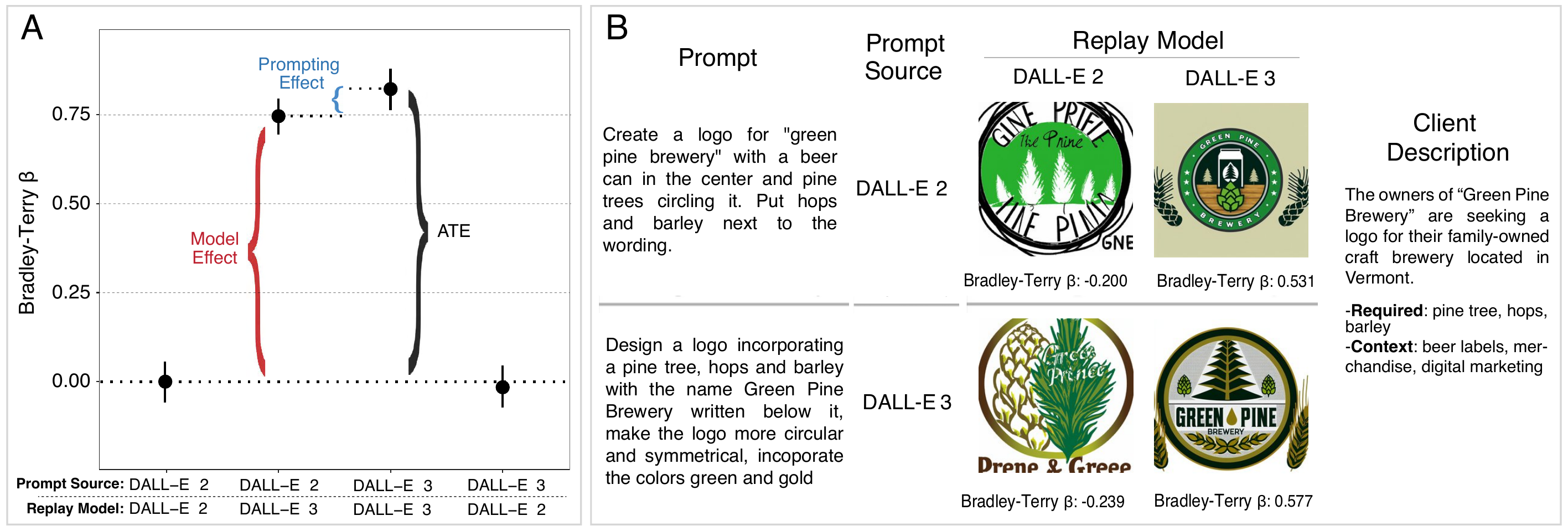}
    \caption*{\small{\textit{\emph{(a)} Average performance of prompts evaluated across four prompt-model combinations. Comparing DALL-E 2 prompts evaluated on DALL-E 2 versus DALL-E 3 isolates the model effect. Comparing reused DALL-E 2 prompts on DALL-E 3 to original DALL-E 3 prompts on DALL-E 3 isolates the prompting effect. Error bars represent 95\% confidence intervals based on bootstrapped standard errors clustered at the participant level. \emph{(b)} A single client description with two submitted prompts: one written by a DALL-E 2 participant (top row) and one by a DALL-E 3 participant (bottom row). Images show how each prompt performs on both models, illustrating the model and prompting effects qualitatively.}}}
\label{fig:fig2new}
\end{figure}

Next, we apply the same replay-based decomposition used in the image replication task to separate the overall performance gains in the logo generation experiment into components attributable to model improvements and to user prompt adaptation. Panel~A of Figure~\ref{fig:fig2new} presents the resulting decomposition.

The overall average treatment effect of upgrading from DALL-E~2 to DALL-E~3 is $0.8075$ in Bradley--Terry $\beta$. In contrast to the image replication task, the vast majority of this gain is attributable to the model effect: replaying prompts written by DALL-E~2 participants on DALL-E~3 increases performance by $0.748$ ($p < 10^{-10}$), accounting for approximately $92.7\%$ of the total effect. The corresponding prompting effect is small in magnitude ($0.0577$), statistically indistinguishable from zero ($p = 0.101$), and accounts for only $7.2\%$ of the total effect. Consistent with this pattern, replaying prompts written for DALL-E 3 on DALL-E 2 yields no performance improvement ($\Delta \beta = -0.0125$, $p = 0.62$).

Panel B of Figure~\ref{fig:fig2new} provides a qualitative illustration of this pattern for a representative logo brief. Although the prompt written by the DALL-E~3 participant is longer and more detailed—consistent with the prompt adaptation patterns shown in Panel~C of Figure~\ref{fig:1abcnew}—it does not perform substantially better on DALL-E~3 than the prompt written by the DALL-E~2 participant. Instead, the primary performance difference arises from evaluating either prompt on the more capable model. While this comparison is illustrative rather than representative, it highlights how prompt adaptation in this task does not translate into large additional gains beyond those delivered by the model itself.

Taken together, these results indicate that in the logo generation task, realized performance gains are driven primarily by improvements in model capability, with prompt adaptation playing a comparatively limited role.

\subsubsection{Distributional Effects}

\begin{table}[t]
\centering
\small
\caption{Distributional Effects of Model Upgrades in the Logo Generation Task}
\begin{tabular}{@{}c@{\quad \quad \quad}c@{\quad \quad}c@{}}
\toprule
\textbf{Effect} & \textbf{Estimate (SE)} & \textbf{\emph{p}-value}\\
\midrule
\addlinespace
Total & 1.135 (0.050) & {$<$ $10^{-5}$}\\
Total $\times$ Outcome Quantile (Percentile) & $-$0.0065 (0.0008) & {$<$ $10^{-5}$}\\
\addlinespace
\midrule
\addlinespace
Model & 1.012 (0.035) & {$<$ $10^{-5}$}\\
Model $\times$ Outcome Quantile (Percentile) & $-$0.0053 (0.0005) & {$<$ $10^{-5}$}\\
\addlinespace
\midrule
\addlinespace
Prompting & 0.122 (0.050) & 0.014\\
Prompting $\times$ Outcome Quantile (Percentile) & $-$0.0013 (0.0008) & 0.117\\
\addlinespace
\bottomrule
\end{tabular}
\caption*{\small{\textit{Negative interactions with outcome quantile indicate that gains from model upgrades are larger at lower points in the realized outcome distribution, implying compression of performance outcomes. As in the image replication task, distributional compression is driven by the model effect, while we do not detect a statistically significant distributional pattern for the prompting effect. Note: the un-interacted total effect coefficient gives the QTE estimate at the 0th percentile.}}}
\label{table:qte_logo}
\end{table}

We also apply the same QTE approach used in the image replication task to examine how the model upgrade reshapes the distribution of performance outcomes in the logo generation experiment. Table~\ref{table:qte_logo} reports estimates of the total, model, and prompting effects at different points of the realized outcome distribution, with outcome quantiles indexed by percentile.

As was the case in the image replication task, the total effect of upgrading from DALL-E~2 to DALL-E~3 exhibits a negative and statistically significant interaction with outcome quantile, indicating that performance gains are larger at lower points in the outcome distribution and that the upgrade compresses the distribution of outcomes. This pattern is again driven by the model effect: holding prompts fixed, the interaction between the model effect and outcome quantile is negative and statistically significant. By contrast, we do not detect a statistically significant interaction between the prompting effect and outcome quantile. Although this lack of statistical significance does not rule out the compressive impact of prompt adaptation predicted by the unbounded formulation of the model, it places an upper bound on the magnitude of any such effect in this setting.

Taken together, these results indicate that, as in the image replication task, distributional compression in the logo generation experiment arises primarily from improvements in model capability, with prompt adaptation playing a comparatively limited role in shaping the distribution of outcomes.

\subsubsection{Prompt Revision}

We conclude analyses of this experiment by examining the effects of automated prompt rewriting in the logo generation task. As in the image replication task, participants assigned to DALL-E~3 with automatic prompt revision outperform those using DALL-E~2, with an average treatment effect of $0.8337$ in BT $\beta$ ($p < 10^{-10}$). However, in contrast to the image replication task, automated prompt rewriting does not reduce performance relative to DALL-E~3 without rewriting; instead, it is associated with a $3.2\%$ increase in BT $\beta$ on average (95\% CI: $[-5.6\%, 12.1\%]$).

One possible interpretation is that in an open-ended, creative task, the specific prompt rewriting used here, which primarily embellishes and elaborates user inputs, can enrich prompts in ways that facilitate exploration, introduce useful variation, or highlight design elements that improve perceived quality. While we do not directly observe these mechanisms, the positive effect of rewriting in this task is consistent with the idea that automated prompt modification can complement user prompting when it is aligned with users’ goals and the structure of the task.


\section{Discussion}

In this paper, we show that realized performance gains from generative AI upgrades reflect an interaction between model capability, user adaptation, and task structure. In a bounded, steerable task with an objective target (image replication), improvements in model quality translated into performance gains through two channels: roughly half of the improvement arose mechanically from the stronger model, while the other half reflected users adapting their prompts to better exploit the model’s capabilities. In this setting, prompt adaptation is a substantive and economically meaningful complement to model improvement. By contrast, in an open-ended creative task without a natural performance ceiling (logo generation), performance gains were driven primarily by improvements in model capability. Although users again modified their prompting behavior when interacting with the stronger model, replay analyses indicate that prompt adaptation accounted for only a small share of the overall improvement. Thus, we argue that prompt adaptation functions as a dynamic complement to generative AI models, with task structure emerging as a key boundary condition: it plays a first-order role when success is defined against a fixed objective, but a more limited role when quality is open-ended and comparative.


A related question we address is whether prompt adaptation must occur through direct user experimentation or whether it can be partially automated by the system itself. In practice, many generative AI systems implement automated prompt rewriting to reduce user effort, implicitly positioning automation as a substitute for user-driven adaptation. Our findings suggest that the effectiveness of such automation depends critically on task structure and alignment with user goals. In the bounded image replication task, the DALL-E~3 endpoint’s automated prompt rewriting reduced realized gains, consistent with misalignment between the rewriting logic and users’ objective of faithfully reproducing a target image. By contrast, in the open-ended logo generation task, the same rewriting mechanism modestly improved performance. Together, these findings underscore the importance of careful design in automated prompt rewriting systems. When system prompts are well aligned with user goals and task structure, automated rewriting can improve performance. but poorly aligned system prompts, especially when hidden, can degrade performance and/or limit users’ ability to discover or exploit emergent but useful ways of interacting with a generative AI system. Such complexities are highlighted by related work, such as \citet{yao2024clinicians}, which studies automated prompt optimization in the context of clinical documentation.

More generally, our results both align with and help clarify themes in the literature on human–AI collaboration and the economics of technology, particularly the role of dynamic complements such as prompt adaptation that co-evolve with technical change \citep{bohm2023co-evolution}. As generative AI models advance, often substantially from month to month, organizations that fail to adapt their prompting strategies in task settings where precision and steerability matter may forgo a meaningful share of the economic value these upgrades make possible. This co-evolutionary pattern is not unique to AI. It echoes dynamics observed in earlier general-purpose technologies, where technical improvements often yielded modest returns until complementary skills and practices evolved to match \citep{brynjolfsson1993productivity, david1990dynamo}. However, the pace of generative AI advancement introduces a distinctive challenge: the window for adaptation is far shorter. Organizations that treat prompting as a one-time investment rather than an ongoing capability risk failing to capture the full value of model upgrades---echoing the productivity paradox observed when complementary assets lag behind technological potential \citep{brynjolfsson2000beyond, brynjolfsson2021productivity}. Our findings complement this stream of work by offering direct empirical evidence that a specific dynamic complement---user adaptation, specifically through prompt refinement---can account for a substantial share of realized performance gains from model upgrades in task settings where precision and steerability matter.

Seen from a broader organizational perspective, these findings also speak to long-standing work on IT-enabled dynamic capabilities \citep{bharadwaj2000resource, joshi2010changing, teece1997dynamic} and post-adoptive IT use \citep{jasperson2005comprehensive}. Prompt adaptation, as we observe it, is not grounded in formal training or specialized skill. Participants in our study, drawn from a general population rather than a specialized group of prompt engineers, improved performance through trial-and-error within a single session. This contrasts with earlier IT transitions, where complementary capabilities often required extensive training \citep{attewell1992technology, von2006democratizing}. The relative accessibility of prompt adaptation highlights one potential pathway toward more broadly distributed productivity gains, provided that users are supported by scaffolds and interfaces that enable iterative refinement \citep{rogers2003diffusion}. At the same time, this accessibility introduces risk. Over-optimizing for a particular model version may reduce users’ ability to adapt as systems evolve. This behavioral lock-in resembles challenges in architectural innovation, where tightly coupled routines inhibit flexibility when key system components change \citep{henderson1990architectural}. Supporting long-term adaptation may require not just prompt training, but workflows and learning mechanisms that encourage ongoing experimentation.

Our study has several limitations. First, our experimental interface was intentionally memoryless: each prompt was processed independently, and participants were informed that prior interactions would not influence subsequent generations. This simplification was necessary to isolate prompt-level adaptation in a controlled setting, but it departs from the conversational context that characterizes many real-world AI systems; our estimates therefore characterize performance and adaptation conditional on a memoryless interaction structure. Second, our focus was restricted to a single transition (from DALL-E 2 to DALL-E 3) and one type of generative AI (text-to-image generation). Although our conceptual framework suggests these mechanisms may generalize to other domains, further research is needed to assess how these dynamics play out in text generation, programming, scientific research, and other high-stakes settings. Third, we observe short-run adaptation behavior in a controlled setting, whereas longer-term learning dynamics, organizational feedback structures, or team-based workflows may shape prompting strategies differently in real-world environments. Finally, while our replay analysis helps isolate the contribution of prompt adaptation relative to model improvements, it does not fully disentangle the causal effects of specific prompt modifications (e.g., lengthening, lexical substitution, structural rephrasing), particularly given the path-dependent nature of user adaptation.

Building on these limitations, several promising research directions emerge. One important next step is to examine prompt adaptation in conversational, memory-enabled settings, where prior interactions persist and can be leveraged for cumulative learning by both users and models, and to study how this co-evolution impacts performance. Relatedly, future studies might explore prompt adaptation longitudinally, observing how users develop durable heuristics that generalize across domains and respond to shifting model behavior over time. Another line of inquiry involves organizational-level complements to prompting, such as shared prompt repositories, collaborative refinement practices, or analytics dashboards that surface effective patterns. A third area concerns interface design—specifically, how features like auto-complete, prompt scoring, or real-time feedback influence users’ ability to experiment and adapt. Finally, researchers might explore how organizations balance standardization (which streamlines processes) with adaptability in prompting workflows, in order to avoid creating rigid routines that lag behind technical advances.

For researchers and practitioners interested in the economics of AI, our findings reinforce the idea that complements play a central role in shaping performance. In some task settings, prompt adaptation can account for as much as half of the realized performance gains from model upgrades, underscoring that skill development and technological evolution must be treated as interdependent elements of innovation trajectories \citep{arthur2009nature, dosi1982technological}. As generative AI continues to advance, organizations that invest in \textit{adaptive}, not static, complementary skills will be better positioned to realize its full value.

\bibliographystyle{informs2014}
\bibliography{prompting}

\newpage


\appendix
\renewcommand{\thefigure}{A.\arabic{figure}}
\renewcommand{\thetable}{A.\arabic{table}}
\setcounter{figure}{0}
\setcounter{table}{0}
\setcounter{proposition}{0}

\section{Conceptual Framework and Derivations} \label{sec:framework_appendix}.

This appendix develops the formal derivations and proofs underlying the conceptual framework presented in the main text. Whereas the main text emphasizes intuition and empirical implications, the goal here is to make the underlying mechanics transparent.

We first analyze a \emph{bounded quality function}, where output quality is capped at 1, and then extend the analysis to unbounded tasks. The exposition below focuses on the bounded case; the unbounded case follows by analogy.

\subsection{Model Setup}

Consider a task where users attempt to replicate a target object (e.g., an image) using a generative model. Perfect replication yields a quality of 1.  
Define the \emph{quality function}
\[
Q(\theta,s,x)=1-\exp[-\theta s x],
\]
where $\theta\in(0,1]$ denotes \emph{model capacity}; $s\in(0,1]$ denotes \emph{user skill} at writing prompts; and $x\ge0$ denotes the \emph{user’s effort} in generating prompts.

\noindent This specification has two desirable properties:
\begin{enumerate}
    \item \emph{Boundedness.} Even for very large $x$, $Q<1$; quality cannot exceed perfection.
    \item \emph{Diminishing returns.} The marginal gain from effort, $Q_x=\theta s e^{-\theta s x}$, declines as $x$ increases.
\end{enumerate}

\noindent Prompting entails a linear \emph{cost function}
\[
c(x)=k x,\qquad k>0,
\]
interpreted as the time or cognitive cost per unit of effort.  
The user’s utility from a given combination of model capacity, skill, and effort is
\[
U(\theta,s,x)=1-e^{-\theta s x}-k x.
\]

\subsection{User Optimization}

The user chooses effort $x\ge0$ to maximize $U(\theta,s,x)$. The first-order condition is:
\[
\theta s e^{-\theta s x}=k
\;\;\Rightarrow\;\;
x^*(\theta,s)=\frac{1}{\theta s}\ln\!\Bigl(\frac{\theta s}{k}\Bigr).
\]

\noindent An interior optimum requires $\theta s>k$, ensuring that $x^(\theta,s)>0$ and the logarithm is well defined. Intuitively, the user’s effective productivity $\theta s$ must exceed the marginal cost of effort $k$. This inequality is readily satisfied when prompting costs are small (i.e., $k\approx0$).

Substituting the optimal effort into the quality function yields the user’s maximum attainable quality:
\begin{equation}
Q^*(\theta,s)=1-\frac{k}{\theta s}.
\end{equation}
This closed-form expression will serve as the basis for all subsequent comparative statics and distributional analysis.

\subsection{Comparative Statics}

The partial derivatives of $Q^*$ are straightforward:
\[
\frac{\partial Q^*}{\partial \theta}=\frac{k}{\theta^{2}s}>0,\qquad
\frac{\partial Q^*}{\partial s}=\frac{k}{\theta s^{2}}>0,\qquad
\frac{\partial Q^*}{\partial k}=-\frac{1}{\theta s}<0.
\]

Hence, output quality increases in both model capacity and user skill, and decreases with the cost of effort. Improvements in model quality or prompting skill therefore unambiguously raise output quality, whereas higher marginal costs of effort reduce it.

\subsection{Distributional Impact of Model Improvements (Overall Equalizing Effect)}
\label{sec:model_improvement_heterogeneity_si}

We now examine how increases in model capacity $\theta$ reshape the \emph{distribution of optimal quality} $Q^*$ across users. Because $Q^*$ is strictly increasing in skill $s$, dispersion in $Q^*$ reflects variation in skill, since $Q^*$ is a monotone transformation of $s$ (and hence of $G_s$). Let $G_{Q^*}(\pi)$ denote the inverse CDF of $Q^*$—that is, the quality achieved by the user at the $\pi$-th percentile of the quality distribution—and let $G_s(\pi)$ denote the inverse CDF of the skill distribution. Then:
\[
G_{Q^*}(\pi)=1-\frac{k}{\theta G_s(\pi)}.
\]

\begin{proposition}[Equalizing Effect of Model Improvements]
For any strictly increasing skill distribution $G_s(\pi)$, we have
\[
\frac{\partial^2 G_{Q^*}(\pi)}{\partial\theta\,\partial\pi}<0.
\]
\end{proposition}

\begin{proof}
Differentiating once gives
\[
\frac{\partial G_{Q^*}(\pi)}{\partial\theta}
   =\frac{k}{\theta^{2}G_s(\pi)}>0,
\]
so all users benefit from higher $\theta$. Differentiating again with respect to $\pi$,
\[
\frac{\partial^2 G_{Q^*}(\pi)}{\partial\theta\,\partial\pi}
   =-\frac{k}{\theta^{2}G_s^{2}(\pi)}\,G_s'(\pi)<0,
\]
because $G_s'(\pi)>0$. \qedhere
\end{proof}

Model improvements therefore \textit{compress} the distribution of output quality, reducing dispersion across users. Intuitively, greater model capacity raises quality throughout the distribution but shifts the lower portion upward more strongly in relative terms, since outputs at the top are already close to the performance ceiling.

\subsection{Decomposing Total Improvements in Quality} \label{sec:app_decompose}

The total change in optimal quality with respect to model capacity, $\theta$, can be expressed as a combination of a direct effect of the model itself and an indirect effect that operates through re-optimization of effort. To distinguish between these two direct and indirect channels, we introduce a \emph{counterfactual quality function} that treats the model capacity determining effort and the model capacity entering the quality function as distinct parameters.
\[
Q^{cf}(\theta_2,s,x^*(\theta_1,s))
  = 1 - \exp[-\theta_2 s\,x^*(\theta_1,s)]
  = 1 - \Bigl(\frac{k}{\theta_1 s}\Bigr)^{\!\theta_2/\theta_1},
\]
which represents the quality achieved when a user employs the effort that would be optimal for model~$\theta_1$ while using a model with capacity~$\theta_2$. The \emph{model} and \emph{prompting} effects are obtained by differentiating this counterfactual quality function with respect to $\theta_2$ and $\theta_1$, respectively, and evaluating the results at $\theta_1=\theta_2=\theta$:
\begin{align*}
M(\theta,s) &= \frac{\partial Q^{cf}(\theta_2,s,x^*(\theta_1,s))}{\partial\theta_2}
  = \frac{k}{\theta^{2}s}\ln\!\Bigl(\frac{\theta s}{k}\Bigr) \\
P(\theta,s) &= \frac{\partial Q^{cf}(\theta_2,s,x^*(\theta_1,s))}{\partial\theta_1}
  = \frac{k}{\theta^{2}s}\!\left[1-\ln\!\Bigl(\frac{\theta s}{k}\Bigr)\right].
\end{align*}
$M(\theta,s)$ captures the \emph{direct mechanical improvement} in quality when model capacity increases but effort is held fixed. In contrast, $P(\theta,s)$, captures the \emph{behavioral response}—the gain that results from users adjusting their effort once the model improves. By construction,
\[
\frac{dQ^*(\theta,s)}{d\theta} = M(\theta,s) + P(\theta,s).
\]

\subsection{Distributional Implications of the Model and Prompting Effects}

Although the decomposition in Appendix~\ref{sec:app_decompose} holds pointwise for each user, the \emph{distributional} consequences of model and prompting effects require characterizing how these effects reshape the distribution of outcomes.
Let $G_{Q^{cf}}(\pi)$ denote the inverse CDF of the counterfactual quality:
\[
G_{Q^{cf}}(\pi)=1-\Bigl(\frac{k}{\theta_1 G_s(\pi)}\Bigr)^{\!\frac{\theta_2}{\theta_1}}.
\]

\begin{proposition}[Model Effect: Equalizing Channel] Holding prompting effort fixed, an increase in model capacity compresses the distribution of output quality: the spread of outcomes narrows as performance at the lower end of the distribution rises proportionally more. More formally, 
\[
\frac{\partial^2 G_{Q^{cf}}(\pi)}{\partial\theta_2\,\partial\pi}<0.
\]
\end{proposition}

\begin{proof}
Differentiating with respect to $\theta_2$,
\[
\frac{\partial G_{Q^{cf}}(\pi)}{\partial\theta_2}
 =-\Bigl(\frac{k}{\theta_1 G_s(\pi)}\Bigr)^{\!\frac{\theta_2}{\theta_1}}
  \frac{1}{\theta_1}
  \ln\!\Bigl(\frac{k}{\theta_1 G_s(\pi)}\Bigr).
\]
Differentiating again and setting $\theta_1=\theta_2=\theta$,
\[
\frac{\partial^2 G_{Q^{cf}}(\pi)}{\partial\theta_2\,\partial\pi}
 =\frac{k}{\theta^{2}G_s^{2}(\pi)}G_s'(\pi)
   \Bigl[\ln\!\Bigl(\frac{k}{\theta G_s(\pi)}\Bigr)+1\Bigr].
\]
For sufficiently small $k$, the bracketed term is negative, implying that the entire expression is negative. \qedhere
\end{proof}

\begin{proposition}[Prompting Effect: Expanding Channel] Allowing users to re-optimize their prompting effort expands the distribution of output quality: performance differences across users widen as the upper portion of the distribution shifts further upward. More formally, 
\[
\frac{\partial^2 G_{Q^{cf}}(\pi)}{\partial\theta_1\,\partial\pi}>0.
\]
\end{proposition}

\begin{proof}
Differentiating with respect to $\theta_1$,
\[
\frac{\partial G_{Q^{cf}}(\pi)}{\partial\theta_1}
 =\frac{\theta_2}{\theta_1^{2}}
  \Bigl(\frac{k}{\theta_1 G_s(\pi)}\Bigr)^{\!\frac{\theta_2}{\theta_1}}
  \Bigl[\ln\!\Bigl(\frac{k}{\theta_1 G_s(\pi)}\Bigr)+1\Bigr].
\]
Differentiating again and setting $\theta_1=\theta_2=\theta$,
\[
\frac{\partial^2 G_{Q^{cf}}(\pi)}{\partial\theta_1\,\partial\pi}
 =-\frac{1}{\theta}
   \Bigl(\frac{k}{\theta G_s(\pi)}\Bigr)
   \frac{G_s'(\pi)}{G_s(\pi)}
   \Bigl[\ln\!\Bigl(\frac{k}{\theta G_s(\pi)}\Bigr)+2\Bigr].
\]
For sufficiently small $k$, the term in brackets is negative, so the derivative is positive. \qedhere
\end{proof}

\subsection{Summary of Distributional Impacts}

In this exponential-bounded specification, the distributional consequences of all effects are analytically unambiguous:

\begin{center}
\begin{tabular}{lll}
\toprule
\textbf{Effect} & \textbf{Cross-partial sign} & \textbf{Distributional interpretation} \\
\midrule
Model effect & $<0$ & Equalizing (Compression) \\
Prompting effect & $>0$ & Expanding (Dispersion) \\
Total effect & $<0$ & Equalizing (Compression) \\
\bottomrule
\end{tabular}
\end{center}

\noindent The overall equalizing pattern established in Section~\ref{sec:model_improvement_heterogeneity_si} thus arises because the compressive model channel dominates the dispersive prompting channel in magnitude. Intuitively, as model capacity increases, gains in output quality are largest at the lower end of the distribution, where outcomes remain far from the quality ceiling. Improvements at the upper end are more limited, leading to an overall compression in the distribution of performance. Behavioral adaptation through prompting has the opposite tendency, stretching the upper tail, but not enough to offset the dominant equalizing shift.

\subsection{Unbounded Quality} \label{app:unbounded_quality}
We now consider a concave but unbounded quality specification,
\[
Q(\theta,s,x) = \log\!\big(1 + (\theta + s)x\big)
\]
with the same linear cost $c(x)=k x$. This formulation preserves monotonicity and concavity in $x$, $\theta$, and $s$, but relaxes the upper bound on attainable quality. The user’s utility function is therefore
\begin{align*}
U(\theta,s,x) = \log\!\big(1 + (\theta + s)x\big) - kx
\end{align*}

\paragraph{User optimization.}
The first-order condition is
\[
\frac{\theta + s}{1 + (\theta + s)x} = k
\]
Assuming an interior solution ($\theta + s > k$), the optimal effort, quality, and utility are
\begin{align*}
x^*(\theta,s) &= \frac{1}{k} - \frac{1}{\theta + s}\\[4pt]
Q^*(\theta,s) &= \log\!\left(\frac{\theta + s}{k}\right)\\[4pt]
U^*(\theta,s) &= \log\!\left(\frac{\theta + s}{k}\right) - 1 + \frac{k}{\theta + s}
\end{align*}

\paragraph{Comparative statics.}
Optimal quality increases with both model capacity and user skill, and decreases with prompting cost:
\begin{align*}
\frac{\partial Q^*}{\partial \theta} &= \frac{1}{\theta + s} > 0 \qquad
\frac{\partial Q^*}{\partial s} = \frac{1}{\theta + s} > 0 \qquad
\frac{\partial Q^*}{\partial k} = -\frac{1}{k} < 0
\end{align*}

\paragraph{Distributional Impact of Model Improvements (Overall Equalizing Effect)}
We now analyze how increases in model capacity $\theta$ reshape the \emph{distribution of optimal quality} $Q^*$ across users.
Let $G_{Q^*}(\pi)$ denote the inverse CDF of $Q^*$ and $G_s(\pi)$ the inverse CDF of skill. Then:
\begin{align*}
G_{Q^*}(\pi)=\log\!\left(\frac{\theta + G_s(\pi)}{k}\right)
\end{align*}

\begin{proposition}[Equalizing Effect of Model Improvements, Unbounded Case]
For any strictly increasing skill distribution $G_s(\pi)$,
\[
\frac{\partial^2 G_{Q^*}(\pi)}{\partial\theta\,\partial\pi}
   =-\frac{G_s'(\pi)}{(\theta+G_s(\pi))^2}<0
\]
\end{proposition}

\begin{proof}
Differentiating once gives
\[
\frac{\partial G_{Q^*}(\pi)}{\partial\theta}
   =\frac{1}{\theta+G_s(\pi)}>0
\]
so all users gain from higher $\theta$. Differentiating again with respect to $\pi$,
\begin{align*}
\frac{\partial^2 G_{Q^*}(\pi)}{\partial\theta\,\partial\pi}
   =-\frac{G_s'(\pi)}{(\theta+G_s(\pi))^2}<0
\end{align*}
\end{proof}

Model improvements therefore \emph{compress} the quality distribution, raising performance throughout the distribution but proportionally more at lower quantiles---mirroring the equalizing effect found in the bounded model.

\paragraph{Counterfactual quality and decomposition.}
As in the bounded model, we decompose total quality improvements into direct (model) and indirect (prompting) components using a counterfactual formulation:
\begin{align*}
    Q^{cf}(\theta_2,s,x^*(\theta_1,s))
      = \log\!\Big(1+(\theta_2+s)\,x^*(\theta_1,s)\Big)
\end{align*}
where $\theta_2$ governs the model capacity in the quality function and $\theta_1$ determines the user’s optimal effort. Differentiating with respect to $\theta_2$ and $\theta_1$ and evaluating at $\theta_1=\theta_2=\theta$ gives
\begin{align*}
    M(\theta,s)=\frac{(\theta+s)-k}{(\theta+s)^2},\qquad
    P(\theta,s)=\frac{k}{(\theta+s)^2}
\end{align*}
so that the total effect of model improvement equals the sum of model and prompting effects: $\frac{dQ^*(\theta,s)}{d\theta}=M(\theta,s)+P(\theta,s)$ as before.

\paragraph{Distributional implications.}
Let $G_{Q^{cf}}(\pi)$ denote the inverse CDF of counterfactual quality. It can be shown that $Q^{cf}$ is strictly increasing in $s$ if $k < \theta_1 + s$, a condition needed for an interior solution. Thus:

\begin{align*}
G_{Q^{cf}}(\pi;\theta_1,\theta_2)
= \log\!\Bigg( 1 + \frac{\theta_2 + G_s(\pi)}{k}
- \frac{\theta_2 + G_s(\pi)}{\theta_1 + G_s(\pi)} \Bigg)
\end{align*}

\begin{proposition}[Model Effect: Equalizing Channel, Unbounded Case]
Holding prompting effort fixed, an increase in model capacity compresses the distribution of output quality: the spread of outcomes narrows as performance at the lower end of the distribution rises proportionally more. Formally,
\begin{align*}
\frac{\partial^2 G_{Q^{cf}}(\pi)}{\partial\theta_2\,\partial\pi}
   = \frac{2k - \big( \theta + G_s(\pi) \big)}{(\theta+G_s(\pi))^3} G_s'(\pi) <0
\end{align*}
which is negative for sufficiently small $k$.
\end{proposition}

\begin{proposition}[Prompting Effect: Equalizing Channel, Unbounded Case]
Allowing users to re-optimize their prompting effort compresses the distribution of output quality: the spread of outcomes narrows as performance at the lower end of the distribution rises proportionally more. Formally,  
\begin{align*}
\frac{\partial^2 G_{Q^{cf}}(\pi)}{\partial\theta_1\,\partial\pi}
   = -\frac{2k}{(\theta+G_s(\pi))^3} G_s'(\pi) < 0.
\end{align*}
\end{proposition}

\newpage

\section{Experiment Design Details} \label{sec:exp_design}

In this appendix, we provide details on the design of both experiments reported in the main text.

\subsection{Randomization}

For both experiments, we randomized participants across two dimensions: the target task (target image for the image replication experiment and target brief for the logo generation experiment) and the text-to-image generative AI model that participants had access to. We randomized participants across both dimensions simultaneously using complete randomization, generating 45 possible target item-model cells per experiment. We conducted a balance check after the conclusion of each experiment with a $\chi^2$ test across all cells. For the image replication experiment, we obtained $\chi^2 = 7.056, \; df = 44, \; p > 0.999$, thus we cannot reject the null hypothesis that the proportions are equal across all 45 groups. For the logo generation experiment, we obtained $\chi^2 = 5.383, \; df = 44, \; p > 0.999$, thus we cannot reject the null hypothesis that the proportions are equal across all 45 groups. In both experiments, participants were unaware of the underlying randomization.

\subsubsection{Balance Checks}

Table \ref{tab:pval_balance_main_combined} reports covariate balance checks for the image replication experiment across prior GPT usage, gender, and age. We find no statistically significant evidence of imbalance across any of these dimensions. Table \ref{tab:pval_balance_combined} reports the corresponding balance checks for the logo generation experiment. While one pairwise comparison yields a $p$-value below 0.05, this is not unexpected given that we conduct 18 balance tests across the two experiments.

Importantly, the statistically significant comparison arises in a contrast involving the DALL-E 3 (Revised) arm. As described in the main text, this arm is included primarily as a secondary, diagnostic condition to assess the consequences of automated system-side prompt rewriting, and it is not central to the paper’s primary treatment contrasts or replay-based identification strategy. We do not rely on direct comparisons between DALL-E 3 (Revised) and DALL-E 3 (Verbatim) for our headline results, and inspection of the balance table indicates that the deviation is driven by the DALL-E 3 (Revised) group rather than broad differences across the core arms of interest.

Finally, to assess whether this pattern could have been induced by sample restrictions or data processing, in Table \ref{tab:pval_balance_combined_full} we re-conduct the logo-generation balance checks on the full sample prior to applying any exclusions. The same low $p$-value persists, indicating that it is not created by filtering and is consistent with random variation in assignment rather than some systematic bias in our treatment of the data.

\begin{landscape}
\begin{table}[p]
\centering
\caption{Covariate Balance Across All Three Arms (Image Replication Experiment)}
\label{tab:pval_balance_main_combined}
\begin{threeparttable}
\footnotesize
\begin{tabular}{lccccccccc}
\toprule
\textbf{Covariate / Level} & \textbf{N(D2)} & \textbf{Prop(D2)} & \textbf{N(D3r)} & \textbf{Prop(D3r)} & \textbf{N(D3v)} & \textbf{Prop(D3v)} & \textbf{D2 vs. D3r} & \textbf{D2 vs. D3v} & \textbf{D3r vs. D3v} \\
 &  &  &  &  &  &  & \textbf{p-val, $\chi^2$(df)} & \textbf{p-val, $\chi^2$(df)} & \textbf{p-val, $\chi^2$(df)} \\
\midrule
\textbf{GPT Usage} &  &  &  &  &  &  & \multirow{5}{*}{\begin{tabular}[c]{@{}c@{}} \\ p = 0.745, \\ $\chi^2$ = 1.948 (4)\end{tabular}} & \multirow{5}{*}{\begin{tabular}[c]{@{}c@{}} \\ p = 0.654, \\ $\chi^2$ = 2.449 (4)\end{tabular}} & \multirow{5}{*}{\begin{tabular}[c]{@{}c@{}} \\ p = 0.629, \\ $\chi^2$ = 2.587 (4)\end{tabular}} \\
\hspace{1em}None & 215 & 0.329 & 186 & 0.305 & 186 & 0.296 &  &  &  \\
\hspace{1em}Little & 190 & 0.291 & 178 & 0.292 & 195 & 0.311 &  &  &  \\
\hspace{1em}Some & 179 & 0.274 & 173 & 0.284 & 170 & 0.271 &  &  &  \\
\hspace{1em}A lot & 70 & 0.110 & 72 & 0.118 & 76 & 0.121 &  &  &  \\
\hspace{1em}NA/Missing & 1 & 0.002 & 0 & 0.000 & 2 & 0.003 &  &  &  \\
\addlinespace
\textbf{Gender} &  &  &  &  &  &  & \multirow{4}{*}{\begin{tabular}[c]{@{}c@{}} \\ p = 0.727, \\ $\chi^2$ = 1.309 (3)\end{tabular}} & \multirow{4}{*}{\begin{tabular}[c]{@{}c@{}} \\ p = 0.319, \\ $\chi^2$ = 3.517 (3)\end{tabular}} & \multirow{4}{*}{\begin{tabular}[c]{@{}c@{}} \\ p = 0.231, \\ $\chi^2$ = 4.297 (3)\end{tabular}} \\
\hspace{1em}Female & 354 & 0.541 & 332 & 0.545 & 341 & 0.544 &  &  &  \\
\hspace{1em}Male & 279 & 0.427 & 261 & 0.428 & 271 & 0.432 &  &  &  \\
\hspace{1em}Other or prefer not to say & 21 & 0.032 & 16 & 0.026 & 13 & 0.021 &  &  &  \\
\hspace{1em}NA/Missing & 1 & 0.002 & 0 & 0.000 & 4 & 0.006 &  &  &  \\
\addlinespace
\textbf{Age Group} &  &  &  &  &  &  & \multirow{6}{*}{\begin{tabular}[c]{@{}c@{}} \\ p = 0.847, \\ $\chi^2$ = 2.013 (5)\end{tabular}} & \multirow{6}{*}{\begin{tabular}[c]{@{}c@{}} \\ p = 0.649, \\ $\chi^2$ = 3.334 (5)\end{tabular}} & \multirow{6}{*}{\begin{tabular}[c]{@{}c@{}} \\ p = 0.250, \\ $\chi^2$ = 6.631 (5)\end{tabular}} \\
\hspace{1em}[18,26) & 112 & 0.171 & 109 & 0.179 & 99 & 0.158 &  &  &  \\
\hspace{1em}[26,32) & 136 & 0.208 & 122 & 0.200 & 122 & 0.195 &  &  &  \\
\hspace{1em}[32,39) & 139 & 0.213 & 132 & 0.217 & 147 & 0.235 &  &  &  \\
\hspace{1em}[39,49) & 127 & 0.194 & 127 & 0.208 & 119 & 0.190 &  &  &  \\
\hspace{1em}[49,83] & 140 & 0.214 & 119 & 0.195 & 138 & 0.220 &  &  &  \\
\hspace{1em}NA/Missing & 1 & 0.002 & 0 & 0.000 & 4 & 0.006 &  &  &  \\
\bottomrule
\end{tabular}
\vspace{0.5em}
\begin{tablenotes}[flushleft]
\fontsize{9pt}{11pt}\selectfont
\item[] \parbox{0.95\linewidth}{Sample sizes: DALL-E 2 (D2) = 655, DALL-E 3 Revised (D3r) = 609, DALL-E 3 Verbatim (D3v) = 629. \\
$N(\cdot)$ columns report the count of observations in each treatment arm belonging to that category, $\text{Prop}(\cdot)$ gives the proportion of respondents in that category for the arm. Chi-squared test columns show p-values and $\chi^2$ test statistics with degrees of freedom for each pairwise comparison.}
\end{tablenotes}
\end{threeparttable}
\end{table}
\end{landscape}
\newpage

\begin{landscape}
\begin{table}[p]
\centering
\caption{Covariate Balance Across All Three Arms (Logo Generation Experiment)}
\label{tab:pval_balance_combined}
\begin{threeparttable}
\footnotesize
\begin{tabular}{lccccccccc}
\toprule
\textbf{Covariate / Level} & \textbf{N(D2)} & \textbf{Prop(D2)} & \textbf{N(D3r)} & \textbf{Prop(D3r)} & \textbf{N(D3v)} & \textbf{Prop(D3v)} & \textbf{D2 vs. D3r} & \textbf{D2 vs. D3v} & \textbf{D3r vs. D3v} \\
 &  &  &  &  &  &  & \textbf{p-val, $\chi^2$(df)} & \textbf{p-val, $\chi^2$(df)} & \textbf{p-val, $\chi^2$(df)} \\
\midrule
\textbf{GPT Usage} &  &  &  &  &  &  & \multirow{5}{*}{\begin{tabular}[c]{@{}c@{}} \\ p = 0.098, \\ $\chi^2$ = 7.843 (4)\end{tabular}} & \multirow{5}{*}{\begin{tabular}[c]{@{}c@{}} \\ p = 0.165, \\ $\chi^2$ = 6.497 (4)\end{tabular}} & \multirow{5}{*}{\begin{tabular}[c]{@{}c@{}} \\ p = 0.032, \\ $\chi^2$ = 10.542 (4)\end{tabular}} \\
\hspace{1em}None & 67 & 0.107 & 42 & 0.069 & 70 & 0.113 &  &  &  \\
\hspace{1em}Little & 117 & 0.186 & 123 & 0.201 & 100 & 0.162 &  &  &  \\
\hspace{1em}Some & 205 & 0.327 & 227 & 0.371 & 234 & 0.379 &  &  &  \\
\hspace{1em}A lot & 237 & 0.378 & 219 & 0.358 & 213 & 0.345 &  &  &  \\
\hspace{1em}NA/Missing & 2 & 0.003 & 1 & 0.002 & 0 & 0.000 &  &  &  \\
\addlinespace
\textbf{Gender} &  &  &  &  &  &  & \multirow{4}{*}{\begin{tabular}[c]{@{}c@{}} \\ p = 0.901, \\ $\chi^2$ = 0.581 (3)\end{tabular}} & \multirow{4}{*}{\begin{tabular}[c]{@{}c@{}} \\ p = 0.563, \\ $\chi^2$ = 2.046 (3)\end{tabular}} & \multirow{4}{*}{\begin{tabular}[c]{@{}c@{}} \\ p = 0.417, \\ $\chi^2$ = 2.838 (3)\end{tabular}} \\
\hspace{1em}Female & 323 & 0.516 & 326 & 0.533 & 319 & 0.520 &  &  &  \\
\hspace{1em}Male & 297 & 0.474 & 277 & 0.453 & 293 & 0.480 &  &  &  \\
\hspace{1em}Other / Prefer not to say & 6 & 0.010 & 7 & 0.011 & 5 & 0.008 &  &  &  \\
\hspace{1em}NA/Missing & 2 & 0.003 & 2 & 0.003 & 0 & 0.000 &  &  &  \\
\addlinespace
\textbf{Age Group} &  &  &  &  &  &  & \multirow{6}{*}{\begin{tabular}[c]{@{}c@{}} \\ p = 0.465, \\ $\chi^2$ = 4.610 (5)\end{tabular}} & \multirow{6}{*}{\begin{tabular}[c]{@{}c@{}} \\ p = 0.501, \\ $\chi^2$ = 4.348 (5)\end{tabular}} & \multirow{6}{*}{\begin{tabular}[c]{@{}c@{}} \\ p = 0.392, \\ $\chi^2$ = 5.203 (5)\end{tabular}} \\
\hspace{1em}[18,29) & 126 & 0.201 & 120 & 0.196 & 113 & 0.185 &  &  &  \\
\hspace{1em}[29,35) & 103 & 0.164 & 126 & 0.206 & 118 & 0.193 &  &  &  \\
\hspace{1em}[35,43) & 145 & 0.231 & 129 & 0.211 & 134 & 0.219 &  &  &  \\
\hspace{1em}[43,52) & 123 & 0.196 & 105 & 0.172 & 129 & 0.211 &  &  &  \\
\hspace{1em}[52,87] & 129 & 0.206 & 130 & 0.213 & 123 & 0.201 &  &  &  \\
\hspace{1em}NA/Missing & 2 & 0.003 & 2 & 0.003 & 0 & 0.000 &  &  &  \\
\bottomrule
\end{tabular}
\vspace{0.5em}
\begin{tablenotes}[flushleft]
\fontsize{9pt}{11pt}\selectfont
\item[] \parbox{0.95\linewidth}{Sample sizes: DALL-E 2 (D2) = 628, DALL-E 3 Revised (D3r) = 612, DALL-E 3 Verbatim (D3v) = 617. \\
$N(\cdot)$ columns report the count of observations in each treatment arm belonging to that category, $\text{Prop}(\cdot)$ gives the proportion of respondents in that category for the arm. Chi-squared test columns show p-values and $\chi^2$ test statistics with degrees of freedom for each pairwise comparison.}
\end{tablenotes}
\end{threeparttable}
\end{table}
\end{landscape}
\newpage

\begin{landscape}
\begin{table}[p]
\centering
\caption{Covariate Balance Across All Three Arms (Logo Generation Experiment, Pre-filtering; N = 1924)}
\label{tab:pval_balance_combined_full}
\begin{threeparttable}
\footnotesize
\begin{tabular}{lccccccccc}
\toprule
\textbf{Covariate / Level} & \textbf{N(D2)} & \textbf{Prop(D2)} & \textbf{N(D3r)} & \textbf{Prop(D3r)} & \textbf{N(D3v)} & \textbf{Prop(D3v)} & \textbf{D2 vs. D3r} & \textbf{D2 vs. D3v} & \textbf{D3r vs. D3v} \\
 &  &  &  &  &  &  & \textbf{p-val, $\chi^2$(df)} & \textbf{p-val, $\chi^2$(df)} & \textbf{p-val, $\chi^2$(df)} \\
\midrule
\textbf{GPT Usage} &  &  &  &  &  &  & \multirow{5}{*}{\begin{tabular}[c]{@{}c@{}} \\ p = 0.134, \\ $\chi^2$ = 7.044 (4)\end{tabular}} & \multirow{5}{*}{\begin{tabular}[c]{@{}c@{}} \\ p = 0.267, \\ $\chi^2$ = 5.203 (4)\end{tabular}} & \multirow{5}{*}{\begin{tabular}[c]{@{}c@{}} \\ p = 0.034, \\ $\chi^2$ = 10.441 (4)\end{tabular}} \\
\hspace{1em}None & 67 & 0.103 & 44 & 0.070 & 73 & 0.114 &  &  &  \\
\hspace{1em}Little & 121 & 0.186 & 123 & 0.196 & 101 & 0.158 &  &  &  \\
\hspace{1em}Some & 212 & 0.326 & 235 & 0.375 & 240 & 0.374 &  &  &  \\
\hspace{1em}A lot & 248 & 0.382 & 223 & 0.356 & 227 & 0.354 &  &  &  \\
\hspace{1em}NA/Missing & 3 & 0.005 & 2 & 0.003 & 5 & 0.008 &  &  &  \\
\addlinespace
\textbf{Gender} &  &  &  &  &  &  & \multirow{4}{*}{\begin{tabular}[c]{@{}c@{}} \\ p = 0.876, \\ $\chi^2$ = 0.690 (3)\end{tabular}} & \multirow{4}{*}{\begin{tabular}[c]{@{}c@{}} \\ p = 0.825, \\ $\chi^2$ = 0.900 (3)\end{tabular}} & \multirow{4}{*}{\begin{tabular}[c]{@{}c@{}} \\ p = 0.762, \\ $\chi^2$ = 1.163 (3)\end{tabular}} \\
\hspace{1em}Female & 330 & 0.508 & 332 & 0.530 & 332 & 0.518 &  &  &  \\
\hspace{1em}Male & 311 & 0.478 & 285 & 0.455 & 304 & 0.474 &  &  &  \\
\hspace{1em}Other / Prefer not to say & 7 & 0.011 & 7 & 0.011 & 5 & 0.008 &  &  &  \\
\hspace{1em}NA/Missing & 3 & 0.005 & 3 & 0.005 & 5 & 0.008 &  &  &  \\
\addlinespace
\textbf{Age Group} &  &  &  &  &  &  & \multirow{6}{*}{\begin{tabular}[c]{@{}c@{}} \\ p = 0.568, \\ $\chi^2$ = 3.873 (5)\end{tabular}} & \multirow{6}{*}{\begin{tabular}[c]{@{}c@{}} \\ p = 0.877, \\ $\chi^2$ = 1.793 (5)\end{tabular}} & \multirow{6}{*}{\begin{tabular}[c]{@{}c@{}} \\ p = 0.636, \\ $\chi^2$ = 3.418 (5)\end{tabular}} \\
\hspace{1em}[18,29) & 130 & 0.200 & 122 & 0.195 & 118 & 0.184 &  &  &  \\
\hspace{1em}[29,35) & 111 & 0.171 & 130 & 0.208 & 121 & 0.189 &  &  &  \\
\hspace{1em}[35,43) & 147 & 0.226 & 130 & 0.208 & 140 & 0.218 &  &  &  \\
\hspace{1em}[43,52) & 128 & 0.197 & 109 & 0.174 & 133 & 0.207 &  &  &  \\
\hspace{1em}[52,87] & 132 & 0.203 & 133 & 0.212 & 129 & 0.201 &  &  &  \\
\hspace{1em}NA/Missing & 3 & 0.005 & 3 & 0.005 & 5 & 0.008 &  &  &  \\
\bottomrule
\end{tabular}
\vspace{0.5em}
\begin{tablenotes}[flushleft]
\fontsize{9pt}{11pt}\selectfont
\item[] \parbox{0.95\linewidth}{Sample sizes: DALL-E 2 (D2) = 651, DALL-E 3 Revised (D3r) = 627, DALL-E 3 Verbatim (D3v) = 646. Includes 7 users missing survey crosswalk.\\
$N(\cdot)$ columns report the count of observations in each treatment arm belonging to that category, $\text{Prop}(\cdot)$ gives the proportion of respondents in that category for the arm. Chi-squared test columns show p-values and $\chi^2$ test statistics with degrees of freedom for each pairwise comparison.}
\end{tablenotes}
\end{threeparttable}
\end{table}
\end{landscape}
\newpage

\subsection{Generative Models} For both experiments, we randomly assigned participants to 1 of 3 generative models:

\begin{enumerate}
    \item DALL-E 2
    \item DALL-E 3 (Verbatim) (sometimes simply referred to as DALL-E 3)
    \item DALL-E 3 (Revised)
\end{enumerate}

\noindent Both the ``verbatim'' and ``revised'' versions of the DALL-E 3 treatment utilize the same underlying image-generating model; the distinction lies in the pre-processing applied before submitting user prompts to OpenAI's image-generating API. OpenAI's DALL-E 3 system, by design, employs a GPT-4 model to rewrite user prompts, adding more detail before processing the modified prompt using the DALL-E 3 image-generating model. During our experiment, it was not possible to explicitly disable this prompt rewriting feature of the DALL-E 3 system. To manage this behavior, we defined two treatments utilizing the DALL-E 3 model.

In the DALL-E 3 (Revised) treatment arm, we submit the participant's prompt directly to OpenAI's API and do not interfere with the default prompt rewriting process. In the DALL-E 3 (Verbatim) treatment arm, we prepend text to the participant's prompt instructing the GPT-4 model to refrain from modifying it before passing it forward to DALL-E 3. This prepended text was never visible to participants and was modeled after a prefix specifically suggested in OpenAI's online documentation for the DALL-E 3 endpoint.\footnote{See \href{https://platform.openai.com/docs/guides/images/usage?context=node}{here} and \href{https://community.openai.com/t/api-image-generation-in-dall-e-3-changes-my-original-prompt-without-my-permission/476355}{here} for online documentation.} We modified the recommended prefix slightly to account for the fact that we did not expect our participants to always submit ``extremely simple prompts.'' The exact text we prepended to participant prompts is copied below:

\begin{quote}
``I NEED to test how the tool works with my prompt as it is written. DO NOT add any detail; just use it AS IS:''
\end{quote}
Prepending this text to participants' prompts did reduce the rate at which OpenAI's endpoint modified prompts, but compliance was not perfect. Thus, we view the ``verbatim'' treatment arm as more of an intent-to-treat intervention. The GPT model still modified 59\% of participant prompts. The average token sort ratio (TSR) between the original prompt and the modified prompt was 77 for the DALL-E 3 (Verbatim) arm, compared to an average token sort ratio of 44 across the entire DALL-E 3 (Revised) treatment arm (a TSR of 100 denotes an exact string match). Conditional on any modification (any observations with TSR $<$ 100), the average TSR between the original prompt and the modified prompt was 61 for the DALL-E 3 (Verbatim) arm, compared to an average TSR of 44 across the entire DALL-E 3 (Revised) treatment arm.

\subsection{Model Endpoints}
We used the following model endpoints and parameters to generate images from prompts:
\begin{enumerate} 
    \item \textbf{OpenAI JavaScript API:} For both the image replication and logo generation experiments, we used the image generation endpoint of the official OpenAI JavaScript library (release version 3.1.0) to generate images for user prompts in the experimental interface. For all treatment arms, we set the image size parameter to be 1024 x 1024 pixels. For the DALL-E 3 (Revised) and DALL-E 3 (Verbatim) treatment arms, we set the quality parameter to \texttt{standard} and the style parameter to \texttt{natural}. 
    \item \textbf{Azure OpenAI Service:} For the image replication experiment, we used the image generation endpoints in the Python implementation of Azure OpenAI Service to generate all replay images based off the original user prompts. For prompts replayed through the DALL-E 2 treatment arm, we deployed a set of DALL-E 2 models on Azure OpenAI Service and set the API version for each to the \texttt{2023-06-01-preview} version. For prompts replayed through the DALL-E 3 (Revised) and DALL-E 3 (Verbatim) treatment arms, we deployed a set of  DALL-E 3 models on Azure OpenAI Service and set the API version for each to the \texttt{2023-12-01-preview} version. The parameter values for image size, and quality and style for the DALL-E 3 treatment arms, were set to the same values as above.
    \item \textbf{OpenAI Python API:} For the logo generation experiment, we used the image generation endpoints in the official OpenAI Python library (release version 0.27.8) to generate all replay images based off the original user prompts. The parameter values for image size, and quality and style for the DALL-E 3 treatment arms, were set to the same values as above.
\end{enumerate}

\subsection{Image Replication Experiment}

In this subsection, we describe design elements specific to the image replication experiment.

\subsubsection{Task Design}

Participants were asked to reproduce a single target image as closely as possible using a text-to-image generative AI model (i.e., DALL-E 2, DALL-E 3 without prompt revision, or DALL-E 3 with prompt revision, all developed by OpenAI). They did so by successively submitting prompts. In response to each submitted prompt, the model would generate an image, which was then displayed to the participant next to their assigned target image. Participants were instructed to make at least 10 attempts to recreate the target image within a 25-minute window, with no upper limit on their number of attempts.

All interactions between participants and the generative AI models occurred on a custom-built online interface designed to resemble OpenAI's ChatGPT interface but with some adjustments related to our task (e.g., displaying the target image and the total number of attempts so far to the user). On the right-hand side of the interface, participants were shown the target image they were randomly assigned to recreate. On the left-hand side, participants were shown their previously submitted prompts as well as the resulting generated images. We placed the text box where participants were able to write and submit their prompts at the bottom of the interface. Prompts were limited to a maximum of 1,000 characters (approximately 200-250 words). Participants were informed that their interactions with their assigned model would be memoryless, i.e., the model retained no memory of previous prompts and only used the current prompt to generate each image. Before the task, participants were provided with written and video instructions on how to interact with our experiment interface. Our task did not assume nor require prior experience with \emph{any} generative AI tools.

After the task, we surveyed participants' opinions and preferences regarding generative AI tools. We also inquired about their self-assessed occupational skills and how often they 1) engaged in creative writing, 2) wrote specific instructions, and 3) engaged in any sort of computer programming. Finally, we collected socio-demographic data, such as age and gender.

\subsubsection{Target Images} We randomly assigned participants to 1 of 15 target images. The set of target images consisted of 5 images each from 3 different broad categories: business and marketing, graphic design, and architectural photography. We chose these to represent the use cases suggested by the prompt categories on \href{https://promptbase.com/}{https://promptbase.com/}, a leading marketplace for image generation prompts. The images vary in color, style, content, and complexity within and across categories. These images can be found online linked to the pre-registration document: \textbf{URL removed to maintain anonymity}.
Performance, and variability of performance varied substantially across images. In other words, some images were much easier than others to replicate with the generative models, which we view as additional evidence that the set of 15 images was reasonably diverse.

\subsubsection{Subjects}

Our Prolific-recruited US sample for the image replication task (N = \initialNRep{}) was limited to fluent English speakers, and we prevented participants from completing the task more than once. We also prevented users from completing the task on mobile devices or tablets. Data was collected between December 12, 2023 and December 19, 2023.
Participants were guaranteed a payment of \$4 USD for completing the task and could earn an additional \$8 USD (a 200\% bonus) if they ranked in the top 20\% of participants in terms of the image most similar to the target (this was operationalized  with the Dreamsim score between each image and the targe, which we describe in section \ref{sec:image-similarity}). The median (mean) time to complete our entire task, including a demographic survey, was 23.5 (26.4) minutes. Given that 20\% of subjects received a bonus, the average compensation for participants in our study was \$5.60 USD per person, or about \$13 USD per hour. We explained the payment and incentive scheme to participants in full multiple times during the onboarding phase of the experiment, and asked participants to confirm their understanding before they were allowed to complete the task. The onboarding process also included multiple attention checks; participants who failed the first check were immediately disqualified. For subsequent checks, participants were required to retry until they demonstrated understanding.

\subsubsection{Pre-registration} \label{sec:pre-reg-short}

This study was pre-registered. The pre-registration document included our hypotheses, planned analyses, and sample size justification. The pre-registration document can be found at \textbf{URL removed to maintain anonymity}.

\subsection{Logo Generation Experiment}

\subsubsection{Task Design}

The logo generation task design was identical in format to that of the image replication experiment.
The only difference was that participants were instructed to make the best logo possible according to the instructions presented on the right-hand side of the interface.
They were similarly required to make at least 10 attempts in 25 minutes and to answer the same set of post-task survey questions.

\subsubsection{Logo Instructions} \label{sec:logos}

Participants were randomly assigned to one of fifteen logo-generation tasks. 
Each task described an organization that needed a logo and included the instruction: 
``Please create the best possible design based on the client's description and listed requirements.'' 
All briefs provided (i) a title, (ii) one sentence of background, (iii) required visual elements, and (iv) the intended context of use. 
They were written with three goals in mind: to capture a wide variety of realistic use cases, to include a few objective elements so independent evaluators could assess them consistently, and to allow for open-ended creative possibilities.

In no particular order, the fifteen logo tasks were:

\paragraph{1. Craft Pine Brewery Logo}
The owners of Green Pine Brewery are seeking a logo for their family-owned craft brewery located in Vermont.
\textbf{Required visual elements:} pine tree, hops, barley. \textbf{Context for use:} beer labels, merchandise, digital marketing.

\paragraph{2. Logo for Community Park Sign}
The City of Springfield Parks Department is seeking a sign for the new Elmwood Community Park, a welcoming outdoor family space.
\textbf{Required visual elements:} trees, playground equipment, children playing. \textbf{Context for use:} park entrance, leaflets, community posters.

\paragraph{3. Soccer Team Logo}
The coach of Riverdale FC is seeking a logo for their soccer team to build team spirit and local pride.
\textbf{Required visual elements:} the city’s iconic bridge over the river, soccer ball. \textbf{Context for use:} uniforms, merchandise, stadium banners/signage.

\paragraph{4. University Logo}
Prestwick University is seeking a logo emphasizing academic excellence.
\textbf{Required visual elements:} open book, torch. \textbf{Context for use:} digital marketing, printed materials, university merchandise.

\paragraph{5. Coffee Shop Logo}
Downtown Brew, a trendy independent coffee shop, is seeking a logo to attract people as a place to work.
\textbf{Required visual elements:} coffee cup, city skyline. \textbf{Context for use:} cups, signage, digital marketing.

\paragraph{6. Artisan Bakery Logo}
The Daily Loaf, an artisan bakery known for sourdough breads, wants a logo highlighting craftsmanship.
\textbf{Required visual elements:} loaf of bread, wheat stalk. \textbf{Context for use:} packaging, storefront, digital marketing.

\paragraph{7. Fitness Studio Logo}
Peak Performance Gym wants a logo to attract wealthy young professionals.
\textbf{Required visual elements:} dumbbell, mountain peak. \textbf{Context for use:} gym signage, apparel, digital marketing.

\paragraph{8. Music Festival Logo}
The organizers of the annual Summer Beats Festival seek a logo to celebrate electronic music and summer vibes.
\textbf{Required visual elements:} sound waves, the sun. \textbf{Context for use:} posters, tickets, merchandise.

\paragraph{9. Veterinary Clinic Logo}
The Happy Tails Vet Clinic seeks a friendly logo for their animal health practice.
\textbf{Required visual elements:} dog, cat, heart. \textbf{Context for use:} clinic signage, uniforms, magazine ads/paper leaflets.

\paragraph{10. Local Farmers Market Logo}
The Green Valley Weekend Farmers Market seeks a logo emphasizing fresh, local produce and artisan crafts.
\textbf{Required visual elements:} basket of vegetables, shovel. \textbf{Context for use:} market banners, tote bags, digital marketing.

\paragraph{11. Ice Cream Shop Logo}
Super Scoops, an ice cream parlor, desires a playful and attractive logo.
\textbf{Required visual elements:} ice cream cone, tongue. \textbf{Context for use:} storefront, menus, ice cream packaging.

\paragraph{12. Bookstore Logo}
Pages \& Co., a cozy independent bookstore, seeks a logo to attract book lovers.
\textbf{Required visual elements:} smiling person holding a book. \textbf{Context for use:} signage, bookmarks, stickers.

\paragraph{13. Mountain Resort Logo}
The owners of Highland Retreat, a mountain resort, want a logo emphasizing outdoor adventure.
\textbf{Required visual elements:} mountain range, cabin. \textbf{Context for use:} resort signage, merchandise, digital marketing.

\paragraph{14. Tech Startup Logo}
MegaSurf is a tech startup developing a mobile app that connects surfers in a neighborhood.
\textbf{Required visual elements:} surfboard, ocean wave. \textbf{Context for use:} app icon, digital marketing, business cards.

\paragraph{15. Nonprofit Organization Logo}
The leaders of Happy Housing seek a compassionate-looking logo representing their mission to ensure all children have a home.
\textbf{Required visual elements:} smiling child, a roof with a chimney. \textbf{Context for use:} fundraising leaflets, t-shirts, event signage.

\subsubsection{Subjects}

Our Prolific-recruited US sample for the logo generation task (N = \initialNLogo{}) was limited to fluent English speakers, and we prevented participants from completing the task more than once. We also prevented users from completing the task on mobile devices or tablets. Data was collected between June 30, 2025 and July 1, 2025.
Participants were guaranteed a payment of \$4 USD for completing the task and could earn an additional \$8 USD (a 200\% bonus) if they ranked in the top 20\% of participants in BT score of their best-performing image. The median (mean) time to complete our entire task, including a demographic survey, was 26.1 (29.5) minutes. Given that 20\% of subjects received a bonus, the average compensation for participants in our study was \$5.60 USD per person, or about \$11.50 USD per hour. We explained the payment and incentive scheme to participants in full multiple times during the onboarding phase of the experiment, and asked participants to confirm their understanding before they were allowed to complete the task. The onboarding process also included multiple attention checks; participants who failed the first check were immediately disqualified. For subsequent checks, participants were required to retry until they demonstrated understanding.

\subsubsection{Pre-registration}

This study was pre-registered. The pre-registration document included our hypotheses, planned analyses, and sample size justification. The pre-registration document can be found at \textbf{URL removed to maintain anonymity}.

\newpage

\section{Data} \label{sec:vars}

This appendix provides detailed information about the data collection, measurement, and variables used in our analyses, expanding on the overview provided in the main text.

\subsection{Survey Data}

We collected additional participant information via a Qualtrics survey that included:

\paragraph{Demographics.} We collect the following information from each participant: Ethnicity, Gender, Age, Highest level of education attained (some high school, high school, some college, associate's degree, bachelor's degree, master's degree, doctoral degree, professional degree, other), Years of work experience, Annual Income (0-\$25k, \$25.001k-\$50k, \$50.001k-75k, \$75.001k-\$100k, \$100.001k-\$150k, \$150k+), and elicitation of sets of O*NET job skills that participants used in their occupation (reading comprehension, active listening, writing, speaking, critical thinking, social perceptiveness, coordination, instructing, programming, judgment and decision making, systems evaluations, science, active learning, learning strategies, monitoring, complex problem analysis, technology design, troubleshooting, quality control analysis, systems analysis).

\paragraph{Opinions and Skills.} The opinions and skills we collect information on are: computer programming proficiency and usage frequency (self-reported), Structured and creative writing proficiency and usage frequency (self-reported), Generative AI tool proficiency and usage frequency (self-reported), Attitudes towards net social impact of Generative AI (self-reported), Advice for (hypothetical) future participants on how to perform well on the task.

\subsection{Prompt Data}
For each prompt, we recorded the text of the participant's prompt, the order in which it was submitted, the timestamp of submission, and for the DALL-E 3 treatment arms, the revised prompt returned by the model.

\subsection{Image Data} \label{sec:image-data}

We collected three different sets of images:
\begin{enumerate}
    \item \textbf{The participant-facing images (image replication experiment and logo generation experiment):} These images were created directly by the participants, generated by the model to which they were randomly assigned using the prompt they submitted. For the image replication experiment, these were generated from December 12 to December 19, 2023. For the logo generation experiment, these were generated from June 30 to July 1, 2025. \label{bullet:original_images}
    
    \item \textbf{Post-hoc resampled images (image replication experiment only):} For any given prompt, the output of the text-to-image model is stochastic. To better approximate the expected image from a given prompt, we generated 20 additional images for each prompt after the image replication experiment concluded. We provide full details on this procedure in Section~\ref{sec:methods}. These images were generated from December 26, 2023 to January 27, 2024. These images are not used for analyses presented in the main text, but were used for other pre-registered analyses. These additional analyses are discussed in Section \ref{sec:pre-reg}.\label{bullet:resampled_original}
    
    \item \textbf{Post-hoc replayed images (image replication experiment and logo generation experiment):} For both experiments, to decompose our overall effects into model and prompting effects, we generated ``counterfactual images'' for each prompt written under the DALL-E 2 and DALL-E 3 (Verbatim) treatments. In other words, we submitted all prompts written under both the DALL-E 2 and DALL-E 3 (Verbatim) treatments to both the DALL-E 2 and DALL-E 3 (Verbatim) endpoints.

    For the image replication experiment, similar to the resampling procedure outlined above, we generated 10 images per prompt per model: we generated a single replay for each prompt-model pair from March 16 to 18, 2024, and then, to increase power, generated the replications for these replay images from June 14 to 27, 2024. This replay process produced a total of 20 images per prompt---10 under the original model, 10 under the counterfactual model. We re-submitted prompts to their original model to account for potential model drift, as this exploratory analysis was conducted multiple months after our initial data collection. For consistency, this replay data is used throughout the main text of our paper.
    
    For the logo generation experiment, we generated a single replay for each prompt-model pair. This took place from July 3 to 9, 2025. In this case, given the negligible gap in time between participant data collection and participant image replay, there was no need to account for potential model drift as we did in the image replication experiment. 
\end{enumerate}

\subsection{Image Replication Sample Construction} \label{sec:sample_construction}

The image replication experiment sample analyzed in the main text was constructed as follows.

The initial ``raw'' dataset collected during the experiment is comprised of 24,672 rows of raw prompt data (one prompt per row) generated by \initialNRep{} participants. We first removed rows with blank prompt entries, invalid prolific IDs, and unsuccessful attempts (logging errors). These exclusion criteria were pre-registered. This left us with 2,029 participants and 24,123 prompts. We next removed participants from our sample if they failed to submit at least 10 prompts. This exclusion criterion was pre-registered and also explained to participants, who were told that payment was contingent on submitting at least 10 successful prompts and a ``good-faith effort.'' This left us with 1,914 participants and 23,566 prompts.

Although participants were allowed to submit as many prompts as they desired in the 25-minute time span, we limited all analyses to each participant's first 10 prompts---the minimum required to receive payment for the task. This exclusion criteria was not pre-registered, and is noted in the list of deviations from pre-registration in Section~\ref{sec:pre-reg}.\ref{sec:deviations}. We restrict our analysis dataset in this way because participants who chose to submit more than 10 prompts may have been systematically different than those who did not. Excluding any prompt beyond the \nth{10} attempt allows us to alleviate selection bias concerns. This left us with 1,914 participants and 19,140 prompts. 

We next removed participants who submitted the same prompt at least five times in a row at any point during the task. This exclusion criterion was pre-registered. To avoid reward hacking, we did not specify to participants the ``no more than 5 repeated prompts'' criterion for ``good-faith effort.'' This left us with 1,899 participants and 18,990 prompts. We then removed participants who failed to complete the Qualtrics survey. This exclusion criteria was pre-registered. This left us with 1,893 participants and 18,930 prompts. We also removed prompts from our dataset according to a number of post-hoc, non-pre-registered exclusion criteria to ensure data quality and avoid selection bias. These criteria include the following:
\begin{enumerate}
    \item Prompts sometimes trigger errors in OpenAI's safety system because they contain language that might be deemed unsafe under OpenAI's policies. The specific language that triggers these errors is constantly changing and not available publicly. If a prompt triggered a safety error during the replication or replay process, we re-submitted the prompt up to 50 times or until the 10 original arm replications/replay samples had been collected. We removed prompts if they failed to generate 10 replications on the original model or 10 replay samples under the counterfactual model during the replication/replay process. This affected 343 prompts across all three treatment arms.
    \item Due to rare database synchronization latency issues, some prompts were assigned duplicate attempt numbers by the MongoDB database that we used to collect our data. This data collection error led to issues in the data analysis process. Thus, we excluded prompts with duplicate attempt numbers. This affected 34 prompts across all three treatment arms, and 20 prompts between the DALL-E 2 and DALL-E 3 (Verbatim) treatment arms, approximately 0.1\% of the original data.
\end{enumerate}

In total, these post-hoc, non-pre-registered exclusion criteria affected 370 prompts across all remaining participants. Our final sample thus included \finalNRep{} participants and \finalPromptNRep{} prompts.

\subsection{Logo Generation Sample Construction} \label{sec:sample_construction_rep}

The logo generation experiment sample analyzed in the main text was constructed as follows.

The initial ``raw'' dataset collected during the experiment is comprised of 23,159 rows of raw prompt data (one prompt per row) generated by \initialNLogo{} participants. We first removed rows with blank prompt entries, invalid prolific IDs, and unsuccessful attempts (logging errors). These exclusion criteria were pre-registered. This left us with 2,011 participants and 22,721 prompts. We next removed participants from our sample if they failed to submit at least 10 prompts. This exclusion criterion was pre-registered and also explained to participants, who were told that payment was contingent on submitting at least 10 successful prompts and a ``good-faith effort.'' This left us with 1,924 participants and 22,312 prompts.

Although participants were allowed to submit as many prompts as they desired in the 25-minute time span, we limited all analyses to each participant's first 10 prompts---the minimum required to receive payment for the task. This exclusion criteria was not pre-registered, and is noted in the list of deviations from pre-registration in Section~\ref{sec:pre-reg}.\ref{sec:deviations}. We restrict our analysis dataset in this way because participants who chose to submit more than 10 prompts may have been systematically different than those who did not. Excluding any prompt beyond the \nth{10} attempt allows us to alleviate selection bias concerns. This left us with 1,924 participants and 19,240 prompts. 

We next removed participants who submitted the same prompt at least five times in a row at any point during the task. This exclusion criterion was pre-registered. To avoid reward hacking, we did not specify to participants the ``no more than 5 repeated prompts'' criterion for ``good-faith effort.'' This left us with 1,892 participants and 18,920 prompts. We additionally removed a small number of participants and their prompts for whom we discovered post-hoc that the treatment condition to which they were exposed in the experimental interface did not match up with their assigned task label in the MongoDB database that we used to collect our data. This exclusion criterion was not pre-registered. This left us with 1,864 participants and 18,640 prompts.

We then removed participants who failed to complete the Qualtrics survey. This exclusion criteria was pre-registered. This left us with \finalNLogo{} participants and 18,570 prompts. We also removed prompts from our dataset according to a number of post-hoc, non-pre-registered exclusion criteria to ensure data quality and avoid selection bias. These criteria include the following:
\begin{enumerate}
    \item Prompts sometimes trigger errors in OpenAI's safety system because they contain language that might be deemed unsafe under OpenAI's policies. The specific language that triggers these errors is constantly changing and not available publicly. If a prompt triggered a safety error during the replay process, we re-submitted the prompt up to 3 times. We removed prompts if they failed the complete set of replays under the counterfactual model during the replay process.
    \item Due to rare database synchronization latency issues, some prompts were assigned duplicate attempt numbers by the MongoDB database that we used to collect our data. This data collection error led to issues in the data analysis process. Thus, we excluded prompts with duplicate attempt numbers.
\end{enumerate}

In total, these post-hoc, non-pre-registered exclusion criteria affected 145 prompts across all remaining participants. Our final sample thus included \finalNLogo{} participants and \finalPromptNLogo{} prompts.

\subsubsection{Removing ``Off-Topic'' Submissions (Image Replication Experiment Only)} While analyzing our data, we found that our sample contained a number of ``off-topic'' prompts that did not seem related to the task. As a robustness check on our main results, we used the following process to systematically identify and remove ``off-topic'' prompts. First, we generated embeddings for each prompt using OpenAI's \texttt{text-embedding-3-small} model. We then calculated the mean embedding for each target image. Next, we calculated the Euclidean distance between each prompt's embedding vector and the mean embedding vector for prompts corresponding to the focal prompt's assigned target image. Finally, we removed the 2.5\% of prompts that were most distant from the mean image-level prompt embedding vector. This led to the removal of 481 prompts across all three treatment arms, and 338 prompts between the DALL-E 2 and DALL-E 3 (Verbatim) treatment arms. All of our main text results are robust to the exclusion of these ``off-topic'' prompts.

\newpage

\section{Dependent Variables} \label{sec:dep-var}

\subsection{Image Similarity} \label{sec:image-similarity} 

We pre-registered two quantitative measures of image similarity: the cosine similarity of CLIP embedding vectors and a recently developed measure called `DreamSim' \citep{fu2023dreamsim}. In the main text, we present analyses using CLIP embedding cosine similarity, since it is likely more familiar to readers. Our results are qualitatively and quantitatively similar using DreamSim instead.

\paragraph{CLIP Embedding Cosine Similarity.} To calculate CLIP embedding cosine similarity, we first generated CLIP embedding vectors \citep{radford2021learning} using the version of CLIP available on Hugging Face \citep{CLIP-embeddings} for each participant-generated image and for each target image. Unlike traditional image embeddings that only encode visual features, CLIP embeddings also capture semantic relationships between images and descriptive text. We then calculated the cosine similarity between each participant-generated image's CLIP embedding and the relevant target image's CLIP embedding.

\paragraph{DreamSim.} DreamSim is an image similarity measure proposed recently by \citep{fu2023dreamsim}. The authors claim that relative to a measure such as CLIP embedding cosine similarity, DreamSim measures image similarity in a way that more effectively captures human visual perceptions of similarity. Because the original DreamSim metric outputs a distance measure, we invert this score $\tilde{D} = 1 - (\text{original DreamSim})$ to recast it as a similarity score. After doing so, both the inverted DreamSim and CLIP embedding cosine similarity are closer to 1 when two images are more similar and closer to 0 when two images are more dissimilar.

We find that these two measures of image similarity are highly correlated in our sample ($\rho_{pearson} = 0.763$, 95\% CI: [0.755 0.770]), and our main results are robust to the use of either measure. We present the results obtained when conducting our main text analyses using DreamSim in Section~\ref{sec:dreamsim_robust}.

\subsection{Bradley-Terry Strength} \label{sec:bradley-terry}

To generate a Bradley-Terry strength value for each original and replayed image in the logo generation experiment, we first collected all images and enumerated all pairwise comparisons within each of the 15 tasks. We then randomly sampled, without replacement, 20\% of the pairwise comparisons within each task to feed into our LLM-based evaluation approach, resulting in approximately 20 million pairwise comparisons. For the LLM-based evaluation, for each pairwise comparison, we called the response endpoint of the official OpenAI Python library (release version 1.97.0) with the following prompt, with the task brief and the two images appended after:

\begin{quote}
    ``You are provided a logo design request from a client, as well as two images (Image A and Image B). Please evaluate both images carefully and determine which one better fulfills the client's specific requirements. Please respond with only the letter label of the image, either `A' or `B'.
\end{quote}

For all requests, we used the multi-modal \texttt{gpt-4.1-nano} model and set the temperature parameter to be 0. The images were first base64-encoded before being included in the request input. We further randomized the order in which the images were presented in the prompt to avoid the effects of prompt position bias.
To convert the resulting pairwise preferences into continuous strength values, we used the Bradley--Terry model. In this model, each item $i$ is assumed to have an associated latent strength parameter $\beta_i \in \mathbb{R}$, and the probability that item $i$ is preferred to item $j$ is given by
\begin{align*}
    \Pr(i \succ j) \;=\; \frac{\exp(\beta_i)}{\exp(\beta_i) + \exp(\beta_j)}.
\end{align*}
Our goal is to estimate the set of strength parameters $\{\beta_i\}$ for all images. For each task, once we collected the LLM-generated pairwise comparison outcomes and converted them into win--loss matrices, we estimated the $\beta_i$ parameters by fitting the Bradley--Terry model via maximum likelihood using a standard iterative procedure.

\subsection{Prompt Length}
We measure the lengths of prompts written by participants in our sample, both in terms of the number of \emph{words} in a given prompt and in terms of the number of \emph{characters} in a given prompt. In our main text analysis, we present results only in terms of the number of words, since the two outcomes are highly correlated ($\rho_{pearson} = 0.9954$, 95\% CI: [0.99528, 0.99560]).

\subsection{Embedding-based Prompt Similarity} 
\label{sec:prompt_similarity}

We calculate two measures of embedding-based prompt similarity: successive similarity and aggregate similarity. Both measures use the vector embedding representation of each prompt in our sample, which we obtained using OpenAI's \texttt{text-embedding-3-small} model \citep{neelakantan2022text}. The two similarity measures are defined as follows:

\paragraph{Successive similarity:}  The successive similarity ($ss$) is a measure of the similarity of a participant's prompt to their immediately preceding prompt. We define the successive similarity of a prompt $p_{i,n}$ written by user $i$ to the their immediately preceding prompt $p_{i,n-1}$ as:

\begin{equation}
ss_{i,n,n-1} = \frac{\mathbf{E(p_{i,n})} \cdot \mathbf{E(p_{i,n-1})}}{||\mathbf{E(p_{i,n})}||\, ||\mathbf{E(p_{i,n-1})}||},
\end{equation}

\noindent where $\mathbf{E(p_{i,n})}$ is the vector embedding representation of participant $i$'s $n^{\text{th}}$ prompt, $p_{i,n}$. This measure starts with participant $i$'s \nth{2} attempt, as the calculation requires a previous attempt. 

\paragraph{Aggregate similarity:} The aggregate similarity ($as$) is a measure of how dispersed each user's prompts are around their ``average prompt'' (calculated by taking the element-wise average of all prompt embeddings produced by the user). We define the aggregate similarity for the 10 prompts written by a given user as:

\begin{equation}
as_i = \frac{1}{10} \sum_{n=1}^{10} \lVert \mathbf{E(p_{i,n})} - \overline{\mathbf{E(p_{i,n})}} \rVert_2^2,
\end{equation}

\noindent where $\mathbf{E(p_{i,n})}$ is again the vector embedding representation of participant $i$'s $n^{\text{th}}$ prompt, $p_{i,n}$, and $\overline{\mathbf{E(p_{i,n})}}$ is the element-wise mean of all 10 of participant $i$'s prompts.

\subsection{Successive Prompt Token Sort Ratio}

Starting with each participant's second prompt, we also calculated the token sort ratio (TSR) of each prompt $p_{i,n}$ to the immediately preceding prompt $p_{i,n-1}$. TSR is a fuzzy string-matching technique \citep{singla2012string} that provides a continuous measure of how similar two strings are. 

\subsection{Successive Prompt `Contains Previous Prompt' Dummy}

Starting with each participant's second prompt, we record whether each prompt $p_{i,n}$ contains the immediately preceding prompt $p_{i,n-1}$ as an exact substring.

\subsection{Prompt Composition}
We use the \texttt{spaCy v3.7.4} Python package's \texttt{en\_core\_web\_sm} model to tag the parts of speech (POS) in each prompt. SpaCy's models utilize the "universal POS tags" from the Universal Dependencies framework for grammar annotation \cite{posuniversaldep}. These tags encompass parts of speech such as adjectives, adverbs, nouns, and verbs. The model tags each word in a prompt according to this framework, after which we count the total number of words corresponding to each part of speech for each prompt.

\subsection{Strategic Shifts}
In addition to calculating the successive and aggregate similarity of prompts written by particular users, we also attempt to identify particular moments when participants shift their approach to prompting. In order to do so, we adapt a method proposed in \citep{torricelli2023role} (because they are conducting research in a different context, \citet{torricelli2023role} refer to these shifts as ``topical transitions'' as opposed to ``strategic shifts''). To identify these strategic shifts, we first calculate the mean cosine similarity ($MCS$) for the embedding vectors of every possible pair of prompts submitted in response to a given target image, $t$:

\begin{equation}
MCS_t = \frac{2}{P_t(P_t-1)}\sum_{a = 1}^{P_t} \sum_{b = a+1}^{P_t} \textrm{CosineSim}(\mathbf{E(p_{a,t})}, \mathbf{E(p_{b,t})}).
\end{equation}
where $P_t$ is the total number of prompts submitted in response to a given target image, and $a$ and $b$ are indices representing individual prompts for that target.

We then label any given prompt as a strategic shift ($SS$) if the cosine similarity of its embedding vector with that of the previous prompt is lower than this target-image-level mean:

\begin{equation}
\begin{aligned}
SS(\mathbf{p}_{i,t}) = 
\begin{cases}
1 & \text{if } \textrm{CosineSim}\big( \mathbf{E(p}_{i,t}), \mathbf{E(p}_{i-1,t}) \big) \\
  & \quad < MCS_t \\
0 & \text{otherwise}
\end{cases}
\end{aligned}
\end{equation}

\noindent It is worth noting that \cite{torricelli2023role} uses the participant-level mean, as opposed to the task-level mean, as the cutoff for a topical shift. We use the task-level mean because in our setting, it did not seem appropriate that half of each participant's submitted prompts would be strategic shifts.

\newpage

\section{Methods} \label{sec:methods}

This appendix provides additional methodological details to supplement the analyses presented in the main text.

\subsection{Stratification} \label{sec:stratification}
The results shown in the main text and in the supplementary information are typically stratified by reference image and iteration. In some analyses, we have stratified only on the reference image (e.g., for analyses presented at the iteration level). The exact stratification for each finding is indicated in section \ref{sec:analysis}. To stratify our results, we calculate a weighted average across $j = 1,...,J$ cells defined by our stratification variables:

\begin{align*}
    \overline{Y}_{strat} = \sum_{j=1}^J \frac{N_j}{N} \bar{Y}_j
\end{align*}

\noindent To calculate the variance (and standard error) of this sample mean, we apply the following:

\begin{align*}
    \widehat{\mathrm{Var}}(\overline{Y}_{strat}) = \widehat{\mathrm{Var}}\left( \sum_{j=1}^J \frac{N_j}{N} \bar{Y}_j \right) = \sum_{j=1}^J \left(\frac{N_j}{N}\right)^2 \frac{s_j^2}{N_j}
\end{align*}

\(N_j\) is the population size of stratum \(j\). \(N\) is the total population size across all strata. \(\bar{Y}_j\) is the sample mean for stratum \(j\). \(J\) is the total number of strata. \(s_j\) is the sample standard deviation of stratum \(j\). Therefore, \(s_j^2\) is sample variance stratum \(j\).

\subsection{Z-Scoring} \label{sec:zscoring}

Our analysis found statistically significant differences in performance variability across the 15 target images used in our experiments, as discussed in Section~\ref{sec:pre-reg}. The main text also demonstrated that performance increases across successive attempts. To ensure our results are not driven by this image-level or attempt-level variation, we replicated all analyses using the within-image-attempt Z-score of CLIP-cosine similarity for each image produced by participants. Formally, this is:

\begin{equation}
\begin{aligned}
Z(\text{Sim}_{i,n,t}) = 
\frac{ \text{Sim}_{i,n,t} - \text{Mean}_{n,t}\left( \text{Sim}_{i,n,t} \right) }
     { \text{SD}_{n,t}\left( \text{Sim}_{i,n,t} \right) }
\end{aligned}
\end{equation}
where $\textrm{Sim}_{i,n,t}$ is the cosine similarity of user $i$'s image in attempt $n$ to target image $t$. The mean and standard deviation are computed for each image-attempt pair, but across both the DALL-E 2 and DALL-E 3 treatment arms. We also applied this rescaling to test the robustness of our DreamSim-based analyses. 

Nearly all robustness analyses reported here use the Z-score scaled measure of performance within each image-attempt set. The only exception is analyses that examine improvements and prompts across attempts (e.g., Figure 1B), where Z-scores are computed only within each target image and across all attempts for that image.

We use the same Z-scoring technique for robustness checks of the logo generation task. Instead of $t$ indexing the target image in these cases, it indexes the brief/vignette the participant was assigned to.

\subsection{Accounting for Model Stochasticity (Image Replication Experiment Only)} \label{sec:stochasticity} 

Generative AI models produce stochastic outputs in response to a given prompt. This stochasticity is controlled by a parameter called temperature, which could not be modified using the DALL-E API at the time of our experiment. To account for this model stochasticity, we generated 10 images for each prompt submitted by participants across all treatment arms. We then calculated the similarity between each replication and its corresponding target image, and computed an "expected" CLIP cosine similarity and DreamSim score for each prompt by averaging across these samples. 

Using these replicated images, we also calculated the standard deviation of cosine similarity induced by this stochasticity. With the expected cosine similarity and its standard deviation per prompt, we computed a normalized Z-score for the observed image relative to its replication distribution. This Z-score measures the extent to which the observed image is better or worse than what's expected for that prompt and is used for further analysis in section \ref{sec:pre-reg}.

We generated these additional samples both for the original prompts on their assigned treatment arms and by replaying prompts on the counterfactual arms, as introduced in Figure 2 in the main text. Importantly, OpenAI updated its content filters between our initial experiment and image re-sampling. As a result, some prompts that originally produced images either generated no images or fewer images than requested during our regeneration attempts. This affected 1.8\% (371 out of 18,990 prompts) of the data in our sample under the ``replaying" procedure (Section~\ref{sec:image-data}).

\newpage

\section{Main Text Analyses} \label{sec:analysis}
This section provides detailed methodological information about the analyses presented in the main text.

\subsection{Image Replication Task}

\subsubsection{Task Performance and ATEs} \label{sec:task_performance_analysis}
The top pane of Figure 1B compares the average performance across models and attempt numbers (also referred to as iterations). It displays the average cosine similarity score stratified by the reference image. A notable feature in this figure is the performance dip during the second recreation attempt across both treatment arms. This may be due to participants' initial misunderstanding of the model's ``memoryless'' nature. Participants failed to recognize that context from previous prompts was not carried over to new iterations. We observed numerous prompts in the second iteration across users that explicitly referenced the first prompt, a behavior that rarely occurred in subsequent attempts. Alternatively, it may also be indicative of other early exploration by participants. However, from the third prompt onward, participants appeared to grasp the independence of each attempt, as evidenced by a marked decrease in cross-prompt references and a corresponding rebound in performance.

The bottom pane of Figure 1B shows the average treatment effect (ATE) per iteration, which is the difference between the stratified averages of DALL-E 3 and DALL-E 2 in the top pane.\footnote{In Section~\ref{sec:analysis}, when we refer to ``DALL-E 3'', we mean ``DALL-E 3 (Verbatim)'' unless otherwise specified.}
To test the widening impact of using DALL-E 3 on performance relative to DALL-E 2, we run the following fixed effects linear model with participant-level ($i$) clustered standard errors where iteration is treated as a numeric variable:

\begin{equation} \label{eq:ate-per-attempt-1}
\begin{aligned}
Y_{i,n,t} = \; & \beta_0 + \beta_1 \, \text{iteration} + \beta_2 \, \mathbb{I}[\text{dalleVersion = 3}]_i \\
              & + \beta_3 \, \text{iteration} \times \mathbb{I}[\text{dalleVersion = 3}]_i + \gamma_t + \epsilon_{i,n,t}
\end{aligned}
\end{equation}

\noindent The coefficient estimates generated by this model are: $\hat{\beta_1} = 0.0011 \; (0.0003)$, $p = 0.0004$; $\hat{\beta_2} = 0.0120 \; (0.0037)$, $p = 0.0013$; $\hat{\beta_3} = 0.0010 \; (0.0004)$, $p = 0.0227$. The overall ATEs that we report between different pairs of treatment arms (DALL-E 2, DALL-E 3, and DALL-E 3 with revisions) in the main text are estimated from a two-way fixed effect (iteration and target image) model per each pair. Standard errors are cluster robust at the participant level.

\subsubsection{Prompt Characteristics}
Figure 1C compares the prompt length and prompt similarity of the two models. To generate these results, we first remove any prompt that does not constitute a good-faith attempt according to the sample construction procedure detailed in Section~\ref{sec:sample_construction}. The prompt length is the average number of words per model and iteration stratified by the reference image. The prompt similarity is the average cosine similarity between all consecutive pairs of user prompts, which are both determined to be valid attempts, stratified by the reference image (see Section \ref{sec:vars}.\ref{sec:prompt_similarity} for details on similarity calculations). 

The color scale in Figure 1C shows the stratified average similarity to the previous prompt across all users per each model. We find that DALL-E 3 users write prompts that, on average, have $\beta=0.0184$ higher in cosine similarity to their previous prompts using cluster robust standard errors at the participant level ($p=0.0236$).

Comparing the aggregate similarity of all attempts made by a given participant, we also find the prompts from DALL-E 3 participants were, on average, more similar than the prompts of the DALL-E 2 participants. For this analysis, we use the dispersion around the centroid in the prompt embedding space, explained in Section \ref{sec:vars}.\ref{sec:prompt_similarity}, as the dependent variable.
When we average across all participants by model, we find that the average distance of prompts written by DALL-E 3 participants to their centroid is $\beta=0.0191$ smaller than inferior users ($p=0.0083$). Standard errors are cluster-robust at the participant level.

\subsubsection{ATE Decomposition}
\label{sec:ate_decomposition}
Figure 2 in the main text decomposes the ATE into the model and prompting effects. This decomposition is conceptually similar to a simple mediation analysis, with an important difference being that we can observe counterfactual outcomes (e.g., prompting DALL-E 3 as if it is DALL-E 2). This is not typically the case in mediation analysis, and makes causal identification rely on fewer assumptions. 

To obtain counterfactual outcomes, we fed or ``replayed'' the participant prompts when interacting with one model (e.g., inferior) on another model (e.g., superior). The notation \textit{(prompt, model)} specifies which treatment arm the prompts were written under and which model was used in the replay. For example, (2,3) indicates replaying prompts written under DALL-E 2 on DALL-E 3. To be clear, (2,2) and (3,3) correspond to the original observed treatment arms, while (2,3) and (3,2) are the counterfactual outcomes of interest.\footnote{To avoid problems with model drift for the image replication experiment, we regenerated images for all four possible combinations at the same time and used these images for all analyses in the main text.}

The left-most point in Figure \ref{fig:fig2} corresponds to the average CLIP cosine similarity to the target image of (2,2). To make the interpretation of the results clearer, we have subtracted this quantity from all average quality scores and added a dashed line throughout. The second point from the left corresponds to average similarity to the target of (2,3), the third point from the left to (3,3), and the rightmost points to (3,2). All average similarity scores are stratified by iteration and reference image, and the standard errors are bootstrapped and cluster-robust at the participant level. 

The model effect, as shown by the red braces in Figure \ref{fig:fig2}, corresponds to the average increase in quality of (2,3) relative to (2,2). In the terminology of mediation analysis, the model effect would be referred to as the direct effect. The prompting effect, as shown by the blue braces in Figure 2, corresponds to the average increase in quality of (3,3) relative to (2,3). In the terminology of mediation analysis, the prompting effect would be referred to as the indirect effect. We can also test the difference in average quality between (3,3) and (3,2), as well as the difference between (3,2) and (2,2). Both of these differences are visible in Figure 2; the second is small and not statistically significant. 

The point estimates for the total, model, and prompting effects, along with their significance values reported in the main text, are obtained from two-way fixed-effects regressions that compare average outcomes across the two relevant conditions. Specifically, each effect is estimated using a model with iteration and target-image fixed effects, where an indicator for the effect of interest serves as the main independent variable:
\begin{equation}
    \label{eq:ind-dir-model}
    Y_{i,n,t} = \beta_0 + \beta_1 \text{effect} + \alpha_n + \gamma_t  + \epsilon_{i,n,t}
\end{equation}
where $\beta_1$ is the coefficient on the effect type in question (i.e., model or prompting).
To estimate the different effects, we simply use the above model and filter the data as appropriate. For example, to estimate the ATE, the data contains all (2,2) and (3,3) scores, and in this case effect=1 for observations in (3,3) group. Similarly, to estimate the direct or model effect, the data contains all (2,2) and (2,3) scores and effect=1 for observations in (2,3) group. Finally, to estimate the indirect or prompting effect, the data contains all (2,3) and (3,3) scores and effect=1 for observations in (3,3) group. The standard errors for each estimated model are cluster robust at the participant level, and p-values are adjusted accordingly.

\subsubsection{Distributional Impacts}
\label{sec:distribution_impact}
Table \ref{table:qte} in the main text demonstrates how total, model, and prompting QTEs vary across different score quantile levels. The total effect compares the cosine similarity of outputs from DALL-E 3 users replayed on DALL-E 3 to those from DALL-E 2 users replayed on DALL-E 2. The model effect compares outputs from DALL-E 2 users replayed on DALL-E 3 against the same users replayed on DALL-E 2. The prompting effect compares outputs from DALL-E 3 users replayed on DALL-E 3 to those from DALL-E 2 users replayed on DALL-E 3.

To assess user skill, users within each target image, iteration and \emph{replay scenario} are divided into 50 equally sized brackets based on their performance, assigning each bracket a rank corresponding to performance percentiles. For instance, when evaluating DALL-E 2 users replayed on DALL-E 2, each user's skill level is determined by their percentile rank among all DALL-E 2 users replayed on DALL-E 2. Conversely, when the same user is replayed on DALL-E 3 (to estimate the model effect or to serve as a baseline in estimating the prompting effect), their performance percentile is determined by their rank relative to all DALL-E 2 users replayed on DALL-E 3, per each iteration and target image.

After assigning performance percentile, we estimate the following linear model with two-way fixed effects (iteration and target image) for each effect type (model, prompting, and total):

\begin{equation}
\label{eq:skill_heter_model}
\begin{aligned}
Y_{i,n,t} = & \beta_0 + \beta_1 \text{effect} + \beta_2 \text{effect} \times \text{Performance Percentile}_{i,n,t} \, + \\
& \alpha_n + \gamma_t  + \epsilon_{i,n,t}
\end{aligned}
\end{equation}
where $\beta_1$ represents the coefficient associated with the effect type being analyzed (model, prompting, or total), and $\beta_2$ captures the interaction between the effect type and user skill level. User skill is measured as the user's rank (binned) within the same iteration, target image, and replay scenario. Standard errors are calculated using cluster-robust methods at the participant level.

This methodology closely follows the procedure described earlier in section \ref{sec:ate_decomposition}, with the additional step of incorporating an interaction term to evaluate skill-level heterogeneity.

\subsection{Logo Generation Task}

\subsubsection{Task Performance and ATEs: Logo Generation Task}
The analytic framework for the logo generation task mirrors that of the image replication task (Section~\ref{sec:task_performance_analysis}). As before, we estimate Equation~\ref{eq:ate-per-attempt-1}, treating iteration as a numeric variable and clustering standard errors at the participant level. The dependent variable here, however, is the average \emph{Bradley--Terry} ($\beta$) strength derived from the pairwise comparison framework described in Section~\ref{sec:bradley-terry}. 

Figure~\ref{fig:1abcnew}B reports the estimated performance trajectory across attempts. Consistent with the image replication task, we again observe a pronounced dip during the second iteration across all model conditions, potentially reflecting participants' temporary misunderstanding of the system's memoryless nature. As before, performance quickly rebounds from the third iteration onward once participants realize that previous prompts are not retained.

The ATE estimates in the main text are from a two-way fixed effect model (iteration and logo taks) with cluster robust standard errors at the participant level. For example, the overall average treatment effect (ATE) between DALL-E~3 and DALL-E~2 participants corresponds to a $0.8075$ increase in Bradley--Terry~$\beta$ (95\%~CI:~[0.7120,~0.9030], $p < 10^{-10}$), equivalent to roughly $0.57$~standard deviations of improvement. Similarly, the estimate comparing DALL-E~3 with revision  and DALL-E~2 is 0.834 (95\%~CI:~[0.7358,~0.9316], $p < 10^{-10}$).

Equation~\ref{eq:ate-per-attempt-1} accounts for variation in ATE over different attempts. The estimated coefficients from that model for the logo generation task are: $\hat{\beta_1} = 0.00493 ; (0.00628)$, $p=0.432$; $\hat{\beta_2} = 0.747 ; (0.0598)$, $p < 10^{-10}$; $\hat{\beta_3} = 0.0135 ; (0.00819)$, $p = 0.099$.
The above estimates suggest a similar pattern: the gap between DALL-E~3 and DALL-E~2 slightly increased over the attempts, although this divergence in the logo task is weaker and lacks sufficient statistical power.

\subsubsection{Prompt Characteristics: Logo Generation Task}
Prompt-level patterns in the logo generation task follow the same structure for image replication and the result is shown in Figure \ref{fig:1abcnew}C. Participants assigned to DALL-E~3 again produced significantly longer prompts (approximately 24.7\% longer on average) and exhibited higher successive prompt similarity relative to DALL-E~2 users.

As in the image replication task, superior-model users produced prompts that were more similar to their previous attempts. On average, successive prompt cosine similarity was $\beta = 0.03689$ higher for DALL-E3 users ($p = 0.00392$). 
Prompt sequences were also more internally consistent for superior-model participants (see section \ref{sec:prompt_similarity}). Their prompts lay $\beta = 0.03642$ closer to their own centroid in embedding space than those of DALL-E~2 users ($p < 10^{-6}$), using cluster-robust standard errors. Using the same analysis approach as in the image replication experiment, we also find that the proportion of nouns and adjectives is statistically indistinguishable across model conditions (43\% for DALL-E 3 vs. 42\% for DALL-E 2; $p = 0.12$)

\subsubsection{ATE Decomposition: Logo Generation Task}
\label{sec:ate_decomposition_logo}
The decomposition procedure for the logo generation task is identical to that described in Section~\ref{sec:ate_decomposition}. We again estimate Equation~\ref{eq:ind-dir-model} with two-way fixed effects for iteration and target item, cluster-robust standard errors at the participant level, and the effect-type indicator defined in the same way. 
The only methodological difference is the outcome: instead of CLIP cosine similarity, we use the Bradley--Terry~$\beta$ strength assigned to each generated logo. As before, replaying participant prompts across models yields estimates of the total, model, and prompting effects.

\subsubsection{Distributional Impacts: Logo Generation Task}
We replicate the analysis of Section \ref{sec:distribution_impact} to assess how total, model, and prompting effects vary across the performance distribution. The same two-way fixed effects model in Equation~\ref{eq:skill_heter_model} is used, substituting the Bradley--Terry~$\beta$ scores in place of cosine similarity outcome. Table~\ref{table:qte_logo} in the main text presents the corresponding regression coefficients.

\newpage

\section{Example Prompts} \label{sec:example_prompts}

\subsection{Image Replication}

Table~\ref{tab:prompts_image} presents the complete set of prompts and their corresponding generated images that were used in our analysis shown in Figure~\ref{fig:1abc}. The table displays three example target images (woman with shopping basket, abstract painting with cross, and online shopping interface), with three different participant-generated versions of each. For each target, we show examples of varying quality, arranged from most similar to least similar. The prompts are presented verbatim from participant submissions, preserving all original formatting and punctuation (or lack thereof).

\begin{table}[htbp]
\centering
\footnotesize
\caption{Participant prompts for the images provided in Figure~\ref{fig:1abc}}
\label{tab:prompts_image}
\begin{tabular}{p{1.5cm}p{10cm}}
\toprule
\textbf{Image} & \textbf{Prompt} \\
\midrule

\raisebox{-1.2cm}{\includegraphics[width=1.5cm]{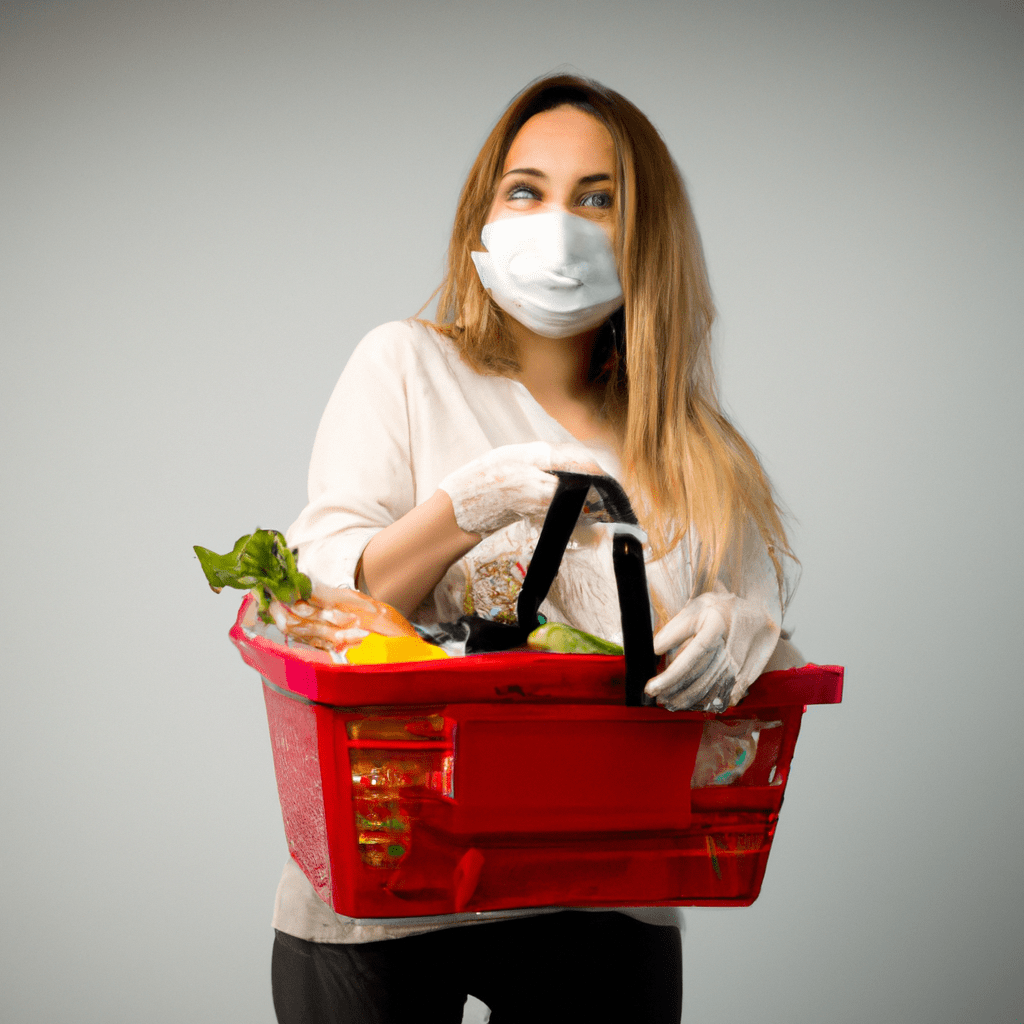}} & 
``create an image of a woman with blond hair, wearing a surgical mask, black gloves, a white t-shirt, and holding a red shopping basket full of groceries''\\
\midrule

\raisebox{-1.2cm}{\includegraphics[width=1.5cm]{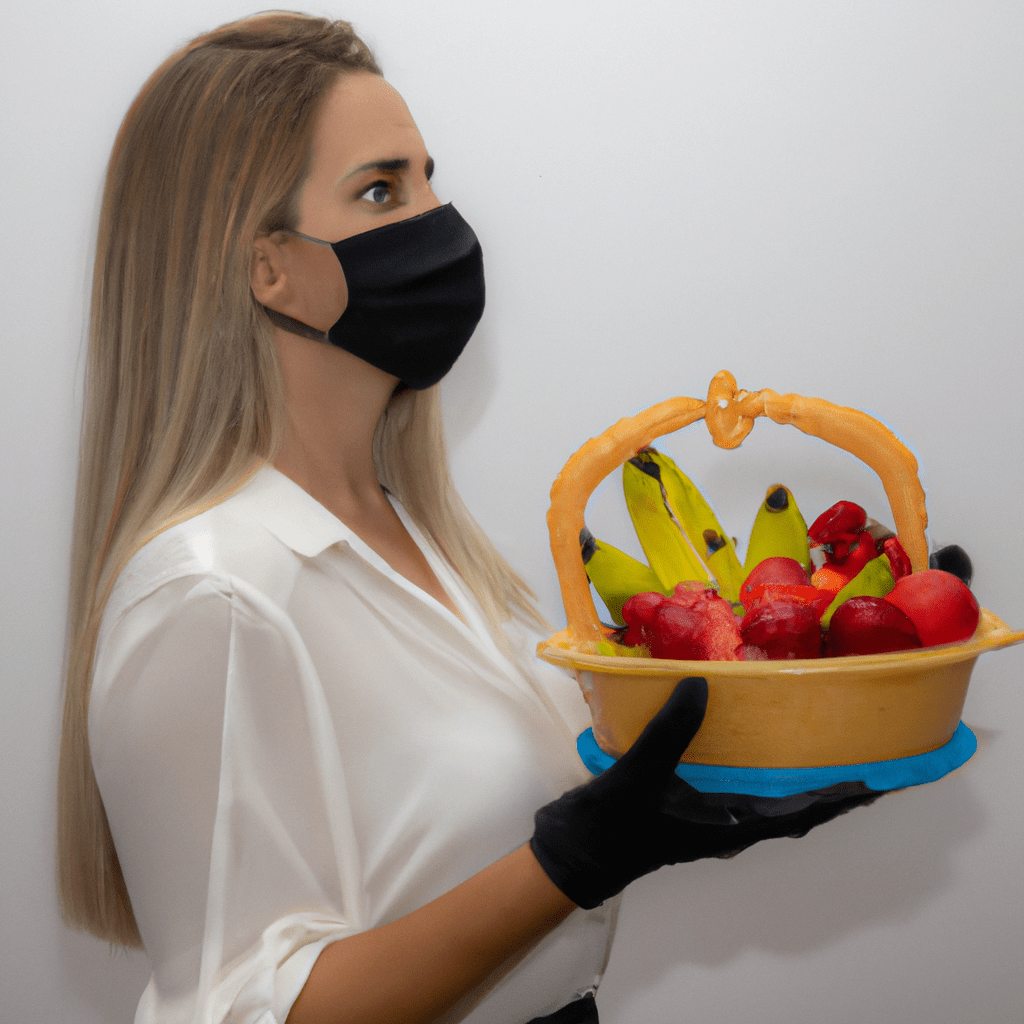}} &
``A lady has long blonde hair, wears BLACK COLORED GLOVES and a BLUE COLORED mask, with a white blouse on. She is holding a RED PLASTIC BASKET filled with different kinds of fruit.''\\
\midrule

\raisebox{-1.2cm}{\includegraphics[width=1.5cm]{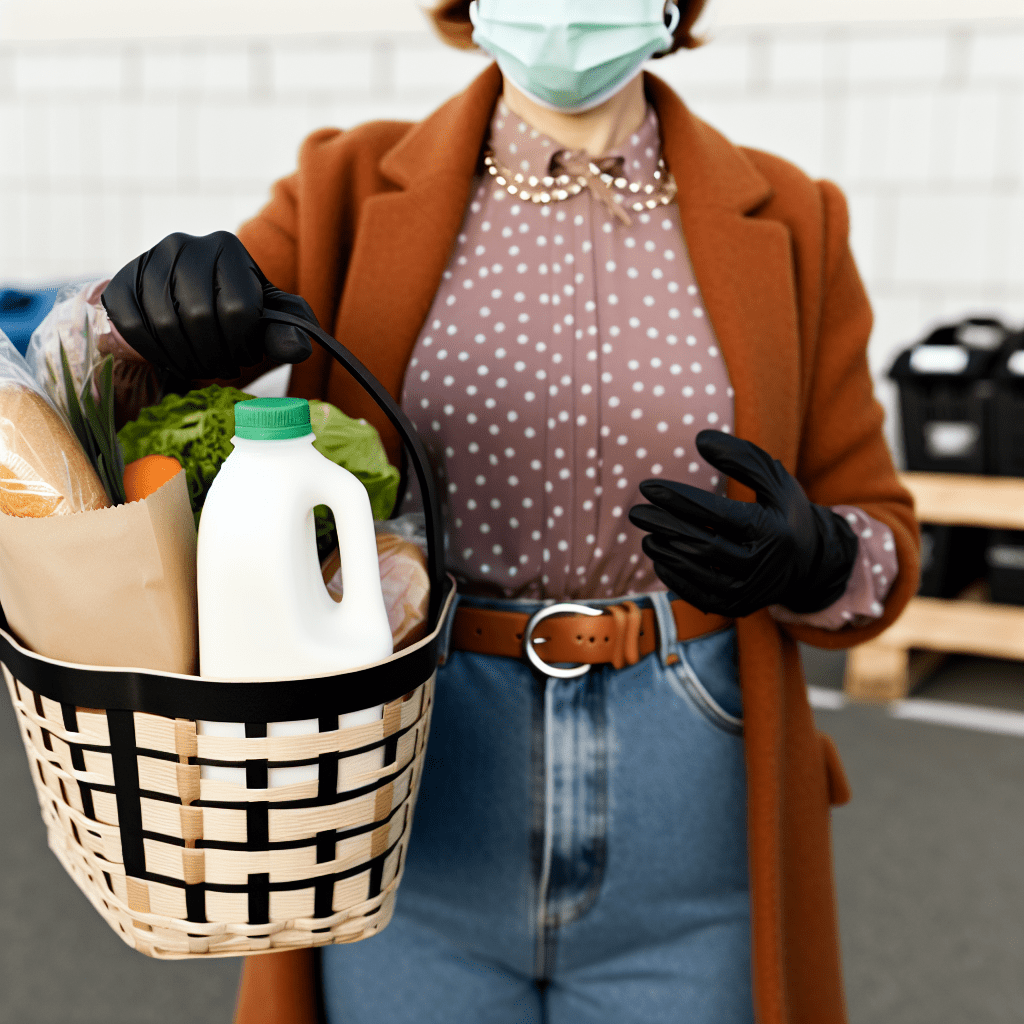}} &
``Generate a stock photo of a woman holding a basket of groceries during the pandemic. Make sure she has on black gloves.''\\
\midrule

\raisebox{-1.2cm}{\includegraphics[width=1.5cm]{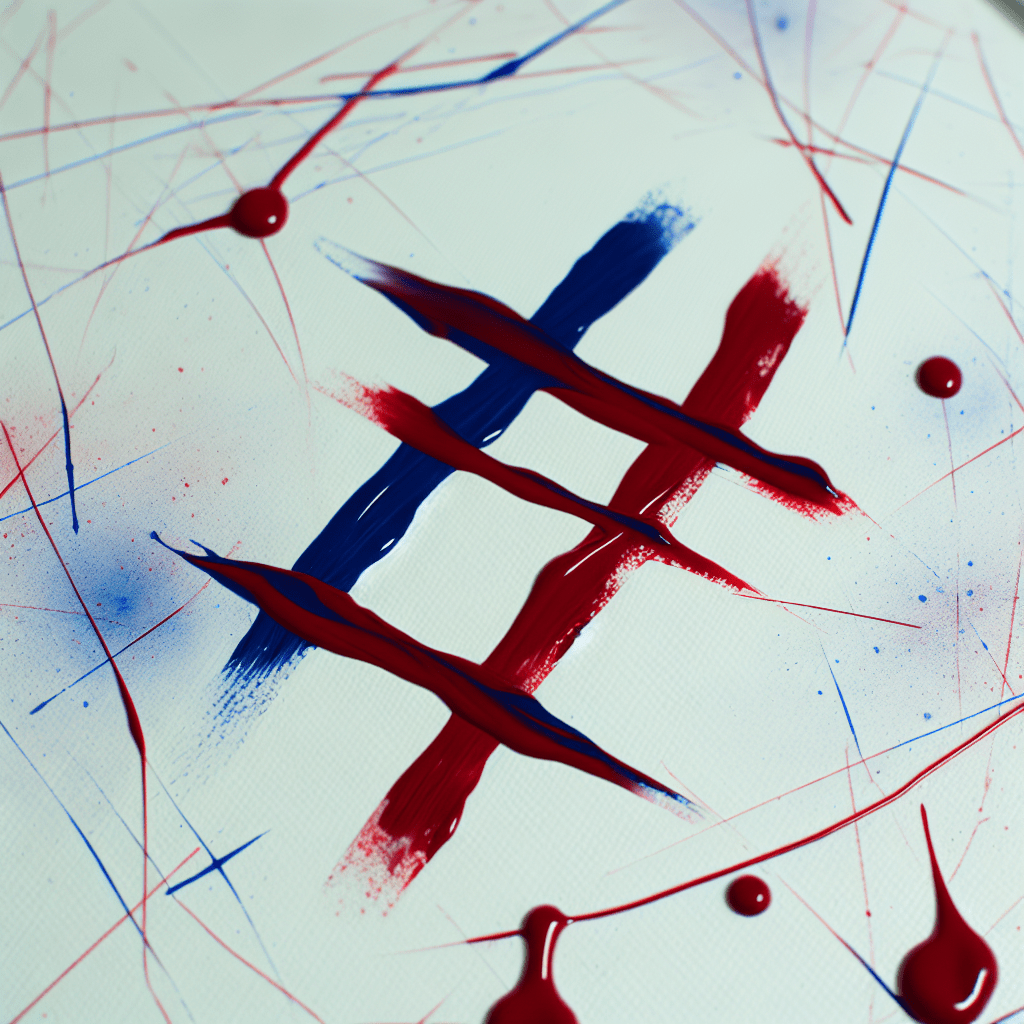}} &
``Thank you! Please create a mostly white painting with red and blue lines in the center, shaped a bit like the \# symbol. Add red and blue splatter as well, but not too much''\\
\midrule

\raisebox{-1.2cm}{\includegraphics[width=1.5cm]{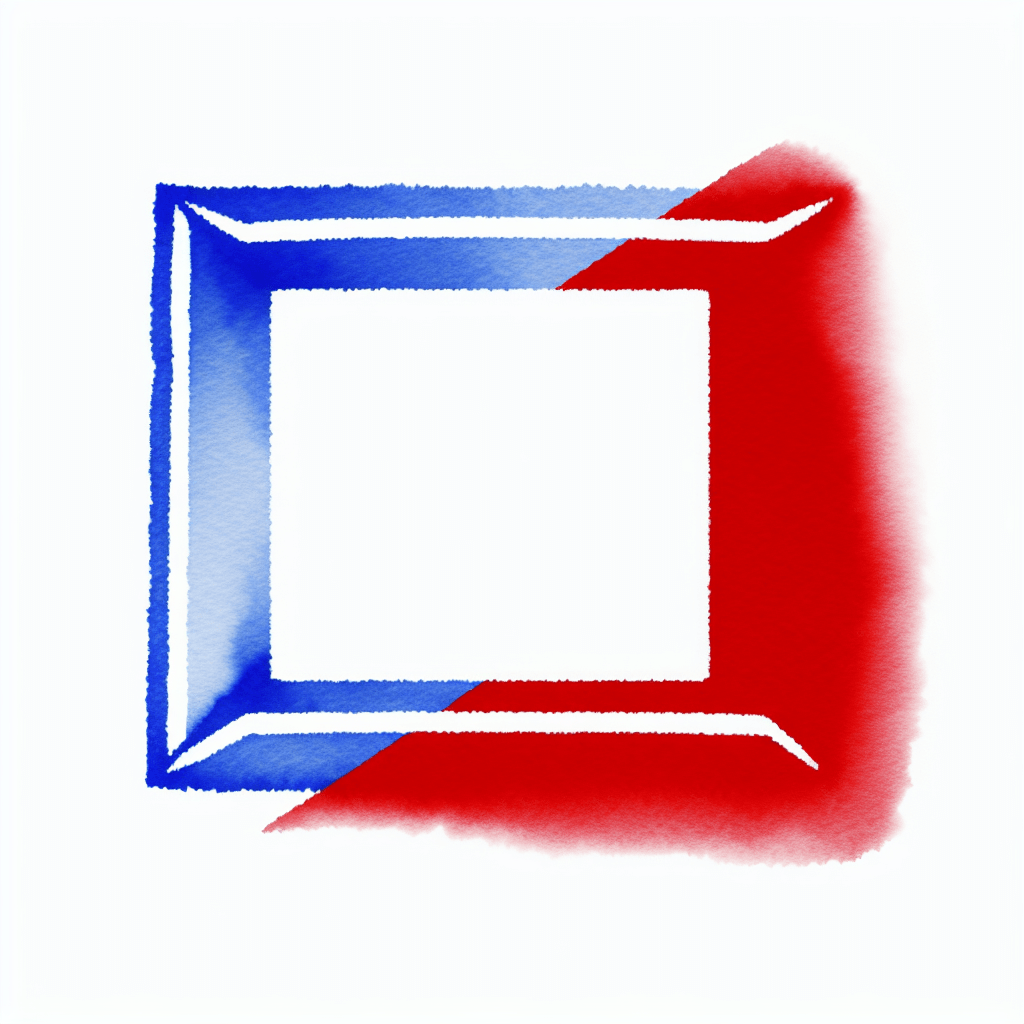}} &
``Water color style. Japanese vibes. White background. Square outline in the middle of the page is standing on its side at a 12 degree angle. The inside of the square is white. Two legs are blue and two are red. T''\\
\midrule

\raisebox{-1.2cm}{\includegraphics[width=1.5cm]{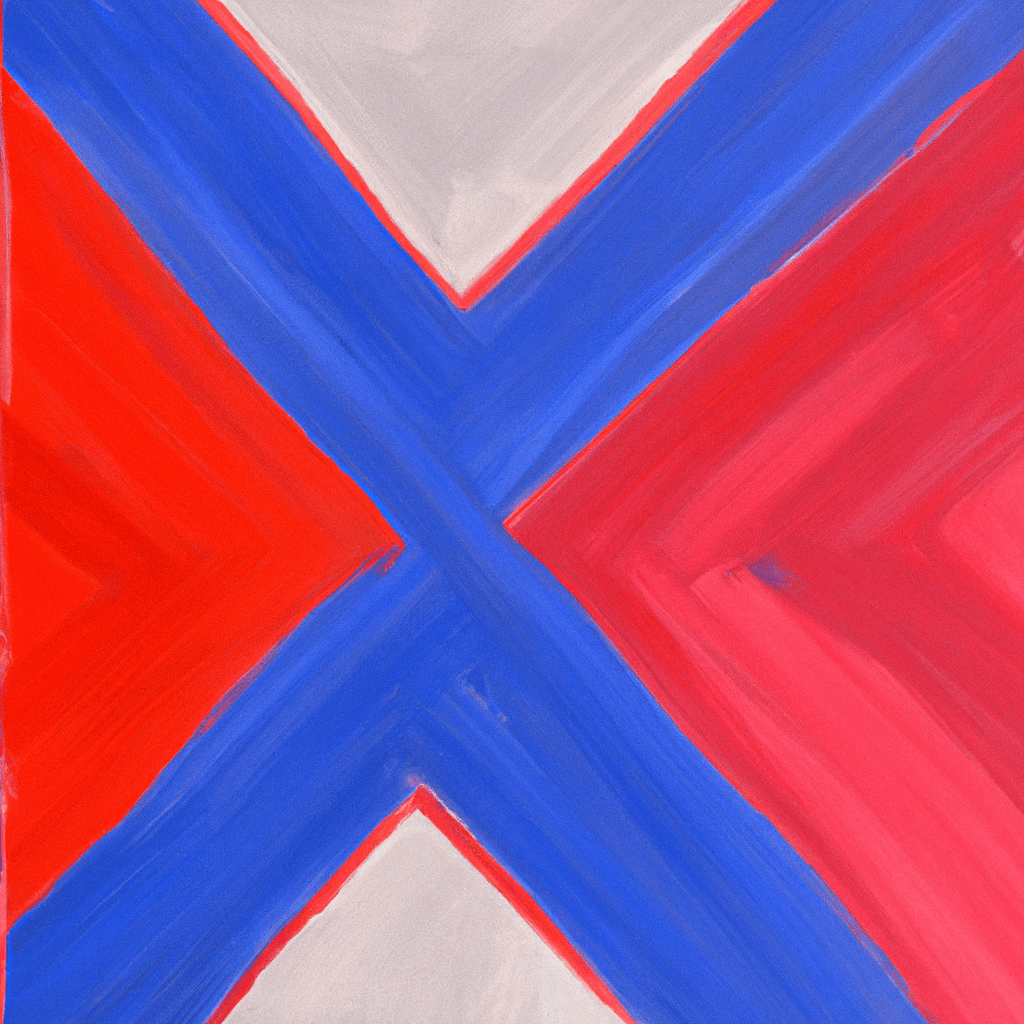}} &
``painting of two x's intersecting to make a slanted square. The square is at a 35 degree angle with the left upper side being two red lines making the x and the lower right side two blue lines. The painting only has red and blue and the square with 2 x's sits in the middle of the painting with light speckles of red and blue drops of paint scattered throughout the painting''\\
\midrule

\raisebox{-1.2cm}{\includegraphics[width=1.5cm]{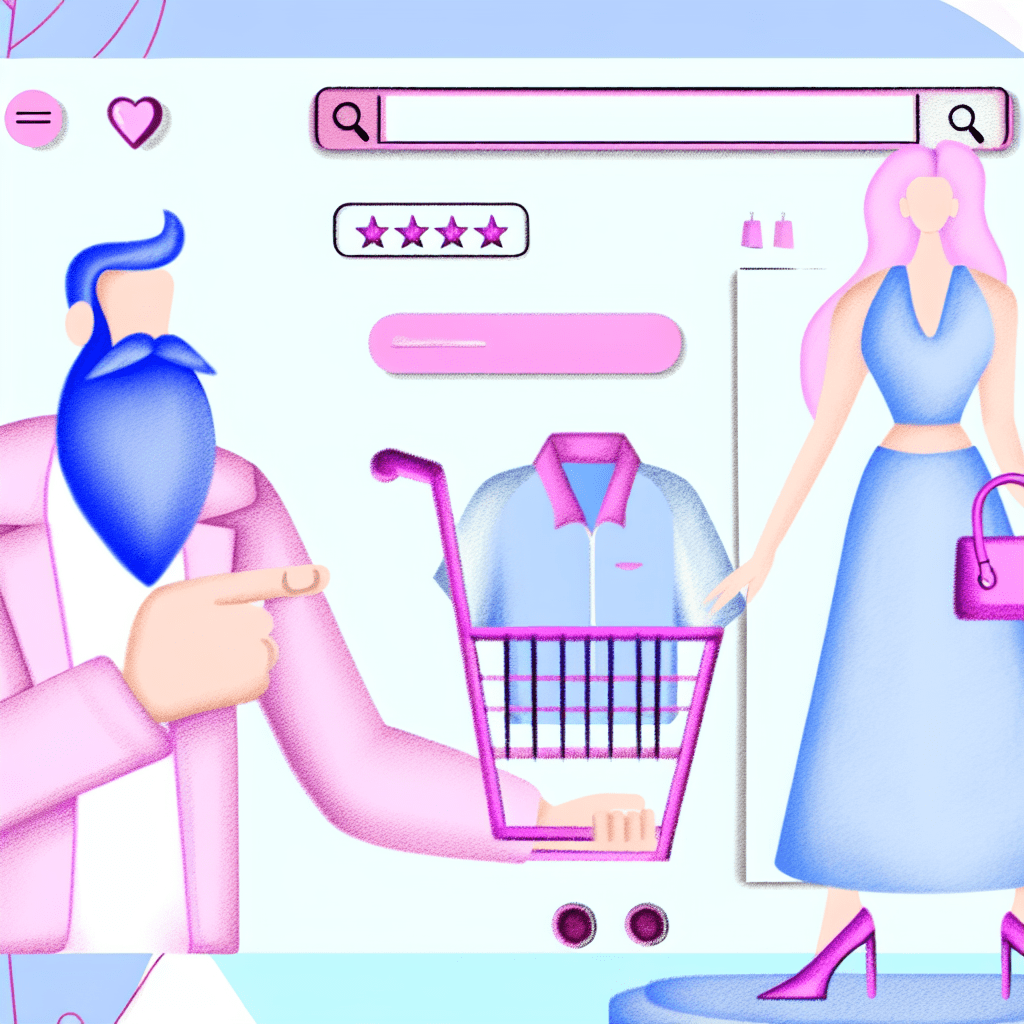}} &
``pastels, abstract, bearded man on left in a pink shirt with blue sleeves pointing to a shirt box image, woman with pink hair in a blue dress and high heels holding the handle of a shopping cart full of purses on right side, pink and white awning in center top, 4 out of 5 stars on left side in pink, animation style. single pink dollar sign on the left. white search bar in the center with a magnifying glass icon on the right side of the search bar.''\\
\midrule
\raisebox{-1.2cm}{\includegraphics[width=1.5cm]{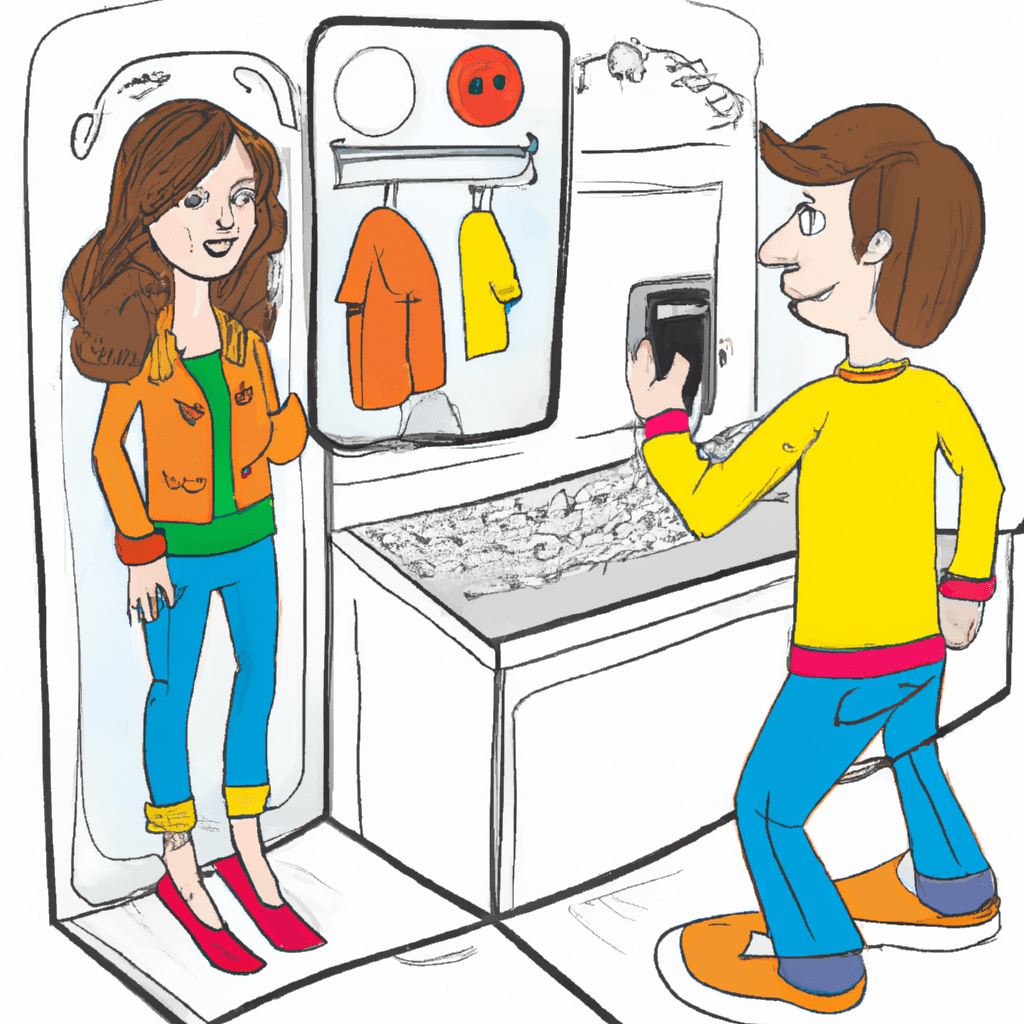}} &
``cartoon-y virtual self checkout. man shopping for clothes. woman pushing cart. phone that looks like a storefront in background.''\\
\midrule

\raisebox{-1.2cm}{\includegraphics[width=1.5cm]{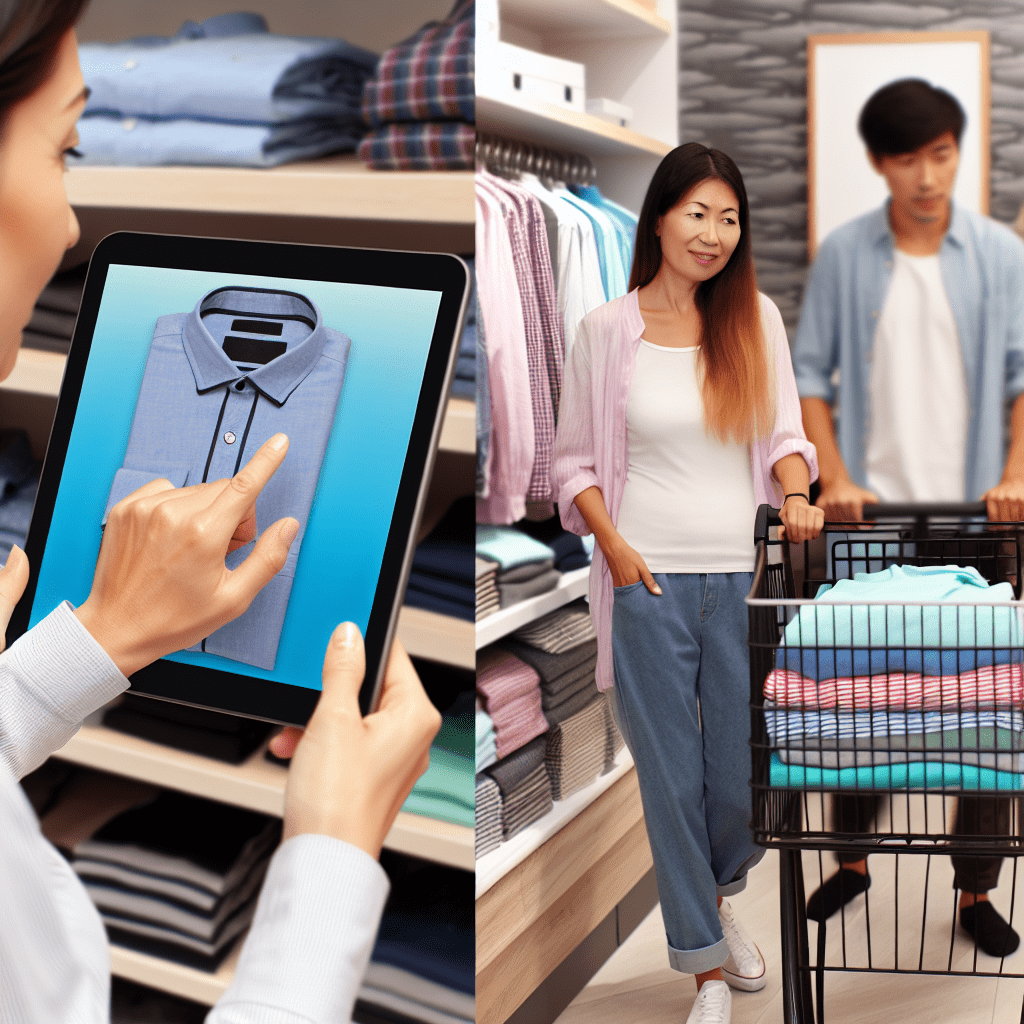}} &
``at the store a man using a touch screen monitor to shop for a shirt and a woman walking by with a store cart full of gift''\\
\bottomrule
\end{tabular}
\end{table}

\subsection{Example Logo Generation Prompts} 

Table~\ref{tab:prompts_logo} presents the complete set of prompts and their corresponding generated images that were used in our analysis shown in Figure \ref{fig:1abcnew} for the logo generation experiment.
It is the exact same format and presents the analogous information as Table~\ref{tab:prompts_image}, but for the logos.

\begin{table}[htbp]
\centering
\footnotesize
\caption{Participant prompts for the images provided in Figure~\ref{fig:1abcnew}}
\label{tab:prompts_logo}
\begin{tabular}{p{1.5cm}p{10cm}}
\toprule
\textbf{Image} & \textbf{Prompt} \\
\midrule

\raisebox{-1.2cm}{\includegraphics[width=1.5cm]{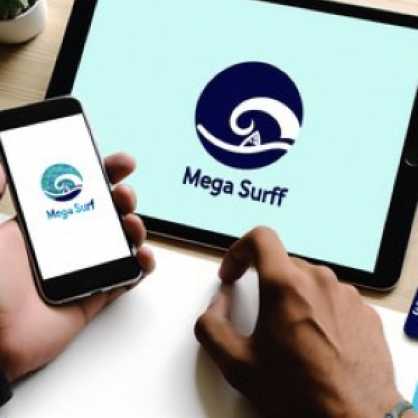}} &
``Acting as a graphic designer design a logo for mega surf. They are a tech startup company developing a mobile app that will connect surfers in a neighborhood. The logo must include a surfboard and an ocean wave ,be very attractive and appeal to new and existing users. It will be used for the app icon, digital marketing and business cards.''\\
\midrule

\raisebox{-1.2cm}{\includegraphics[width=1.5cm]{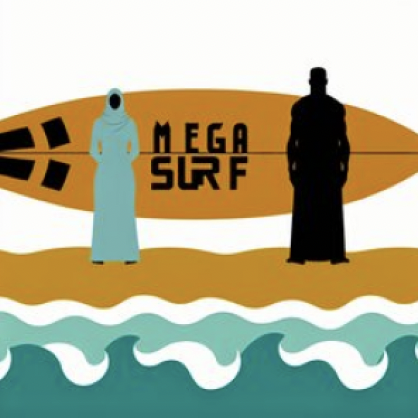}} &
``the words `MegaSurf' are the main focus of the image. a Large simple lightorange surf board ontop of a light teal wave two people holding surf baords can be seen on the shore. the logo is done in a simple sillhoutte style'\\
\midrule

\raisebox{-1.2cm}{\includegraphics[width=1.5cm]{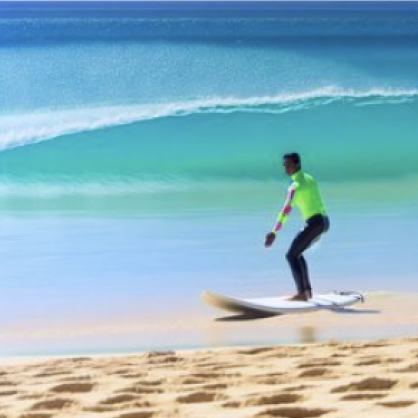}} &
``Create an image from the sandy beach showing a surfer on a surfboard and an ocean wave.''\\
\midrule

\raisebox{-1.2cm}{\includegraphics[width=1.5cm]{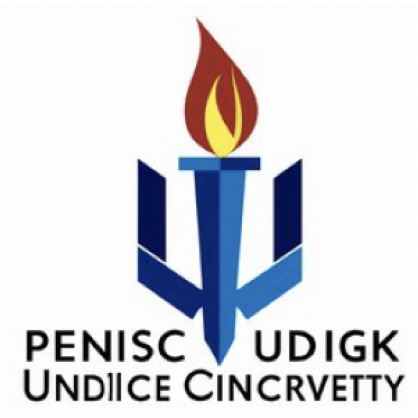}} &
``create a logo for Prestwick University that emphasizes academic excellence, and includes a torch with a red flame, an open book with writing in it, and the color blue''\\
\midrule

\raisebox{-1.2cm}{\includegraphics[width=1.5cm]{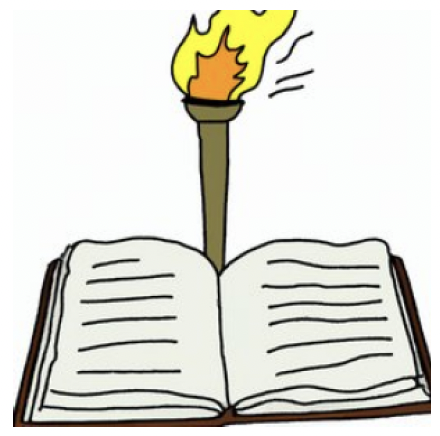}} &
``draw me a simple cartoon drawing of an open book and a torch''\\
\midrule

\raisebox{-1.2cm}{\includegraphics[width=1.5cm]{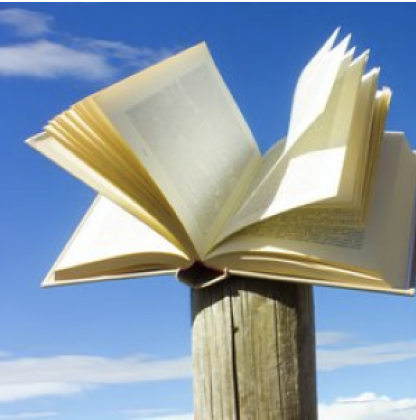}} &
``on the last post make the book open''\\
\midrule

\raisebox{-1.2cm}{\includegraphics[width=1.5cm]{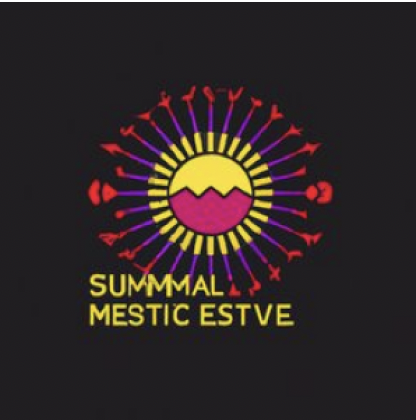}} &
``Create a unique design for a logo for an electronic music event, featuring a sun that emits sound waves instead of sun rays and musical notes. The name of the event is "Summer Beats Festival'' \\
\midrule

\raisebox{-1.2cm}{\includegraphics[width=1.5cm]{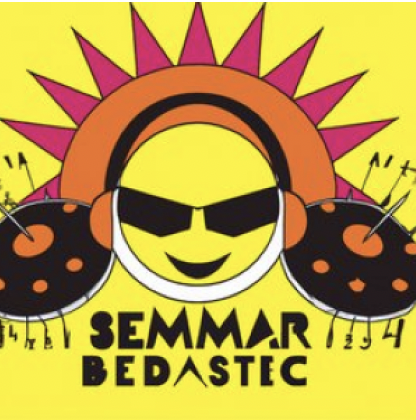}} &
``A logo for Summer Beats Festival. A large sun is the main character and he is the DJ dressed up in rave gear ready to party. Music is bumping loudly from speakers and you can see the sound waves coming off the speakers.''\\
\midrule

\raisebox{-1.2cm}{\includegraphics[width=1.5cm]{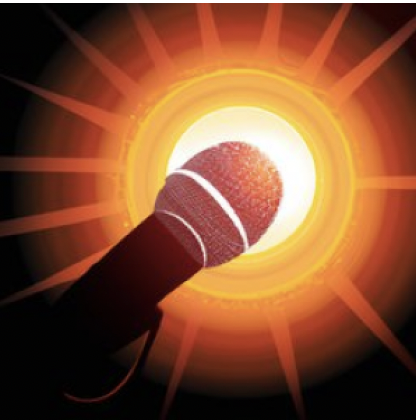}} &
``The sun holding a microphone. Sound waves coming out of the microphone.''\\
\bottomrule
\end{tabular}
\end{table}

\newpage

\section{Robustness Checks} \label{sec:robustness}

This section provides additional analyses that confirm the robustness of our main findings using alternative metrics and statistical approaches.

\subsection{Image Replication Task: DreamSim-based Analysis} \label{sec:dreamsim_robust}

As discussed in Section~\ref{sec:vars}, we repeated all main-text analyses using DreamSim as an alternative image similarity metric. The results, presented below, demonstrate that our findings are not dependent on the specific similarity measure used.

\subsubsection{Overall ATEs} 

In terms of DreamSim, participants using DALL-E 3 produced images that were, on average, $z=0.238$ standard deviations (95\% CI = [0.152, 0.324]) closer to the target image ($\Delta DreamSim = 0.0306$, $p<10^{-7}$) than those produced by participants using DALL-E 2.

\subsubsection{Figure 1} We reran the regression in Section~\ref{sec:analysis}  with $Y_{i,n,t}$ representing the DreamSim outcome instead of cosine similarity:

\begin{equation} \label{eq:ate-per-attempt-2}
\begin{aligned}
Y_{i,n,t} =\; & \beta_0 + \beta_1 \, \text{iteration} + \beta_2 \, \mathbb{I}[\text{dalleVersion = 3}]_i \\
              & + \beta_3 \, \text{iteration} \times \mathbb{I}[\text{dalleVersion = 3}]_i + \gamma_t + \epsilon_{i,n,t}
\end{aligned}
\end{equation}

\noindent The coefficient estimates generated by this analysis are: $\hat{\beta_1} = 0.0034 ; (0.0005)$, $p = 1.3 \times 10^{-12}$; $\hat{\beta_2} = 0.0200 ; (0.0061)$, $p = 0.0011$; $\hat{\beta_3} = 0.0024 ; (0.0007)$, $p = 0.0009$.

\subsubsection{Figure 2}
Decomposing the ATE as measured in terms of DreamSim, we find similar results to those in the main text. The model effect accounts for 54.4\% of the ATE ($\Delta DreamSim = 0.0166$, $p<10^{-7}$), whereas the prompting effect accounts for 45.4\% of the ATE ($\Delta DreamSim = 0.01390$, $p=0.014$).

\subsection{Image Replication Task: Z-score-based Analysis} \label{sec:z_score_robust}

As we discuss in Section~\ref{sec:zscoring}, we repeated all main-text analyses using the within-image-attempt Z-score of CLIP-cosine similarity to account for variation between images and attempts. When comparing across attempts, the Z-score was computed within each image, as mentioned in Section~\ref{sec:zscoring}. The results remain consistent throughout, and in some cases, the differences between DALL-E 3 and DALL-E 2 are even more pronounced than when using raw cosine similarity. 

\subsubsection{Overall ATEs}
\label{si:z_score_robust_ate}
As mentioned in the main text, participants using DALL-E 3 produced images that were, on average, $z=0.19$ standard deviations (obtained from ATE in terms of Z-Scored Cosine Sim = 0.19, 95\% CI = [0.100, 0.271]) closer to the target image ($\Delta CoSim = 0.0164$, $p<10^{-5}$) than those produced by participants using DALL-E 2. Standard errors are clustered at the participant level. 

\subsubsection{Figure 1}
\label{si:z_score_robust_fig1}
On average, participants using DALL-E 3 produced images that were $z = 0.19$ standard deviations closer (the ATE) to the target image than those using DALL-E 2. Like in the main text with CLIP cosine similarity, this treatment effect increased as participants made successive attempts to replicate the target image.

\begin{equation}
\begin{aligned}
\text{ZScore}_{i,n} =\; & \beta_0 + \beta_1 \, \text{iteration} + \beta_2 \, \mathbb{I}[\text{dalleVersion = 3}]_i \\
                        & + \beta_3 \, \text{iteration} \times \mathbb{I}[\text{dalleVersion = 3}]_i + \epsilon_{i,n}
\end{aligned}
\end{equation}

\noindent The coefficient estimates generated by this analysis are: $\hat{\beta_1} = 0.0129 ; (0.0038)$, $p = 0.0007$; $\hat{\beta_2} = 0.1250 ; (0.0457)$, $p = 0.0064$; $\hat{\beta_3} = 0.0128 ; (0.0053)$, $p = 0.015$.

\subsubsection{Figure 2}
\label{si:z_score_robust_fig2}
When we decompose the ATE into the model effect ($z = 0.0791$; $p=8.35\times10^{-6}$) and prompting effect ($z = 0.1046$; $p=0.016$), they account for 43\% and 56\% of the treatment effect, respectively. And when we replay DALL-E 3 prompts on DALL-E 2, there is not a statistically significant increase in performance relative to DALL-E 2 prompts played on DALL-E 2 ($z = -0.033$; $p=0.45$). In short, Figure 2 in the main text is quantitatively and qualitatively unchanged when using the within-image Z-score of the cosine similarity.

\subsection{Logo Generation Task: Z-score-based Analysis}
\label{sec:zscore_logo}

The Z-score procedure for the logo generation task follows the same approach described in  Section~\ref{sec:z_score_robust} for the image replication task. As before, we compute within-logo  Z-scores to normalize Bradley--Terry (BT) strengths, allowing us to remove  baseline variation in difficulty across logo task. The only substantive difference is the outcome:  here the dependent variable is the BT $\beta$ score described in Section~\ref{sec:bradley-terry} rather  than CLIP cosine similarity.

\subsubsection{Overall ATEs}
Applying the same Z-scoring procedure, we find that participants using DALL-E~3 produced logos that  were, on average, $z = 0.569$ standard deviations higher in BT strength than those produced by  DALL-E~2 users (95\% CI = [0.502, 0.636],  $p < 10^{-10}$).
As in the unscaled BT analysis reported in the main text, these results indicate  a substantial and statistically significant advantage for DALL-E 3.

\subsubsection{Figure 3}
Repeating the regression specification used in Section \ref{si:z_score_robust_fig1}, but substituting 
Z-scored BT strengths as the dependent variable, we again observe that the performance gap between 
DALL-E~3 and DALL-E~2 widens modestly across attempts, but we lack sufficient power to detect statistically significant effects. One potential reason is the dip in second–attempt performance, which flattens the across-iteration trend and reduces the estimated slope in a linear model.
The estimated coefficients are: $\hat{\beta_1} = 0.0043 ; (0.0043)$, $p = 0.32$; $\hat{\beta_2} = 0.5104 ; (0.0412)$, $p < 2.2 \times 10^{-16}$; $\hat{\beta_3} = 0.0067 ; (0.0057)$, $p = 0.238$.

\subsubsection{Figure 4}
The replay-based decomposition is unchanged conceptually from Section \ref{si:z_score_robust_fig2}. Using Z-scored BT strengths, we estimate the model effect ($z = 0.522$; $p< 10^{-15}$) and prompting effect ($z = 0.046$; $p=0.167$) to constitute about 91.8\% and 8.1\% of the total treatment effect, respectively.
The relative sizes of model and prompting effects using the z-scaled scores are similar to the unscaled analysis reported in the main text.

The replay of DALL-E~3 prompts on DALL-E~2 shows no meaningful improvement ($z = -0.0028$, $p = 0.934$).

\newpage

\section{Full pre-registered analyses (Image Replication Experiment)} \label{sec:pre-reg}

Prior to conducting the image replication experiment, we developed and pre-registered a comprehensive set of hypotheses and analyses. This pre-registration is deposited at OSF at the following URL: 
\textbf{URL removed to maintain anonymity}.
While the main text of our paper focuses on a subset of our pre-registered analyses, this appendix provides a comprehensive overview of all analyses specified in our pre-registration. Note that for the logo generation task, we limited our pre-registered hypotheses to the analyses appearing in the main text of this paper.

In our pre-registration, we outlined plans to conduct each analysis using eight\footnote{See section \ref{sec:deviations} on deviations from the pre-registration for the remaining two pre-registered outcome variables.} possible outcome variables, all representing different transformations of the same underlying data. These include CLIP embedding cosine similarity and DreamSim, each with three rescaling methods:

\begin{enumerate}
    \item \textbf{No rescaling}: The outcome variable used in its original form.
    \item \textbf{Z-score rescaling}: The outcome variable is transformed into a Z-score following the procedure detailed in Section~\ref{sec:zscoring}.
    \item \textbf{Percentile rank rescaling}: The outcome variable is converted to a percentile rank relative to all other prompts submitted for the same target image.
\end{enumerate}

There is one additional rescaling method that we pre-registered, but do not conduct our analyses with this rescaling. The reason that we do not use it is detailed in the Section discussing deviations from the pre-registration (Section \ref{sec:deviations}).

\subsection{Hypotheses and Results}

Below, we present each hypothesis exactly as stated in our pre-registration document, followed by a summary of our findings. This comprehensive review demonstrates the robustness of the key results highlighted in the main text, while also providing additional insights into the relationship between model capabilities, user behavior, and task performance.

\paragraph*{H1.} \textit{There are differences in prompt engineering ability (as measured through metrics such as average expected prompt quality, initial expected prompt quality, and max expected prompt quality) across demographic attributes and other observables, such as educational background and occupational skills.}

\paragraph{Analysis approach:} For this hypothesis, we conducted multiple ANOVA tests, one for each demographic variable, with the relevant performance measure as the dependent variable and the model (DALL-E version) as a covariate. As a robustness check, we repeated these analyses using the non-parametric Kruskal-Wallis U test. To account for multiple comparisons, we applied the Benjamini-Hochberg adjustment with a false discovery rate of 0.05.

Our analyses revealed several consistent demographic patterns in prompt engineering ability across different outcome measures:

\paragraph{CLIP no rescaling.}
Using ANOVA, we found significant associations between performance and: computer programming frequency, self-reported programming ability, outlook towards generative AI, age, gender, generative AI use, education, imagery writing skill, and self-reported occupational skills (particularly critical thinking, active listening, and quality control).

In linear models examining directionality, we found that participants who reported critical thinking as a job skill and little usage of generative AI or imagery writing skill performed better, on average. Conversely, older participants, men, those with more positive outlooks regarding generative AI, those reporting quality control as an occupational skill, and frequent programmers performed worse on our task.

The more conservative Kruskal-Wallis tests identified fewer significant variables: self-reported programming frequency, outlook towards generative AI, self-reported programming skill, gender, and age.

\paragraph{CLIP Z-score rescaling.}
ANOVA tests revealed significant associations with similar demographic factors as the unscaled measure, plus additional significant associations with self-reported occupational skills in technology design, social perceptiveness, and troubleshooting. Linear models showed that participants reporting critical thinking and troubleshooting as job skills, little generative AI use, little imagery writing skill, some programming skill, and some instruction writing skill performed better. Worse performance was associated with older participants, men, positive generative AI outlook, graduate degrees, self-reported quality control and technology design skills, and frequent programming experience. Kruskal-Wallis tests again identified a smaller set of significant predictors: self-reported programming frequency, generative AI outlook, programming skill, gender, and age.

\paragraph{CLIP Percentile rank rescaling.} 
ANOVA identified significant associations between performance and computer programming frequency, self-reported programming ability and frequency, generative AI outlook, age, gender, generative AI use, and critical thinking as an occupational skill. Linear models showed better performance among those reporting critical thinking as a job skill, little generative AI use, and some programming skill. Worse performance was associated with older participants, men, positive generative AI outlook, and frequent programming experience. Kruskal-Wallis tests revealed the same set of significant predictors as with other scalings: programming frequency, generative AI outlook, programming skill, gender, and age.

\paragraph{DreamSim no rescaling.} 
ANOVA tests identified significant associations between performance and: programming frequency, self-reported programming ability, generative AI outlook, age, gender, generative AI use, imagery and instructional writing skills, and several occupational skills (critical thinking, learning strategies, technology design, and quality control). Linear models showed better performance among those reporting critical thinking and social perceptiveness as job skills, little generative AI use, and little imagery writing or programming skills. Worse performance was associated with older participants, men, positive generative AI outlook, and frequent programming experience. Kruskal-Wallis tests identified significant relationships with: programming frequency, generative AI outlook, imagery writing skill, gender, age, and learning strategies as an occupational skill.

\paragraph{DreamSim Z-score rescaling.}
Using ANOVA, significant associations included all factors identified in the unscaled analysis, plus education and additional occupational skills (social perceptiveness). Linear model directional findings were consistent with the unscaled measure, with additional negative associations for graduate degrees and self-reported technology design, quality control, and learning strategies skills. Kruskal-Wallis tests showed significant relationships with: programming frequency, generative AI outlook, imagery writing skill, gender, age, and occupational skills in learning strategies, social perceptiveness, and technology design.

\paragraph{DreamSim Percentile rank rescaling.}
ANOVA identified significant associations with: programming frequency, programming ability, generative AI outlook, age, gender, generative AI use, imagery writing skill, education, and occupational skills in social perceptiveness, learning strategies, and technology design. Linear models showed better performance among participants reporting social perceptiveness as a job skill, little generative AI use, and little imagery writing or programming skills. Worse performance was associated with older participants, men, positive generative AI outlook, frequent programming experience, and self-reported technology design and learning strategies skills. Kruskal-Wallis tests revealed significant relationships with: programming frequency, generative AI use, generative AI outlook, imagery writing skill, gender, age, and occupational skills in learning strategies, social perceptiveness, and technology design.

\paragraph{Interpretation.}
Across these analyses, several consistent patterns emerge. Interestingly, we found that frequent programming experience was associated with worse task performance, contrary to what might be expected. Men and older participants also performed worse on average. Those reporting critical thinking or social perceptiveness as job skills tended to perform better, while participants with positive outlooks toward generative AI and higher generative AI usage performed worse. These patterns were generally consistent across different outcome measures and scaling approaches, suggesting robust demographic differences in prompt engineering ability.

\paragraph*{H2.} \textit{There are observable differences in the prompting techniques of successful prompt engineers and unsuccessful prompt engineers. Such prompting techniques might include the use of longer prompts, the use of structured prompting techniques, and/or specific patterns in the way that the participant iterates on their prompts over time.}

\paragraph{Analysis approach.} We investigated this hypothesis by examining the relationship between exploration/exploitation strategies in the prompting space and performance outcomes. We characterized participants as more "exploitative" if they wrote prompts that were similar to their previous prompts, and more "exploratory" if their prompts exhibited greater deviation from previous attempts. We operationalized these concepts using several metrics. 

The measures positively associated with exploitation are: average token sort ratio compared to the previous prompt, average cosine similarity between the embeddings of consecutive prompts, and the fraction of times a prompt contains the previous prompt as an exact substring. The 
measures negatively associated with exploitation are: variance of the prompt embedding and the number of topical transitions

We also measured each participant's average prompt length. These variables were calculated for each user across their first 10 attempts, and their association with performance was estimated using linear models with DALL-E version fixed effects.

To examine how exploration/exploitation strategies influence performance across successive attempts, we conducted an additional analysis at the iteration level. We divided user-iteration observations into 6 equal-sized brackets based on performance in the previous iteration, allowing us to explore how prompting behavior might vary depending on the quality of previous attempts. We then estimated the effect of textual similarity to the previous prompt on the quality of the next attempt within each bracket. Our estimates adjusted for other covariates by matching user-iteration observations on target image, DALL-E model, iteration number, and exact quality of the previous attempt \citep{savje2021generalized}.

Our analyses revealed consistent patterns across different outcome measures:

\paragraph{CLIP No rescaling.} At the user level, we found strong and statistically significant associations between performance and prompting strategies, even after Benjamini-Hochberg adjustment for multiple testing. Specifically, token sort ratio, cosine similarity with the previous prompt, and frequency of including the previous prompt were all positively associated with performance. Conversely, embedding variance and number of topical transitions showed negative correlations with performance. Together, these findings indicate that more successful users engaged in greater exploitation, writing prompts that exhibited higher similarity to one another. As noted in the main text, we also found that longer prompts were associated with higher performance.

At the user-prompt level, we observed an interesting pattern: when previous performance was poor, moderate exploration (lower cosine similarity with the previous prompt) was associated with improved subsequent performance, though extremely high levels of exploration did not yield additional benefits. In contrast, when previous performance was high, increased exploitation consistently improved subsequent performance. Similar results emerged when using token sort ratio as the measure of exploration.

With binary measures of exploration (topical transitions or containing the previous prompt), exploitation was associated with higher performance in the next iteration regardless of previous performance bracket. Since these are binary measures, they couldn't capture the non-linear relationship observed with continuous measures in low-performing groups. Nevertheless, we found that topical transitions (more exploration) led to overall performance decreases, with larger decreases for higher previous performance. Similarly, prompts that included the previous prompt showed higher performance, with larger improvements as previous performance increased.

\paragraph{CLIP Z-score rescaling.} Results were consistent with those for unscaled cosine similarity. The primary difference was that the non-linear relationship between performance and continuous measures of exploitation (TSR ratio and cosine similarity with previous prompt) became more pronounced for the bottom two brackets of previous performance. For low-performing prompts in the previous attempt, moderate exploitation levels yielded optimal performance, with significant performance deterioration at both high and low exploitation levels.

\paragraph{CLIP Percentile rank rescaling.} We found the same general patterns described above, with the non-linearity in exploration/exploitation versus performance in the bottom two brackets of previous performance even stronger than observed with Z-score rescaled cosine similarity.

\paragraph{Dream Sim No rescaling.} Results were consistent with those described for unscaled cosine similarity.
\paragraph{DreamSim Z-score rescaling.} Results matched those described for Z-score cosine similarity, with particularly stark non-linearity between performance and continuous measures of exploitation at the user-prompt level for the bottom two brackets of previous performance.
\paragraph{DreamSim Percentile rank rescaling.} Results were consistent with those described for percentile rank rescaled cosine similarity.

\paragraph{Interpretation.} These findings suggest a nuanced relationship between exploration/exploitation strategies and performance. Users who were generally more successful tended to employ more exploitative strategies, refining and building upon previous prompts rather than making dramatic changes. However, at the iteration level, the optimal strategy depended on previous performance: when performance was already high, continued exploitation yielded the best results; when performance was poor, moderate exploration was beneficial. This pattern was consistent across different outcome measures and scaling approaches.

\paragraph*{H3.} \textit{There are differences in prompt engineering techniques (as measured through metrics such as prompt length and iteration-to-iteration token sort ratio) across demographic attributes and other observables, such as educational background and occupational skills.}

\paragraph{Empirical approach:} To test this hypothesis, we estimated linear models with each demographic trait as the independent variable, treatment arm fixed effects as controls, and various prompting behaviors (described in Section~\ref{sec:vars}.\ref{sec:dep-var}) as the dependent variables. To account for multiple testing, we adjusted p-values across all models using the Benjamini-Hochberg procedure. 

Our analysis revealed several significant demographic differences in prompting strategies, with consistent patterns emerging across different measures of prompt similarity and evolution. The detailed findings for each dependent variable are summarized below:

\paragraph{Prompt embedding variance.} We did not find any significant differences across demographic traits in the overall variance of prompt embeddings, suggesting that the breadth of prompt space exploration was relatively consistent across different demographic groups.
    
\paragraph{Strategic shifts.} We observed statistically significant differences based on age, programming frequency, and instructional writing frequency. Specifically, younger participants, those with low programming frequency, and those with some instructional writing experience demonstrated fewer topical transitions across their prompts, indicating a more consistent approach to prompt development.
    
\paragraph{Successive prompt token sort ratio.} Significant differences emerged by age, education, programming frequency, and imagery/instructional writing frequencies. Older users, those with post-graduate degrees, those with high programming frequency, and those with high writing frequency (both for precise instructions and imagery) wrote prompts that were less similar to their previous attempts as measured by token sort ratio. This suggests these groups took a more exploratory approach to prompt iteration.
    
\paragraph{Successive similarity.} We found significant differences across numerous demographic factors: age, gender, education, generative AI outlook, programming skill/frequency, instructional/imagery writing frequency, and certain occupational skills. On average, older users, males, those with post-graduate degrees, frequent programmers, self-reported skilled programmers, those with positive outlooks toward generative AI, frequent generative AI users, those who frequently write instructions or imagery, and those reporting critical thinking and social perceptiveness as occupational skills wrote prompts with lower cosine similarity to their previous attempts. This consistent pattern across multiple demographic factors suggests robust differences in exploration-exploitation tendencies.
    
\paragraph{Successive prompt 'contains previous prompt' dummy.} We found significant differences by age, gender, outlook toward generative AI, and imagery writing frequency. Older users and those with neutral outlooks toward generative AI were less likely to write prompts that contained their previous prompt as an exact substring. In contrast, males and those with some imagery writing skills were more likely to build directly upon their previous prompts by including them in subsequent attempts.

\paragraph{Interpretation.}
These findings reveal a nuanced picture of demographic differences in prompt engineering approaches. Overall, younger users, those with less programming experience, and those with less generative AI experience tended to employ more exploitative strategies, building more consistently upon their previous prompts. In contrast, older users, those with more technical backgrounds, and those with more exposure to generative AI systems exhibited more exploratory behaviors, making larger changes between successive prompts.

Interestingly, while H1 showed that a more exploitative approach was generally associated with better performance, here we find that demographics associated with technical expertise (programming experience, education, etc.) tend toward more exploratory approaches. This apparent contradiction suggests that the relationship between background, prompting strategy, and performance is complex and potentially mediated by other factors not captured in these analyses.

\paragraph*{H4.} \textit{Insofar as the output returned by a generative AI model in response to a prompt is stochastic, the subsequent prompting strategies and prompting outcomes of participants that get lower-than-expected, higher-than-expected, or approximately expected outputs in response to their first prompt are different.} 

\paragraph{Empirical approach:} To test this hypothesis, we examined how the random variation in image quality resulting from model stochasticity affects subsequent user behavior. We first calculated a Z-score for each participant's generated image relative to the expected distribution for that prompt, following the procedure outlined in Section \ref{sec:vars}.\ref{sec:image-data} (with computational details in Section \ref{sec:methods}). This Z-score quantifies how much better or worse the actually generated image was compared to what would be expected for that prompt on average. Higher Z-scores represent instances where the realized image quality exceeded expectations, while lower Z-scores indicate instances where the quality was randomly lower than expected.

To analyze the relationship between this stochasticity and subsequent prompting behavior and performance, we transformed the Z-score into a trichotomous variable with three categories: "lower-than-expected" (Z-score < -0.45), "expected" (-0.45 $\leq$ Z-score $\leq$ 0.45), and "higher-than-expected" (Z-score > 0.45). We then performed two-sample t-tests comparing performance and prompting behavior across these three groups. As a robustness check, we also estimated linear models regressing performance and prompting measures on both the trichotomous Z-score variable and the continuous Z-score, with treatment arm fixed effects. Additionally, we conducted analyses at the user level to determine whether the Z-score of the first prompt affects average performance across all subsequent attempts.

Our analyses revealed several consistent patterns across different outcome measures:

\paragraph{CLIP No rescaling.} We found statistically significant evidence that higher Z-scores for images in previous prompts were associated with increased cosine similarity in subsequent attempts. When comparing performance across the trichotomous Z-score variable, the difference between the top bracket ("higher-than-expected") and bottom bracket ("lower-than-expected") was statistically significant, with the top bracket showing better performance in the next attempt. However, we did not find significant differences between the middle bracket and either the top or bottom brackets.

The linear model did not yield a statistically significant relationship between Z-score and performance in the next iteration, likely due to non-linearity in this relationship across the negative and positive ranges of the Z-score distribution.

At the user level, we found differences between the top and middle brackets of the first prompt's Z-score, though these differences were marginally significant (p=0.048) and did not account for multiple testing. Linear models at the user level did not reveal statistically significant relationships between the first prompt's Z-score and average performance in subsequent attempts.

\paragraph{CLIP Z-score rescaling.} Results were similar to those for unscaled cosine similarity. We found a statistically significant difference between the top and bottom Z-score brackets, with higher performance in the next attempt for the top bracket. No statistically significant effects emerged when comparing the middle bracket with either the top or bottom brackets, or when using linear models. Unlike with the unscaled measure, we found no statistically significant effects at the user level, either when comparing Z-score brackets or when using linear models.

\paragraph{CLIP Percentile rank rescaling:} Results were consistent with those for Z-score rescaled cosine similarity.

\paragraph{DreamSim No rescaling.} Results showed similar patterns to CLIP embedding cosine similarity. Higher Z-scores in previous prompts were associated with better performance in subsequent attempts. The difference between the top Z-score bracket and both the bottom and middle brackets was statistically significant, with the top bracket showing higher performance in the next attempt. However, the difference between the bottom and middle brackets was not significant.

As with cosine similarity, linear models did not yield statistically significant relationships between Z-scores and subsequent performance, likely due to non-linearity. At the user level, we found no statistically significant relationships between the first prompt's Z-score and performance in subsequent attempts.

\paragraph{DreamSim Z-score rescaling.} Results closely matched those for unscaled DreamSim scores, with statistically significant differences between the top bracket and both the bottom and middle brackets, but no significant difference between the bottom and middle brackets. No significant effects emerged from linear models or from user-level analyses.

\paragraph{DreamSim Percentile rank rescaling.} Results aligned with those for Z-score rescaled DreamSim, with one exception: at the user-attempt level, we observed a statistically significant difference only between the top and bottom brackets, not between the top and middle brackets.

\paragraph{Prompt Length.} We found statistically significant evidence that higher Z-scores led to longer subsequent prompts. When using the trichotomous variable, the difference between the top and bottom Z-score brackets was significant, with longer prompts on average in the top bracket. However, we did not observe significant differences between the middle bracket and either the top or bottom brackets. Linear models at the user-attempt level showed a positive and statistically significant relationship, with a one-unit increase in Z-score associated with prompts approximately 0.3 words longer on average. At the user level, the Z-score of the first prompt did not significantly affect the length of subsequent prompts.

\paragraph{Successive Similarity.} Higher Z-scores were associated with increased similarity between consecutive prompts. The top Z-score bracket showed significantly higher prompt similarity compared to both the bottom and middle brackets, though no significant difference emerged between the bottom and middle brackets. Linear models confirmed a positive and statistically significant relationship between Z-scores and successive prompt similarity. No significant effects were found at the user level.

\paragraph{Successive Prompt Token Sort Ratio.} Results mirrored those for successive cosine similarity, with higher Z-scores associated with increased token sort ratios between successive prompts. This pattern was consistent across both trichotomous variable comparisons and linear models. No significant effects emerged at the user level.

\paragraph{Successive prompt 'contains previous prompt' dummy.} Higher Z-scores significantly increased the likelihood that subsequent prompts would contain the previous prompt as an exact substring. The difference between top and bottom Z-score brackets was significant, though differences involving the middle bracket were not. Linear models at the user-attempt level confirmed a positive and statistically significant relationship. No significant effects were found at the user level.

\paragraph{Interpretation}
These findings reveal a consistent pattern: when the stochastic nature of generative AI produces unexpectedly high-quality outputs for a given prompt, users tend to build more directly upon that prompt in their subsequent attempts—writing longer prompts that are more similar to and often directly incorporate the previous prompt. This adaptive behavior leads to better performance in subsequent attempts. However, this effect appears to be local rather than global; the quality of the first prompt affects the immediate next attempt but does not significantly influence overall performance or prompting behavior across all subsequent attempts.

This pattern suggests that users are sensitive to and learn from random variation in model outputs, even when they have no way of knowing whether a particularly good or bad result stems from the quality of their prompt or from model stochasticity. The tendency to build upon prompts that produce better-than-expected results represents an intuitive but potentially suboptimal learning strategy, as it may attribute too much importance to random variations rather than focusing on the underlying quality of the prompt itself.

\paragraph*{H5.} \textit{Average prompt engineering ability (as measured through metrics such as average expected prompt quality, initial expected prompt quality, and max expected prompt quality) and prompting strategies will depend on the capacity of the model that participants are interacting with.}

\paragraph{Empirical approach:} To test this hypothesis, we conducted pairwise comparisons across the three treatment arms using two-sample t-tests to evaluate differences in both performance outcomes and prompting behaviors. For robustness, we also performed ANOVA tests to examine whether any significant differences existed across all three treatment arms simultaneously. P-values were adjusted for multiple testing using the Benjamini-Hochberg procedure.

\textbf{Results:} Our analyses revealed several key differences in both performance and prompting behavior across the different model conditions:

\paragraph{CLIP Z-score rescaling.} We found statistically significant evidence for performance differences across treatment arms when examining participants' first attempts, average performance across all attempts, and best attempts (all measured using taskwide Z-scores rather than task-iteration Z-scores). For first attempt performance, DALL-E 3 (Verbatim) participants significantly outperformed both DALL-E 3 (Revised) and DALL-E 2 participants, while no statistically significant difference emerged between DALL-E 3 (Revised) and DALL-E 2 participants. Average performance across all attempts showed the same pattern: DALL-E 3 (Verbatim) outperformed both other conditions, with no significant difference between DALL-E 3 (Revised) and DALL-E 2. When comparing participants' best attempts, we found statistically significant differences between all three treatment arms in a clear hierarchy: DALL-E 3 (Verbatim) produced better results than DALL-E 3 (Revised), which in turn produced better results than DALL-E 2. ANOVA tests for all three performance measures (first, average, and best attempts) confirmed significant effects of model assignment.

\paragraph{CLIP Percentile rank rescaling.} Results using percentile rank rescaling mirrored those found with Z-score rescaling. DALL-E 3 (Verbatim) participants outperformed both other conditions on first attempt performance and average performance, with no significant differences between DALL-E 3 (Revised) and DALL-E 2 for these metrics. For best attempt performance, all pairwise comparisons revealed significant differences, with DALL-E 3 (Verbatim) outperforming DALL-E 3 (Revised), which outperformed DALL-E 2. ANOVA tests confirmed significant effects of model assignment for all three performance measures.

\paragraph{DreamSim Z-score rescaling.} Results using DreamSim with Z-score rescaling showed the same pattern as CLIP embedding cosine similarity. DALL-E 3 (Verbatim) participants significantly outperformed both other conditions on first attempt and average performance, with no significant differences between DALL-E 3 (Revised) and DALL-E 2. For best attempts, we again found the same hierarchical pattern of performance differences across all three conditions. ANOVA tests confirmed significant effects across all performance measures.

\paragraph{DreamSim Percentile rank rescaling.} Results with percentile rank rescaling of DreamSim scores were consistent with the Z-score findings. DALL-E 3 (Verbatim) participants outperformed both other conditions on first and average performance, with no differences between DALL-E 3 (Revised) and DALL-E 2. For best attempts, all pairwise comparisons were significant, showing DALL-E 3 (Verbatim) > DALL-E 3 (Revised) > DALL-E 2. ANOVA tests confirmed significant effects across all performance measures.

\paragraph{Mean prompt Length.} We found statistically significant differences in prompt length across treatment arms. Participants using DALL-E 2 wrote significantly shorter prompts compared to both DALL-E 3 (Revised) and DALL-E 3 (Verbatim) groups. Interestingly, there was no significant difference in prompt length between the two DALL-E 3 variants. ANOVA results confirmed a significant effect of model version on prompt length.

\paragraph{Aggregate Similarity.} We found no statistically significant differences in the overall variability of prompts across the three treatment arms. ANOVA results confirmed no significant effect of model version on this measure, suggesting that participants explored similar breadths of prompt space regardless of the model they were using. These results contrast with our findings from the main analysis, where prompts written by DALL-E 3 participants were, on average, more consistent than those written by DALL-E 2 participants. This difference reflects statistical power and specification: the main analysis uses a higher-powered model with task and attempt fixed effects, whereas the aggregate-similarity analysis here relies on a simpler cross-sectional comparison.

\paragraph{Successive Similarity.} Analysis of the average cosine similarity between successive prompts revealed no statistically significant differences across treatment arms. ANOVA results confirmed no significant effect of model version on successive prompt similarity, indicating that the tendency to build upon or deviate from previous prompts was similar across models.

\paragraph{Successive prompt token sort ratio.} No statistically significant differences emerged across treatment arms. ANOVA results confirmed no significant effect of model version on this measure of textual similarity between successive prompts.

\paragraph{Successive prompt `contains previous prompt' dummy:} Examining the probability of a current prompt being a superset of the previous prompt showed no statistically significant differences across treatment arms. ANOVA results confirmed no significant effect of model version on this measure.

\paragraph{Interpretation.}
These findings paint an interesting picture of how model capacity influences both performance and prompting behavior. In terms of performance, DALL-E 3 (Verbatim) consistently outperformed the other conditions across all metrics and scaling approaches, demonstrating the clear advantage of the more advanced model when used without automated prompt revision. The DALL-E 3 (Revised) condition generally performed worse than DALL-E 3 (Verbatim) but better than DALL-E 2 on best attempt measures, suggesting that automated prompt revision partially degraded the benefits of the more capable model.

For prompting behavior, the primary difference observed was in prompt length: participants using either version of DALL-E 3 wrote significantly longer prompts than those using DALL-E 2. This suggests that users recognized and adapted to the increased capacity of the more advanced model by providing more detailed instructions. However, we found no significant differences across models in measures of prompt similarity or evolution, suggesting that while participants provided more detail when using more capable models, their overall strategies for iterating on prompts remained relatively consistent regardless of model assignment.

\paragraph*{H6.} \textit{Variability in participants' ability to prompt engineer effectively and prompting strategies will depend on the capacity of the model that participants are interacting with.}

\paragraph{Empirical approach:} To test this hypothesis, we conducted two types of analyses. First, we performed F-tests comparing the variance of participant performance and prompting behaviors between all three pairs of model conditions (DALL-E 2 vs. DALL-E 3 Revised, DALL-E 3 Revised vs. DALL-E 3 Verbatim, and DALL-E 2 vs. DALL-E 3 Verbatim). Second, we estimated quantile treatment effects (QTEs) between all three pairs of models on participant performance and prompting behaviors. We also visually inspected the QTE patterns to determine whether dispersion/"inequality" was being reduced or increased when participants used different models. For example, positive effects for low quantiles and negative/null effects for high quantiles would indicate inequality reduction.

\paragraph{F-tests.} We first present results (with BH Adjusted p-values) for DALL-E 3 (revised) vs. DALL-E 2:
\begin{enumerate}
    \item \textbf{Mean prompt length (words):} DALL-E 3 Revised showed significantly less variance than DALL-E 2 (ratio 0.5217, $p \leq 10^{-4}$).
    \item \textbf{Prompt embedding variance:} DALL-E 3 Revised also showed significantly less variance than DALL-E 2 (ratio 0.7123, $p = 2 \times 10^{-4}$).
    \item \textbf{Cosine similarity with target image, Z-Score:} DALL-E 3 Revised showed significantly more variance than DALL-E 2 (ratio 1.255, $p = 0.0159$).
    \item \textbf{Failed to reject null of no differences in variance:} Mean raw DreamSim score vs. target image (p = 0.1333), number of topical transitions (by token sort ratio) (p = 0.3294), mean token sort ratio with previous prompt (p = 0.6244), mean percentile rank of DreamSim score (p = 0.9035), mean cosine similarity to previous prompt (p = 0.8748), mean raw CosineSim score vs. target image (p = 0.8301), mean proportion of prompts containing previous prompt (p = 0.5461), Z-score of DreamSim score (p = 0.4016), number of topical transitions (cosine similarity) (p = 0.3505), and mean percentile rank of CosineSim (p = 0.0795).
\end{enumerate} 

We next present results (with BH Adjusted p-values) for DALL-E 3 (verbatim) vs. DALL-E 3 (revised):

\begin{enumerate}
    \item \textbf{Mean percentile rank of CosineSim:} DALL-E 3 (verbatim) showed significantly less variance than DALL-E 3 (revised) (ratio 0.7347, $p = 0.0009$).
    \item \textbf{Mean proportion of prompts containing previous prompt:} DALL-E 3 (verbatim) also showed significantly less variance (ratio 0.7352, $p = 0.0009$).
    \item \textbf{Mean percentile rank of DreamSim:} DALL-E 3 (verbatim) showed significantly less variance (ratio 0.7888, $p = 0.0141$).
    \item \textbf{Cosine similarity with target image, Z-Score:} DALL-E 3 (verbatim) showed significantly less variance (ratio 0.7961, $p = 0.0159$).
    \item \textbf{DreamSim with target image, Z-Score:} DALL-E 3 (verbatim) showed significantly less variance (ratio 0.8195, $p = 0.0377$).
    \item \textbf{Number of topical transitions (token sort ratio):} DALL-E 3 (verbatim) showed significantly more variance (ratio 1.2303, $p = 0.0301$).
    \item \textbf{Failed to reject null of no differences in variance:} Mean raw DreamSim score vs. target image (p = 0.1842), mean raw CosineSim score vs. target image (p = 0.5916), prompt embedding variance (p = 0.9771), mean cosine similarity to previous prompt (p = 0.9035), mean prompt length (words) (p = 0.6244), number of topical transitions (cosine similarity) (p = 0.5916), and mean token sort ratio with previous prompt (p = 0.5206).
\end{enumerate}

Finally, the F-tests for DALL-E 2 vs. DALL-E 3 (verbatim):\begin{enumerate}
    \item \textbf{Mean prompt length (words):} DALL-E 3 (verbatim) showed significantly less variance than DALL-E 2 (ratio 0.553, $p \leq 10^{-4}$).
    \item \textbf{Prompt embedding variance:} DALL-E 3 (verbatim) also showed significantly less variance (ratio 0.7157, $p = 2 \times 10^{-4}$).
    \item \textbf{Mean raw DreamSim score vs. target image:} DALL-E 3 (verbatim) showed significantly less variance (ratio 0.7501, $p = 0.0015$).
    \item \textbf{Mean proportion of prompts containing previous prompt:} DALL-E 3 (verbatim) showed significantly less variance (ratio 0.7904, $p = 0.0141$).
    \item \textbf{Mean percentile rank of DreamSim:} DALL-E 3 (verbatim) showed significantly less variance (ratio 0.8005, $p = 0.0159$).
    \item \textbf{Failed to reject null of no differences in variance:} Mean percentile rank of CosineSim (p = 0.1842), Z-score of DreamSim score (p = 0.3294), Mean Raw CosineSim (p = 0.8151), Z-score of CosineSim score with target image (p = 0.9906), mean token sort ratio with previous prompt (p = 0.8748), mean cosine similarity to previous prompt (p = 0.807), number of topical transitions (token sort ratio) (p = 0.3505), and number of topical transitions (cosine similarity) (p = 0.085).
\end{enumerate}

\paragraph{QTE Highlighted Results.} 
Through our quantile treatment effects analysis and visual inspection of QTE plots, we found clear evidence of dispersion reduction. The Z-score of cosine similarity with respect to the target image (calculated within task-iteration) for the DALL-E 3 Revised vs. DALL-E 2 comparison showed patterns consistent with inequality reduction, with larger positive effects at lower quantiles. We also found evidence of dispersion increase: Mean prompt length for all three pairwise model comparisons (DALL-E 3 Verbatim vs. DALL-E 2, DALL-E 3 Revised vs. DALL-E 2, and DALL-E 3 Verbatim vs. DALL-E 3 Revised) showed patterns consistent with increased inequality, with larger positive effects at higher quantiles. Similar patterns emerged for raw DreamSim performance in both DALL-E 3 Verbatim vs. DALL-E 2 and DALL-E 3 Revised vs. DALL-E 2 comparisons.

These findings present a nuanced picture of how model capacity affects variability in performance and prompting behaviors. For prompt length and embedding variance, the more advanced models (both DALL-E 3 variants) showed significantly less variance than DALL-E 2, suggesting more consistent prompting behaviors. For performance measures, the results were mixed: DALL-E 3 (Verbatim) generally showed less variance than DALL-E 3 (Revised), indicating more consistent performance across participants, while DALL-E 3 (Revised) showed higher variance in performance than DALL-E 2.

The QTE analysis suggests that the relationship between model capacity and performance inequality is complex. For some measures (e.g., Z-score of cosine similarity), more advanced models appeared to reduce inequality by disproportionately benefiting lower-performing users. For other measures (e.g., prompt length and raw DreamSim scores), more advanced models appeared to increase inequality by disproportionately benefiting higher-performing users.

\paragraph*{H7} \textit{As participants repeatedly try to complete a task with a given model, the quality of their attempts will increase, and the extent to which the quality increases varies as a function of model capacity.}

\paragraph{Empirical approach:} To evaluate how performance improves across successive attempts and whether this improvement varies by model, we conducted stratified two-sample tests comparing participants' initial scores with their best scores. These comparisons were performed both within each treatment arm (DALL-E 2, DALL-E 3 Verbatim, and DALL-E 3 Revised) and overall across all participants. This approach allowed us to determine both whether participants generally improved with practice and whether the rate of improvement differed across models with varying capabilities.

Our analyses revealed consistent performance improvements across all treatment arms and outcome measures.

\paragraph{CLIP No rescaling.} We found statistically significant improvements from initial to best performance both within each treatment arm and in the overall sample. This indicates that participants were able to improve their performance through iterative attempts regardless of which model they were using. 

\paragraph{CLIP Z-score rescaling.} Results mirrored those of the unscaled analysis, with statistically significant improvements from initial to best performance observed both within each treatment arm and overall. 

\paragraph{CLIP Percentile rank rescaling.} Consistent with other scaling approaches, we observed statistically significant improvements from initial to best performance both within each treatment arm and overall.

\paragraph{DreamSim No rescaling.} DreamSim results aligned with CLIP similarity findings, showing statistically significant improvements from initial to best performance both within each treatment arm and overall.

\paragraph{DreamSim Z-score rescaling.} We again found statistically significant improvements from initial to best performance both within each treatment arm and in the overall sample.

\paragraph{DreamSim Percentile rank rescaling.} Results were consistent with other analyses, showing statistically significant improvements from initial to best performance across all conditions.

\paragraph{Interpretation.}
These findings demonstrate that participants consistently improved their performance through iterative attempts across all model conditions and performance measures. Users were able to learn from feedback and refine their prompts to achieve better results over time, regardless of which model they were using. This learning effect was robust across different outcome measures and scaling approaches, highlighting the importance of iteration and feedback in developing effective prompting strategies.

Although our original hypothesis also posited that the rate of improvement might vary as a function of model capacity, we did not find consistent evidence for differences in improvement rates across models. This suggests that while more capable models yield better absolute performance (as demonstrated in H5), the relative improvement from first to best attempt was similar across model conditions. This finding highlights that the benefits of iterative refinement apply broadly across different model capabilities.

\paragraph*{H8} \textit{The extent to which participants can recreate images using models such as DALL-E 2/3 will vary across images.}

\paragraph{Empirical approach.} To evaluate this hypothesis, we employed two complementary approaches. First, we used GPT-4V (a multimodal generative AI model capable of processing both text and images) to generate optimal prompts for each target image. For each of our 15 target images, we instructed GPT-4V to "Write a DALL-E {2, 3} prompt to recreate this image verbatim as closely and as detailed as possible." We generated two such "AI prompts" per target image—one optimized for DALL-E 2 and one for DALL-E 3. We then submitted these AI-generated prompts to the respective models 20 times each, yielding 60 replicated images per target (as the DALL-E 3 prompts were sent to both DALL-E 3 variants). We measured the cosine similarity between CLIP embeddings of these generated images and their target images, then averaged these similarity scores to quantify GPT-4V's ability to generate effective replication prompts for each target.

Second, we analyzed human performance by measuring the similarity between all participant-generated images and their corresponding target images, averaging these similarities to determine which images were easier or harder for participants to replicate. This two-pronged approach allowed us to assess image difficulty from both AI and human perspectives.

\paragraph{CLIP Results.} When ranking target images by GPT-4V's ability to generate effective prompts, we found substantial variation across images:

\begin{multicols}{3}
\renewcommand{\labelenumi}{\textbf{\arabic{enumi}.}}
\begin{enumerate}[noitemsep]
\item Business Image \#3
\item Business Image \#5
\item Business Image \#1
\item Photography Image \#1
\item Design Image \#2
\item Business Image \#2
\item Photography Image \#5
\item Design Image \#4
\item Business Image \#4
\item Design Image \#1
\item Photography Image \#2
\item Design Image \#5
\item Design Image \#3
\item Photography Image \#3
\item Photography Image \#4
\end{enumerate}
\end{multicols}

The highest average cosine similarity (easiest image for GPT-4V to replicate) was $CoSim = 0.944$ for Business Image \#3, while the lowest score (hardest image to replicate) was $CoSim = 0.734$ for Photography Image \#4.

When ranking target images by participants' ability to generate effective prompts, we observed:

\begin{multicols}{3}
\renewcommand{\labelenumi}{\textbf{\arabic{enumi}.}}
\begin{enumerate}[noitemsep]
    \item Business Image \#3
    \item Business Image \#5
    \item Business Image \#1
    \item Photography Image \#5
    \item Design Image \#2
    \item Design Image \#4
    \item Business Image \#2
    \item Design Image \#5
    \item Design Image \#1
    \item Photography Image \#1
    \item Design Image \#3
    \item Photography Image \#2
    \item Photography Image \#3
    \item Photography Image \#4
    \item Business Image \#4
\end{enumerate}
\end{multicols}

The highest average cosine similarity (easiest image for participants to replicate) was $CoSim = 0.892$ for Business Image \#3, while the lowest score (hardest image to replicate) was $CoSim = 0.669$ for Business Image \#4.

\paragraph{DreamSim Results.}  When using DreamSim as our similarity metric (inverted as 1-DreamSim to create a similarity rather than distance measure), GPT-4V's prompt generation performance ranked as follows:

\begin{multicols}{3}
\renewcommand{\labelenumi}{\textbf{\arabic{enumi}.}}
\begin{enumerate}[noitemsep]
    \item Design Image \#3
    \item Business Image \#2
    \item Business Image \#4
    \item Design Image \#2
    \item Business Image \#1
    \item Business Image \#5
    \item Design Image \#4
    \item Business Image \#3
    \item Photography Image \#1
    \item Photography Image \#5
    \item Photography Image \#2
    \item Design Image \#1
    \item Design Image \#5
    \item Photography Image \#3
    \item Photography Image \#4
\end{enumerate}
\end{multicols}

The highest 1-DreamSim score (easiest image for GPT-4V to replicate) was $\tilde{D} = 0.75$ for Design Image \#3, while the lowest score (hardest image to replicate) was $\tilde{D} = 0.40$ for Photography Image \#4.

When ranking target images by participants' replication ability using DreamSim:

\begin{multicols}{3}
\renewcommand{\labelenumi}{\textbf{\arabic{enumi}.}}
\begin{enumerate}[noitemsep]
    \item Business Image \#3
    \item Business Image \#2
    \item Business Image \#5
    \item Business Image \#1
    \item Design Image \#2
    \item Design Image \#3
    \item Photography Image \#2
    \item Photography Image \#5
    \item Design Image \#5
    \item Design Image \#4
    \item Business Image \#4
    \item Photography Image \#1
    \item Photography Image \#4
    \item Design Image \#1
    \item Photography Image \#3
\end{enumerate}
\end{multicols}

The highest 1-DreamSim score (easiest image for participants to replicate) was $\tilde{D} = 0.575$ for Design Image \#3, while the lowest score (hardest image to replicate) was $\tilde{D} = 0.356$ for Photography Image \#4.

\paragraph{Interpretation}
These results clearly demonstrate substantial variation in image replicability across our target set. Business-related images were generally easier to replicate, while photography images tended to be more challenging. While the exact rankings varied somewhat between GPT-4V and human participants, and between similarity metrics, the overall pattern of relative difficulty remained fairly consistent. This variation in difficulty confirms the importance of our stratified analysis approach and underscores the need to consider image-specific characteristics when evaluating generative AI performance.

\paragraph*{Exploratory Analyses} 

Our pre-registration outlined several exploratory analyses that we planned to conduct beyond our primary hypotheses. Several of these exploratory analyses appear in our main text, particularly the replay analysis that decomposed performance improvements into model and prompting effects. The pre-registered exploratory analyses were described as follows:

\begin{quote}
``We plan to investigate whether differences in prompt engineering ability across demographic and other observed variables will vary depending on the complexity of the task, e.g., the difficulty of the image participants are being asked to replicate. We anticipate power for this analysis will be very low, so we chose to label it as an exploratory analysis rather than a pre-registered hypothesis. 

We anticipate that we may conduct additional analysis of the prompts submitted by participants (and how these prompts evolve over the course of a session). Furthermore, we might explore the tips that participants provide after completing the task on how to prompt engineer effectively.

We also may take original and revised prompts submitted to DALL-E 3 treatment arms and submit them to DALL-E 2 (and vice versa) to see how participants would have counterfactually performed under different treatment assignments than the one to which they were assigned.''
\end{quote}

As noted in the main text, the third exploratory analysis—submitting prompts from one model condition to another model—became central to our investigation of model versus prompting effects. This approach allowed us to isolate how much improvement came from the model's enhanced capabilities versus users adapting their prompting strategies to take advantage of those capabilities.

\subsection{Deviations From Pre-registration} \label{sec:deviations}

We report below all deviations from our pre-registered analysis plan. These deviations primarily resulted from statistical or methodological considerations that became apparent during data analysis, rather than from substantive changes to our research questions or hypotheses.

\paragraph{Statistical tests.} Our pre-registration specified t-tests and Mann-Whitney U tests for many hypotheses. However, this approach proved inappropriate for variables with multiple categories (which characterized most demographic traits). We therefore employed ANOVA and Kruskal-Wallis tests instead, which provide equivalent information for multi-category variables.

\paragraph{Z-score computation.} Our pre-registration was insufficiently precise about computing Z-score performance measures. As detailed in Section \ref{sec:zscoring}, we calculated Z-scores within image-attempt pairs to adjust for variations across target images and attempts, except when comparing performance across attempts, where Z-scores were computed within target images only. The pre-registration had anticipated computing Z-scores only within target images for all analyses.

\paragraph{Additional demographic variables.} We inadvertently omitted Education and Generative AI outlook from our pre-registered list of demographics for measuring task performance heterogeneity. Given their theoretical importance, we included these variables in our analyses despite this accidental omission.

\paragraph{Additional prompt exclusion criteria.} Beyond our pre-registered exclusion criteria, we removed additional prompts that did not appear to constitute "good-faith efforts" based on their text content (see Section \ref{sec:sample_construction} for details). Robustness checks confirmed that our results remain consistent when including these prompts.

\paragraph{T-tests vs. Z-tests.} For testing H5, we used t-tests rather than z-tests because they were easier to implement. As these tests are asymptotically equivalent, this change does not meaningfully affect our results.

\paragraph{Z-score outlier handling.} When testing H4, we observed that Z-scores did not follow a normal distribution and included extreme values (ranging from -21.9 to 7.1). To prevent these outliers from disproportionately influencing our linear models, we excluded observations with absolute Z-scores greater than 3 (2.65\% of user-attempt observations). Robustness checks including these outliers yielded nearly identical results, with minor differences in significance levels for some comparisons.

\paragraph{GPT-4V rescaling omission.} Our pre-registration included a rescaled version of performance metrics based on GPT-4V prompt quality. However, this rescaling would have amounted to subtracting a constant from each unscaled score, and since all our analyses adjusted for target image (either through post-stratification or as a covariate), results would have been identical to those using unscaled measures. We therefore omitted these redundant robustness checks.

\end{document}